\documentclass[manuscript,screen]{acmart}

\AtBeginDocument{%
  \providecommand\BibTeX{{%
    \normalfont B\kern-0.5em{\scshape i\kern-0.25em b}\kern-0.8em\TeX}}}

\usepackage{soul}
\usepackage{graphicx}
\usepackage{booktabs}
\usepackage{float}
\usepackage{overpic}
\usepackage{float}  
\usepackage{subfloat}
\usepackage{stfloats} 
\usepackage[skip=0.5\baselineskip]{caption}
\usepackage{bm}
\usepackage{amsmath}
\usepackage{booktabs}
\usepackage{multirow}
\usepackage{multicol}
\usepackage{enumitem}
\usepackage{subcaption}
\usepackage{threeparttable}
\usepackage{xspace}
\usepackage{color}
\usepackage[normalem]{ulem}
\usepackage{algorithmicx,algorithm}
\usepackage{algpseudocode}
\usepackage{algorithm,algpseudocode}
\usepackage{color, xspace}
\newcommand{\eg}[0]{\emph{e.g.,}\xspace}

\usepackage{color}
\usepackage{tikz}
\usepackage{xcolor}
\usepackage{xspace}
\usepackage{booktabs}
\usepackage{multirow}
\usepackage{makecell}
\usepackage[edges]{forest}

\setcopyright{acmlicensed}
\copyrightyear{2018}
\acmYear{2018}
\acmDOI{XXXXXXX.XXXXXXX}

\acmConference[Conference acronym 'XX]{Make sure to enter the correct
  conference title from your rights confirmation emai}{June 03--05,
  2018}{Woodstock, NY}
\acmISBN{978-1-4503-XXXX-X/18/06}

\begin{document}

\title{A Survey of Personalization: From RAG to Agent}


\author{Xiaopeng Li$^{*}$}
\affiliation{%
  \institution{City University of Hong Kong}
  \country{Hong Kong}
}
\email{xiaopli2-c@my.cityu.edu.hk}

\author{Pengyue Jia$^{*}$}
\affiliation{%
  \institution{City University of Hong Kong}
  \country{Hong Kong}
}
\email{jia.pengyue@my.cityu.edu.hk}

\author{Derong Xu}
\affiliation{%
  \institution{City University of Hong Kong}
  \country{Hong Kong}
}
\affiliation{%
  \institution{University of Science and Technology of China}
  \country{China}
}
\email{derongxu@mail.ustc.edu.cn}

\author{Yi Wen}
\affiliation{%
  \institution{City University of Hong Kong}
  \country{Hong Kong}
}
\email{yiwen23-c@my.cityu.edu.hk}

\author{Yingyi Zhang}
\affiliation{%
  \institution{City University of Hong Kong}
  \country{Hong Kong}
}
\affiliation{%
  \institution{Dalian University of Technology}
  \country{China}
}
\email{yingyizhang@mail.dlut.edu.cn}

\author{Wenlin Zhang}
\affiliation{%
  \institution{City University of Hong Kong}
  \country{Hong Kong}
}
\email{wl.z@my.cityu.edu.hk}

\author{Wanyu Wang}
\affiliation{%
  \institution{City University of Hong Kong}
  \country{Hong Kong}
}
\email{wanyuwang4-c@my.cityu.edu.hk}

\author{Yichao Wang}
\affiliation{%
  \institution{Noah's Ark Lab, Huawei}
  \country{China}
}
\email{wangyichao5@huawei.com}

\author{Zhaocheng Du}
\affiliation{%
  \institution{Noah's Ark Lab, Huawei}
  \country{China}
}
\email{zhaochengdu@huawei.com}

\author{Xiangyang Li}
\affiliation{%
  \institution{Noah's Ark Lab, Huawei}
  \country{China}
}
\email{lixiangyang34@huawei.com}

\author{Yong Liu}
\affiliation{%
  \institution{Noah's Ark Lab, Huawei}
  \country{Singapore}
}
\email{liu.yong6@huawei.com}

\author{Huifeng Guo}
\affiliation{%
  \institution{Noah's Ark Lab, Huawei}
  \country{China}
}
\email{huifeng.guo@huawei.com}

\author{Ruiming Tang$^\dagger$}
\affiliation{%
  \institution{Noah's Ark Lab, Huawei}
  \country{China}
}
\email{tangruiming@huawei.com}

\author{Xiangyu Zhao$^\dagger$}
\affiliation{%
  \institution{City University of Hong Kong}
  \country{Hong Kong}
}
\email{xianzhao@cityu.edu.hk}

\renewcommand{\shortauthors}{X. Li and P. Jia, et al.}

\begin{abstract}
Personalization has become an essential capability in modern AI systems, enabling customized interactions that align with individual user preferences, contexts, and goals. Recent research has increasingly concentrated on Retrieval-Augmented Generation (RAG) frameworks and their evolution into more advanced agent-based architectures within personalized settings to enhance user satisfaction. Building on this foundation, this survey systematically examines personalization across the three core stages of RAG: pre-retrieval, retrieval, and generation. Beyond RAG, we further extend its capabilities into the realm of Personalized LLM-based Agents, which enhance traditional RAG systems with agentic functionalities, including user understanding, personalized planning and execution, and dynamic generation. 
For both personalization in RAG and agent-based personalization, we provide formal definitions, conduct a comprehensive review of recent literature, and summarize key datasets and evaluation metrics. Additionally, we discuss fundamental challenges, limitations, and promising research directions in this evolving field. Relevant papers and resources are continuously updated at the Github Repo\footnote{\url{https://github.com/Applied-Machine-Learning-Lab/Awesome-Personalized-RAG-Agent}}.
\let\thefootnote\relax\footnotetext{* Equal contribution.}
\let\thefootnote\relax\footnotetext{$\dagger$ Corresponding authors.}
\end{abstract}



\received{20 February 2007}
\received[revised]{12 March 2009}
\received[accepted]{5 June 2009}

\maketitle

\section{Introduction}
Large Language Models (LLMs) have revolutionized AI-driven applications by enabling natural language understanding and generation at an unprecedented scale. However, these models often suffer from issues such as outdated responses and hallucinations, which severely hinder the accuracy of information generation. 
Retrieval-Augmented Generation (RAG) has emerged as a promising framework that integrates retrieved information from external corpora, such as external APIs~\cite{google,bing}, scientific repositories~\cite{arxiv,pubmed} or domain-specific databases~\cite{amazon_dataset, espn_dataset}, ensuring more knowledge-grounded and up-to-date outputs. 

\begin{figure}[t]
    \centering
    \includegraphics[width = 0.6\linewidth]{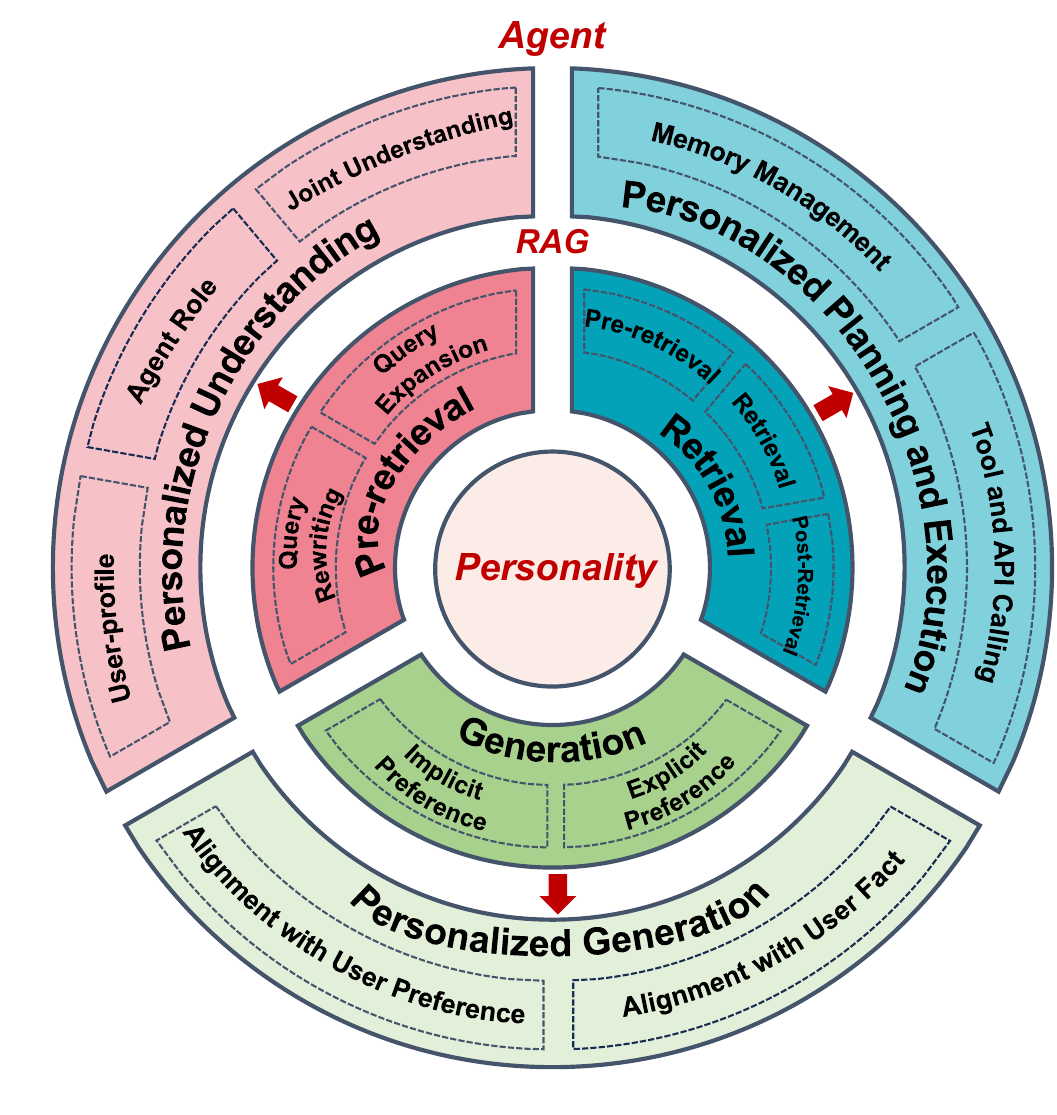}
    \caption{Correlation between personalization and RAG with agent flow.}
    \label{fig:structure}
\end{figure}

Its versatility has led to significant applications across various domains, including question answering~\cite{siriwardhana2023improving}, enterprise search~\cite{bulfamante2023generative} and healthcare~\cite{wu2024medical}, etc. Among these applications, one particularly notable area is in agent workflows, where RAG enhances autonomous systems by providing context-aware, dynamically retrieved, and reliable knowledge. This is because each stage of the RAG process closely mirrors key aspects of an agent’s workflow, as shown in Figure~\ref{fig:structure}. For instance, the query rewriting phase in RAG, which involves semantic understanding and parsing, aligns with the semantic comprehension stage in agent workflows. Likewise, RAG’s retrieval phase, which focuses on extracting the most relevant documents, corresponds to the planning and execution phases of an agent, where decisions are made based on retrieved knowledge. Finally, the generation phase in RAG parallels an agent’s execution stage, where actions are performed based on the given task. This structural alignment suggests that the architecture of RAG is fundamentally converging with agent workflows, solidifying its position as a key facilitator of intelligent and autonomous systems.

Although the structural alignment between RAG and agent workflows highlights their deepening convergence, a critical next step in enhancing these intelligent systems lies in personalization. Personalization is a key driver toward achieving more adaptive and context-aware AI, which is fundamental for the progression toward Artificial General Intelligence (AGI). It plays an essential role in applications such as personalized reasoning~\cite{xu2021transformer, henze2004reasoning}, adaptive decision-making~\cite{lu2011budgeted}, user-specific content generation~\cite{xu2025personalized, shaker2010towards}, and interactive AI systems~\cite{ma2021one, qian2021learning}. However, existing research lacks a comprehensive comparative analysis of personalized RAG and agentic approaches. Current surveys primarily focus on general RAG methodologies~\cite{gao2023retrieval, fan2024survey} or agent-related literature~\cite{li2024personal, wang2024survey, zhang2024survey}, without systematically exploring their implications for personalization. While recent works such as~\cite{zhang2024personalization, liu2025survey} discuss personalization, they predominantly address personalized generation within LLMs or specific downstream tasks, overlooking how personalization can be effectively integrated into RAG and agent workflows.

Motivated by the above issues, this survey aims to provide a comprehensive review of the integration of personalization into RAG and agentic RAG frameworks to enhance user experiences and optimize satisfaction. The key contributions of this work can be summarized as follows:
\begin{itemize}[leftmargin=*] 
\item We provide an extensive exploration of the existing literature on how personalization is integrated into various stages of RAG (pre-retrieval, retrieval, and generation) and agentic RAG (understanding, planning, execution, and generation).
\item We summarize the key datasets, benchmarks, and evaluation metrics used in existing research for each subtask to facilitate future studies in the respective domains. 
\item We also highlight the limitations of current research and suggest future directions for personalized RAG, emphasizing potential advancements to address existing challenges.
\end{itemize}

The outline of this survey is as follows: we introduce what is personalization (Sec.~\ref{sec:what}) and explain how personalization is adopted into RAG pipeline (Sec.~\ref{sec:how}). Then, we present a literature review on where to integrate personalization within different stages of RAG and agentic RAG workflows~(Sec.~\ref{sec:where}) and discuss the key datasets and evaluation metrics used in existing research~(Sec.\ref{sec:evaluation&dataset}). Lastly, we present a discussion on the limitations of current research and future directions~(Sec.~\ref{sec:futuredirection}).

\begin{table*}[t]
\caption{Overview of Personalized RAG and Agent.}
\centering
\resizebox{\textwidth}{!}{
\begin{tabular}{c|c|c|c} 
\toprule
\textbf{Field}                              & \textbf{Sub-field}                                                                                       & \textbf{Subsub-field}                                                                          & \textbf{Papers}                                                                                                                                                                                                                                                                                                                                                                                                                                                                                                                                                                                                                                                                                                                                                                                                                                            \\ 
\midrule
\multirow{7}{*}{\textbf{Pre-retrieval}}     & \multirow{4}{*}{\begin{tabular}[c]{@{}c@{}}Query\\Rewriting\end{tabular}}                       & \begin{tabular}[c]{@{}c@{}}Learning to\\Personalized Query Rewrite\end{tabular}          & CLE-QR \cite{li2022query}, CGF \cite{hao2022cgf}, PEARL \cite{mysore2023pearl}                                                                                                                                                                                                                                                                                                                                                                                                                                                                                                                                                                                                                                                                                                                 \\ 
\cmidrule{3-4}
                                   &                                                                                                 & \begin{tabular}[c]{@{}c@{}}LLM to\\Personalized Query Rewrite\end{tabular} & Least-to-Most Prompting \cite{zhou2022least}, ERAGent \cite{shi2024eragent}, CoPS \cite{zhou2024cognitive}, Agent4Ranking \cite{li2023agent4ranking}, FIG \cite{chen2023graph}, BASES \cite{ren2024bases}                                                                                                                                                                                                                                                                                                                                                                                                                                                                                                                                  \\ 
\cmidrule{2-4}
                                   & \multirow{3}{*}{\begin{tabular}[c]{@{}c@{}}Query\\Expansion\end{tabular}}                       & \begin{tabular}[c]{@{}c@{}}Tagging-based query\\expansion\end{tabular}                & Gossple~\cite{bertier2009toward}, ~\citet{biancalana2009social}, SoQuES~\cite{bouadjenek2011personalized}, ~\citet{zhou2012improving}                                                                                                                                                                                                                                                                                                                                                                                                                                                                                                                                                                         \\ 
\cmidrule{3-4}
                                   &                                                                                                 & Else                                                                                  & ~\citet{lin2006personalized}, ~\citet{bender2008exploiting}, Axiomatic PQEC~\cite{mulhem2016axiomatic}, WE-LM~\cite{wu2017personalized}, PSQE~\cite{bouadjenek2019personalized}, PQEWC~\cite{bassani2023personalized}                                                                                                                                                                                                                                                                                                                                                                                                                       \\ 
\cmidrule{2-4}
                                   & \multicolumn{2}{c|}{Others}                                                                                                                                                             & Bobo~\cite{gao2010utilizing}, ~\citet{kannadasan2019personalized}, PSQE~\cite{baumann2024psqe}                                                                                                                                                                                                                                                                                                                                                                                                                                                                                                                                                                                                                                                 \\ 
\midrule
\multirow{9}{*}{\textbf{Retrieval}}         & \multicolumn{2}{c|}{Indexing}                                                                                                                                                           & PEARL~\cite{mysore2023pearl}, KG-Retriever~\cite{chen2024kg}, EMG-RAG~\cite{wang2024crafting}, PGraphRAG~\cite{au2025personalized}                                                                                                                                                                                                                                                                                                                                                                                                                                                                                                                                                                                                                                                                                                                                                                                                     \\ 
\cmidrule{2-4}
                                   & \multirow{7}{*}{Retrieval}                                                                       & \begin{tabular}[c]{@{}c@{}}Dense\\Retrieval\end{tabular}                               & \begin{tabular}[c]{@{}c@{}}MeMemo \cite{wang2024mememo}, RECAP \cite{liu2023recap}, LAPDOG \cite{huang2024learning}, \citet{gu2021partner}, PersonaLM \cite{mathur2023personalm}, UIA \cite{zeng2023personalized}, XPERT \cite{vemuri2023personalized}, DPSR \cite{zhang2020towards}, \\RTM \cite{bi2021learning}, Pearl \cite{mysore2023pearl}, MemPrompt \cite{madaan2022memory}, ERRA \cite{cheng2023explainable}, MALP \cite{zhang2023llm}, USER-LLM \cite{ning2024user}, PER-PCS \cite{tan2024personalized}\end{tabular}                                                      \\ 
\cmidrule{3-4}
                                   &                                                                                                 & \begin{tabular}[c]{@{}c@{}}Sparse\\Retrieval\end{tabular}                              & OPPU \cite{tan2024democratizing}, PAG \cite{richardson2023integrating}, \citet{au2025personalized}, UniMS-RAG \cite{wang2024unims}, \citet{deng2022toward},                                                                                                                                                                                                                                                                                                                                                                                                                                                                                                                                                                                                  \\ 
\cmidrule{3-4}
                                   &                                                                                                 & \begin{tabular}[c]{@{}c@{}}Prompt-based\\Retrieval\end{tabular}                        & LAPS \cite{joko2024doing}, UniMP \cite{wei2024towards}, \citet{shen2024heart}                                                                                                                                                                                                                                                                                                                                                                                                                                                                                                                                                                                                                                                                                                                  \\ 
\cmidrule{3-4}
                                   &                                                                                                 & Others                                                                                & \citet{salemi2024optimization}, PersonalTM \cite{lian2023personaltm}, \citet{zhang2024personalized}                                                                                                                                                                                                                                                                                                                                                                                                                                                                                                                                                                                                                                                                                            \\ 
\cmidrule{2-4}
                                   & \multicolumn{2}{c|}{Post-retrieval}                                                                                                                                                      & PersonaRAG~\cite{zerhoudi2024personarag}, \citet{pavliukevich2024improving}, UniMS-RAG~\cite{wang2024unims}, \citet{salemi2024learning}, \citet{zhang2025rehearse}, AutoCompressors~\cite{chevalier2023adapting}, FIT-RAG~\cite{mao2024fit}                                                                                                                                                                                                                                                                                                                                                                                                                                                                                                                                                                                                                                                                                                                                                                \\ 
\midrule
\multirow{11}{*}{\textbf{Generation}}        & \multirow{6}{*}{\begin{tabular}[c]{@{}c@{}}Generation from\\Explicit Preferences\end{tabular}}  & \begin{tabular}[c]{@{}c@{}}Direct\\Prompting\end{tabular}                             & P$^2$~\cite{jiang2023evaluating}, Character Profiling~\cite{yuan2024evaluating}  OpinionQA~\cite{santurkar2023whose}, ~\citet{kang2023llms}, ~\citet{liu2023chatgpt}, Cue-CoT~\cite{wang2023cue}, TICL~\cite{cho2025tuning}                                                                                                                                                                                                                                                                                                                                                                           \\ 
\cmidrule{3-4}
                                   &                                                                                                 & \begin{tabular}[c]{@{}c@{}}Profile-Augmented\\Prompting\end{tabular}                  & GPG~\cite{zhang2024guided}, ~\citet{richardson2023integrating}, ONCE~\cite{liu2024once}, LLMTreeRec~\cite{zhang2025llmtreerec}, KAR~\cite{xi2024towards}, Matryoshka~\cite{li2024matryoshka}                                                                                                                                                                                                                                                                                                                                                                                                                                                \\ 
\cmidrule{3-4}
                                   &                                                                                                 & \begin{tabular}[c]{@{}c@{}}Personalized-Prompt\\Prompting\end{tabular}                & \citet{li2024learning}, RecGPT~\cite{zhang2024recgpt}, PEPLER-D~\cite{li2023personalized}, GRAPA~\cite{qu2024graph}, SGPT~\cite{deng2024unlocking}, PFCL~\cite{yu2024personalized}                                                                                                                                                                                                                                                                                                                                                                                                                                                                          \\ 
\cmidrule{2-4}
                                   & \multirow{4}{*}{\begin{tabular}[c]{@{}c@{}}Generation from \\Implicit Preferences\end{tabular}} & \begin{tabular}[c]{@{}c@{}}Fine-tuning-Based\\Methods\end{tabular}                    & \begin{tabular}[c]{@{}c@{}}PLoRA~\cite{zhang2024personalized}, LM-P~\cite{wozniak2024personalized}, MiLP~\cite{zhang2024personalized}, OPPU~\cite{tan2025democratizing}, PER-PCS~\cite{tan2024personalized}, Review-LLM~\cite{peng2024reviewllm},\\UserIdentifier~\cite{mireshghallah2021useridentifier}, UserAdapter~\cite{zhong2021useradapter}, HYDRA~\cite{zhuang2406hydra}, PocketLLM~\cite{peng2024pocketllm}, CoGenesis~\cite{zhang2024cogenesis}\end{tabular}  \\ 
\cmidrule{3-4}
                                   &                                                                                                 & \begin{tabular}[c]{@{}c@{}}Reinforcement\\Learning-Based\\Methods\end{tabular}        & P-RLHF~\cite{li2024personalized}, P-SOUPS~\cite{jang2023personalized}, PAD~\cite{chen2024pad}, REST-PG~\cite{salemi2025reasoning}, \citet{salemi2024optimization}, RewriterSlRl~\cite{li2024learning},\citet{kulkarni2024reinforcement}                                                                                                                                                                                                                                                                                                                                                                                                    \\ 
\midrule
\multirow{13}{*}{\textbf{From RAG to Agent}} & \multirow{6}{*}{\begin{tabular}[c]{@{}c@{}}Personalized\\Understanding\end{tabular}}            & \begin{tabular}[c]{@{}c@{}}In user-profile\\understanding\end{tabular}                & \citet{xu2024penetrative}, \citet{abbasian2023conversational},                                                                                                                                                                                                                                                                                                                                                                                                                                                                                                                                                                                                                                                                                                                                                  \\ 
\cmidrule{3-4}
                                   &                                                                                                 & \begin{tabular}[c]{@{}c@{}}In agent’s role\\understanding\end{tabular}                & RoleLLM~\cite{wang2023rolellm}, Character-LLM~\cite{shao2023character}, \citet{wang2023incharacter},                                                                                                                                                                                                                                                                                                                                                                                                                                                                                                                                                                                                                                                           \\ 
\cmidrule{3-4}
                                   &                                                                                                 & \begin{tabular}[c]{@{}c@{}}In agent’s user-role\\joint understanding\end{tabular}     & SocialBench \cite{chen2024socialbench}, \citet{dai2024mmrole}, \citet{ran2024capturing}, \citet{wang2023enabling}, \citet{tu2024charactereval}, Neeko \cite{yu2024neeko}                                                                                                                                                                                                                                                                                                                                                                                                                                                                                                                                                                    \\ 
\cmidrule{2-4}
                                   & \multirow{2}{*}{\begin{tabular}[c]{@{}c@{}}Personalized Planning\\and Execution\end{tabular}}   & \begin{tabular}[c]{@{}c@{}}Memory\\Management\end{tabular}                            & EMG-RAG \cite{wang2024crafting}, \citet{park2023generative}, \citet{abbasian2023conversational}, RecAgent \cite{wang2023user}, TravelPlanner+ \cite{singh2024personal}, PersonalWAB \cite{cai2025large}, VOYAGER \cite{wangvoyager}, MemoeryLLM \cite{wangmemoryllm}                                                                                                                                                                                                                                                                                                                                                                                                                                                                                                                                                                                      \\ 
\cmidrule{3-4}
                                   &                                                                                                 & Tool and API Calling                                                                  & VOYAGER \cite{wangvoyager}, \citet{zhangbootstrap}, PUMA \cite{cai2025large}, \citet{wang2023enabling}, PenetrativeAI \cite{xu2024penetrative}, \citet{huang2022language}, \cite{park2023generative}, MetaGPT \cite{hong2023metagpt}, OKR-Agent \cite{zheng2023agents}                                                                                                                                                                                                                                                                                                                                                                                                                                                                                                                                                                                                                                                                        \\ 
\cmidrule{2-4}
                                   & \multirow{4}{*}{\begin{tabular}[c]{@{}c@{}}Personalized\\Generation\end{tabular}}               & \begin{tabular}[c]{@{}c@{}}Alignment with \\User Fact\end{tabular}                   & Character-LLM \cite{shao2023character}, \citet{wang2024investigating}, \citet{dai2024mmrole}                                                                                                                                                                                                                                                                                                                                                                                                                                                                                                                                                                                                                                                                                                   \\ 
\cmidrule{3-4}
                                   &                                                                                                 & \begin{tabular}[c]{@{}c@{}}Alignment with User\\Preferences\end{tabular}              & \citet{wang2023rolellm}, \citet{ran2024capturing}, \citet{wang2023incharacter}, \citet{chen2024socialbench}                                                                                                                                                                                                                                                                                                                                                                                                                                                                                                                                                                                                                                                                   \\
\bottomrule
\end{tabular}}
\end{table*}

\section{What is Personalization} \label{sec:what}
Personalization in current research refers to the tailoring of model predictions or generated content to align with an individual's preferences. In the context of RAG and agents, personalization involves incorporating user-specific information at various stages of the RAG pipeline or within agents. User personalization can be categorized into the following types:

\begin{itemize}[leftmargin=*] 
\item Explicit User Profile: Explicitly presented user information, including biographical details, attributes (\eg age, location, gender, education), and social connections (\eg social networks).
\item User Historical Interactions: Behavioral data, including browsing history, clicks, and purchases, which help infer user interests and preferences to improve personalization. 
\item User Historical Content: Implicit personalization derived from user-generated content, such as chat history, emails, reviews, and social media interactions. 
\item Persona-Based User Simulation: The use of LLM-based agents to simulate and generate personalized interactions.
\end{itemize}

Integrating this personalized information at various stages of the RAG and agent workflows enables dynamic alignment with human preferences, thereby making responses more user-centric and adaptive.
\section{How to Adopt Personalization} \label{sec:how}
We define the process of introducing personalization within the RAG pipeline as follows:
\begin{equation}\label{equ:definition}
g = \mathcal{G} \left( \mathcal{R}\left(\mathcal{Q}\left(q,p\right),\mathcal{C},p\right),\text{prompt},p,\theta \right)
\end{equation}
where $p$ denotes personalized information, and the process unfolds in three steps. In the \textbf{pre-retrieval phase}, query processing ($\mathcal{Q}$) refines the query $q$ using personalized information, such as through query rewriting or expansion. During the \textbf{retrieval phase}, the retriever ($\mathcal{R}$) leverages $p$ to fetch relevant documents from the corpus ($\mathcal{C}$). Finally, in the \textbf{generation phase}, the retrieved information, combined with $p$ and structured using the given prompt, id fed into the generator ($\mathcal{G}$) with parameter $\theta$ to produce the final response $g$. It is evident that personalized information directly influences multiple stages of the RAG pipeline. In this survey, we consider the agent system as a specialized application of the RAG framework, where personalization is incorporated in a manner similar to the RAG framework.
\section{Where to Adopt Personalization} \label{sec:where}

\subsection{Pre-retrieval} \label{sec:Pre-retrieval}

\subsubsection{\textbf{Definition}} 
\begin{figure}[t]
    \centering
    \includegraphics[width=1\linewidth]{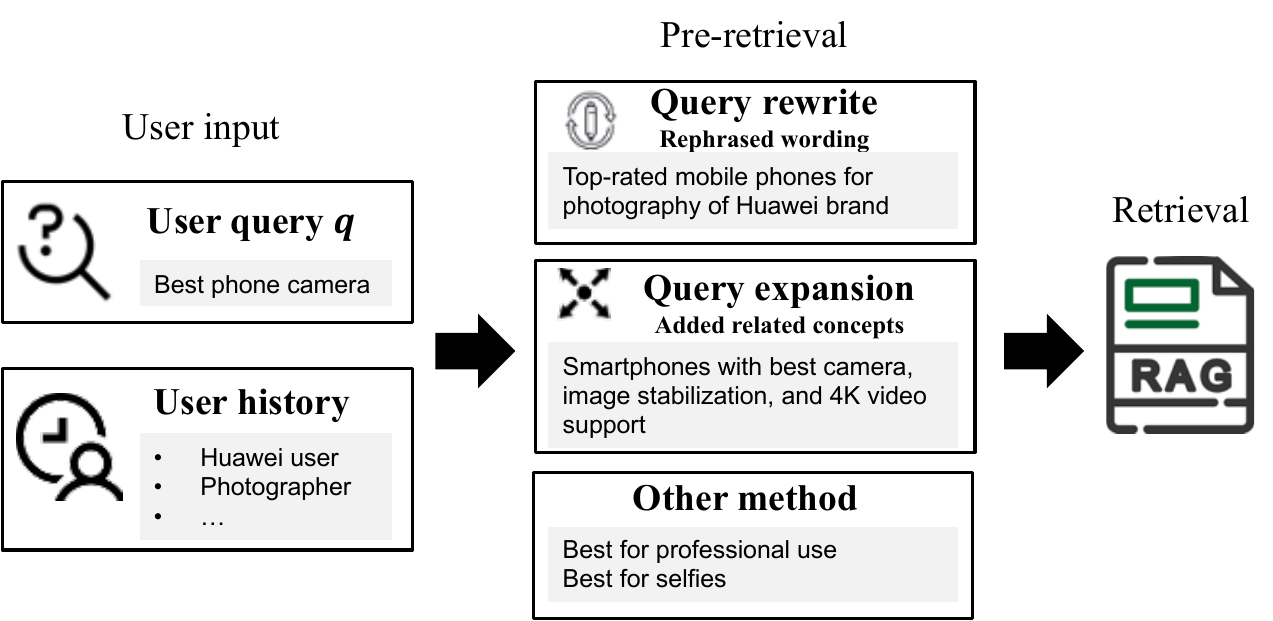}
    \caption{Overview of the personalized pre-retrieval stage.}
    \label{fig:pre_rag}
\end{figure}

Pre-retrieval is a crucial step in information retrieval systems, where the original user query is enhanced or modified before the retrieval process to improve the relevance and quality of the search results, as shown in Figure~\ref{fig:pre_rag}. This process often incorporates additional contextual or personalized information to better align the query with the user's intent. The process can be formalized as follows:
\begin{equation}
    q^{*} = \mathcal{Q}\left(q,p\right)
\end{equation}
where $p$ and $q$ denote the personalized information and original query, and $q^{*}$ is the optimized query after query reformulation.

\subsubsection{\textbf{Query Rewriting}}
Query rewriting in RAG at the pre-retrieval stage refers to the process of reformulating user queries to enhance retrieval effectiveness by improving relevance, disambiguating intent, or incorporating contextual information before retrieving documents from an external knowledge source.
The literature on personalized query rewriting can be broadly classified into two primary categories: (1) Direct Personalized Query Rewriting and (2) Auxiliary Personalized Query Rewriting.

\paragraph{\textbf{\textit{{(1). Direct Personalized Query Rewriting.}}}}

The first category focuses on personalized query rewriting by using direct models. 
For example, \citet{cho2021personalized} 
presents a personalized search-based query rewrite system for conversational AI that addresses user-specific semantic and phonetic errors. 
\citet{nguyen2025rl} apply reinforcement learning techniques to improve query rewriting in online e-commerce systems, leveraging distilled LLMs for personalized performance. 
CLE-QR~\cite{li2022query} explores query rewriting in Taobao's search engine to enhance user satisfaction through customized query adaptation. 
CGF~\cite{hao2022cgf} introduces a constrained generation framework that allows for more flexible and personalized query rewriting in conversational AI. 
\citet{li2024learning} investigate learning methods to rewrite prompts for personalized text generation, improving the relevance and engagement of AI-generated content. 
Additionally, PEARL~\cite{mysore2023pearl} discusses personalizing large language model-based writing assistants through the integration of generation-calibrated retrievers, enhancing AI-generated content.

\paragraph{\textbf{\textit{{(2). Auxiliary Personalized Query Rewriting.}}}} The second category emphasizes personalized query rewriting by using auxiliary mechanisms, such as retrieval, reasoning strategies, and external memory. 
\citet{zhou2022least} propose a least-to-most prompting strategy that aids in complex reasoning within LLMs, which can be adapted for personalized text generation.
ERAGent~\cite{shi2024eragent} enhance retrieval-augmented LLMs to improve personalization, efficiency, and accuracy, indirectly supporting personalized query rewriting for content generation. 
CoPS~\cite{zhou2024cognitive} integrate LLMs with memory mechanisms to create more personalized search experiences, which also influences content generation through better query understanding.
Further,  Agent4Ranking~\cite{li2023agent4ranking} employs multi-agent LLMs to perform semantic robust ranking, including personalized query rewriting to improve search rankings. 
FIG~\cite{chen2023graph} combine graph-based methods with LLMs to query rewrite, improving personalized content generation and conversational interactions. 
Lastly, BASES~\cite{ren2024bases} employ LLM-based agents to simulate large-scale web search user interactions, contributing to the development of personalized query rewriting strategies for content generation.

\subsubsection{\textbf{Query Expansion}}
Query expansion enhances retrieval systems by expanding a user’s original query with additional terms, synonyms, or refined structure to better capture intent. 
This improves the relevance and scope of retrieved documents. Recent advancements in LLMs have reinvigorated this field~\cite{wang2023query2doc,jagerman2023query,jia2024mill}, leveraging their comprehension and generation abilities to expand queries using encoded knowledge or external retrieval, with notable success. Personalized query expansion, a subset, incorporates user-specific data to tailor results, boosting performance and customizing the search experience.

\paragraph{\textbf{\textit{{(1). Tagging-based Query Expansion.}}}}

By 2009, studies began incorporating tagging information to enhance personalized query expansion. For instance, Gossple~\cite{bertier2009toward} introduced the TagMap and TagRank algorithms, which dynamically selected tags from personalized networks constructed using the cosine similarity of user-item tag distances, improving recall performance. Similarly,~\citet{biancalana2009social} recorded user queries and visited URLs, leveraging social bookmarking to extract relevant tags and build a personalized three-dimensional co-occurrence matrix. Based on this, multiple semantically categorized expanded queries were generated to better reflect user interests. Further advancements include SoQuES~\cite{bouadjenek2011personalized}, which integrated tag semantic similarity with social proximity, and a graph-based approach~\cite{zhou2012improving} that utilized Tag-Topic models and pseudo-relevance feedback for term weighting, tailoring the expansion process to individual user preferences.

\paragraph{\textbf{\textit{{(2). Else.}}}}
Apart from tagging-based techniques, early research on Personalized Query Expansion primarily focused on modeling user personalization based on search history~\cite{lin2006personalized}, social networks, or preferences derived from friendship networks~\cite{bender2008exploiting}. The Axiomatic PQEC framework~\cite{mulhem2016axiomatic} formalized expansion rules using both local (user behavior-driven) and social (network-driven) strategies. In 2017, WE-LM~\cite{wu2017personalized} advanced this paradigm by modeling multi-relational networks with word embeddings across tag-word relationships, refining associations through affinity graphs. Later, PSQE~\cite{bouadjenek2019personalized} further improved tagging-based methods using utf-iuf user profiling, integrating a tag similarity graph with user profiles in the online phase to compute expansion terms relevant to user interests in real-time, achieving dynamic personalized expansion. In addition, PQEWC~\cite{bassani2023personalized} leveraged clustering and contextual word embeddings to optimize query expansions dynamically.

\subsubsection{\textbf{Others}}

Besides query rewriting and query expansion, other personalized query-related research focuses on areas like query disambiguation and query auto-completion~\cite{song2024survey}. Bobo~\cite{gao2010utilizing} allows users to input contextual terms reflecting their domain knowledge. In 2019, a method~\cite{kannadasan2019personalized} applied fastText embeddings from recent queries to rank candidates. In addition, PSQE~\cite{baumann2024psqe} employed synthetic user profiles from Wikipedia and word2vec embeddings for query disambiguation.

\subsubsection{\textbf{Discussion}}
While both query rewriting and query expansion aim to align user input with system understanding to enhance retrieval quality, their roles in personalization differ in fundamental ways.Understanding the distinct operational characteristics and application scenarios of each technique is essential for designing effective personalized retrieval systems. The key takeaways are listed as follows: 
\begin{itemize}[leftmargin=*] 
    \item Query rewriting is most beneficial when the original query is \textbf{ambiguous}, underspecified, or misaligned with retrieval intents, particularly in conversational or multi-turn settings. 
    \item Query expansion is most effective when the original query is \textbf{relevant} but incomplete — i.e., when it needs to be semantically broadened to cover additional relevant concepts.
\end{itemize}

\subsection{Retrieval} \label{sec:Retrieval}
\begin{figure}[t]
    \centering
    \includegraphics[width=\linewidth]{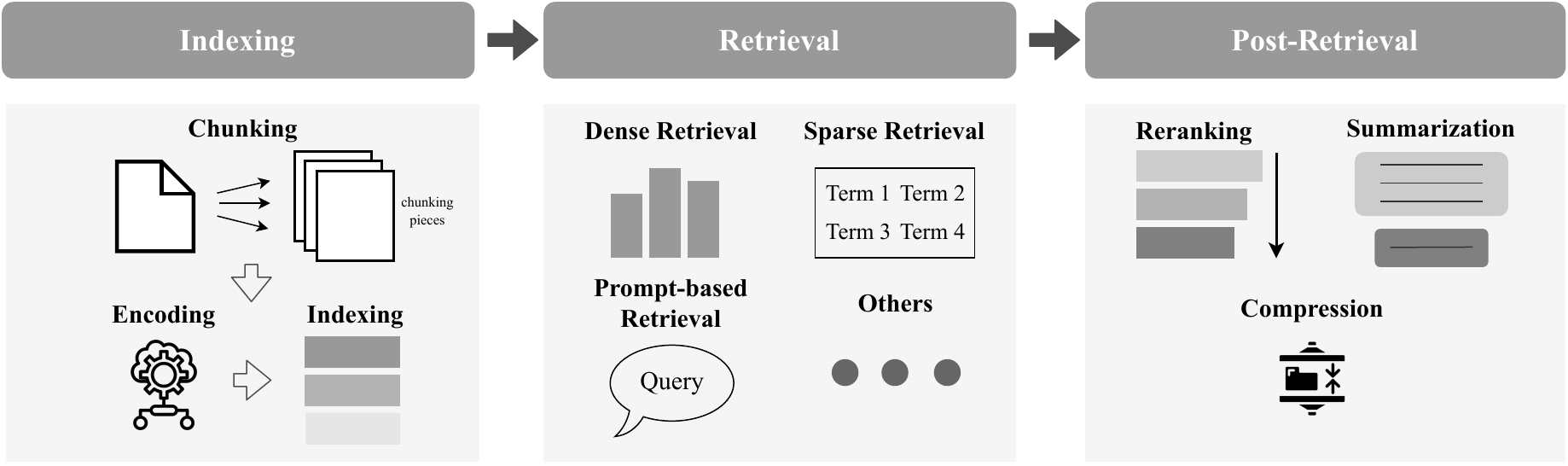}
    \caption{Overview of the personalized retrieval stage.}
    \label{fig:retrieval}
\end{figure}
\subsubsection{\textbf{Definition}}

The retrieval process involves finding the most relevant documents $D^*$ from a corpus $\mathcal{C}$ based on a query $q^*$, as shown in Figure~\ref{fig:retrieval}. To incorporate personalization, additional user-specific information $p$ is integrated into the retrieval function  $\mathcal{R}$. This allows the retrieval process to tailor the selected documents to align with individual user preferences or contexts, thereby enhancing the relevance and personalization of the generated outputs.
\begin{equation}
D^* =  \mathcal{R}\left(q^*,\mathcal{C},p\right)
\end{equation}

In the retrieval process, personalization can primarily be introduced by focusing on three steps: indexing, retrieval, and post-retrieval. These steps ensure efficient and accurate retrieval of relevant documents or knowledge, while tailoring the process to individual user preferences. Below, we provide a detailed explanation of each step.

\subsubsection{\textbf{Indexing}}
Indexing organizes knowledge base data into a structured format to facilitate efficient retrieval. Within the RAG pipeline, documents are either chunked or entirely encoded into representations before being integrated into searchable systems~\cite{douze2024faiss, annoy}. Conventional encoding methods employ either sparse encoding techniques (\eg TF-IDF~\cite{rajaraman2011mining}, BM25~\cite{robertson2009probabilistic}) or dense encoding approaches leveraging pre-trained models, such as BERT~\cite{koroteev2021bert}, Siamese Encoders~\cite{reimers2019sentence}, or LLM-based encoders~\cite{li2023towards, xiao2024c}.

To introduce personalization at the indexing stage, PEARL~\cite{mysore2023pearl} generates user embeddings by encoding personal history data with models like DeBERTa. These embeddings are subsequently clustered to create personalized shared indices. Other approaches integrate knowledge graphs into indexing to enhance retrieval performance. For example, KG-Retriever~\cite{chen2024kg} employs a Hierarchical Index Graph, consisting of a knowledge graph layer and a collaborative document layer, to improve RAG retrieval. EMG-RAG~\cite{wang2024crafting} incorporates personalized memory within an editable knowledge graph, enabling dynamic retrieval. Similarly, PGraphRAG~\cite{au2025personalized} leverages user-centric knowledge graphs to enhance personalization in retrieval tasks.

\subsubsection{\textbf{Retrieval}}
The Retrieval step matches a user query with the indexed knowledge base to fetch relevant candidates. It can be broadly categorized into four different types: (1) Dense Retrieval,  (2) Sparse Retrieval, (3) Prompt-based Retrieval, and  (4) Others.

\paragraph{\textbf{\textit{{(1). Dense Retrieval.}}}}
Dense retrieval methods often use vector embeddings and similarity metrics (\eg cosine similarity) and achieve personalization by encoding user preferences, context, or interactions into query or document embeddings, enabling tailored results through similarity-based matching. For instance, MeMemo \cite{wang2024mememo} retrieves personalized information by matching user-specific embeddings with document vectors, focusing on private, on-device text generation. Similarly, RECAP \cite{liu2023recap} and LAPDOG \cite{huang2024learning} enhance personalized dialogue generation by encoding queries and user profiles as dense vectors and retrieving top-N results, ensuring user-specific context drives the responses. In chatbots, \citet{gu2021partner} integrates conversational context and user profiles to align retrieved responses with user personas. PersonaLM \cite{mathur2023personalm} employs group-wise contrastive learning, training its retrieval model to align user queries with domain-specific text fragments, thereby improving personalization. UIA \cite{zeng2023personalized} employs dual encoders to retrieve documents tailored to user preferences. XPERT \cite{vemuri2023personalized} incorporates temporal events and user interactions into embeddings, enabling large-scale retrieval across millions of items.

Dense retrieval also enhances specific applications like e-commerce, medical assistance, and language models. DPSR \cite{zhang2020towards} and RTM \cite{bi2021learning} encode user queries and product information to personalize product searches dynamically. Pearl \cite{mysore2023pearl} and MemPrompt \cite{madaan2022memory} retrieve personalized content by leveraging historical user data and memory-assisted mechanisms. ERRA \cite{cheng2023explainable} uses review embeddings as dense queries for recommendations. In medical assistance, MALP \cite{zhang2023llm} and USER-LLM \cite{ning2024user} integrate short- and long-term user interactions into embeddings for contextualized, personalized responses. Finally, PER-PCS \cite{tan2024personalized} retrieves relevant information using individual user histories, enhancing the personalization capabilities of large language models.  

\paragraph{\textbf{\textit{{(2). Sparse Retrieval.}}}}
Sparse retrieval methods often rely on term-based matching (\eg BM25) and apply personalization by assigning higher weights to terms or keywords that are more relevant to the user.  OPPU \cite{tan2024democratizing} uses the BM25 algorithm to select the k most relevant records from the user's historical data for the current query. Similarly, PAG \cite{richardson2023integrating} incorporates user input and profiles to enhance summarization and retrieval, aligning sparse representations with personalization objectives for large language models. \citet{au2025personalized} uses BM25 search algorithms to find entries related to the target user or neighboring users through the graph structure. UniMS-RAG \cite{wang2024unims} combines sparse and dense retrieval by leveraging multi-source knowledge, such as dialogue context and user images, to refine personalized responses in dialogue systems. Lastly, \citet{deng2022toward} apply sparse retrieval to support fact-based queries, considering user queries and preferences to enhance answer generation for e-commerce applications. 

\paragraph{\textbf{\textit{{(3). Prompt-based Retrieval.}}}}
Prompt-based retrieval leverages prompts to guide retrieval from the model or external sources and introduces personalization by crafting user-specific prompts that guide the retrieval process. These prompts may include explicit user preferences, historical interactions, or detailed instructions that reflect the user’s unique requirements. By embedding this personalized context directly into the prompt, the retrieval process can dynamically adjust to capture and return results that are most relevant to the user. LAPS \cite{joko2024doing} focuses on multi-session conversational search by storing user preferences and dialogue context, then using prompts to retrieve relevant information tailored to the user's biases and categories of interest. UniMP \cite{wei2024towards} employs user interaction histories as input to prompt-based retrieval, enabling personalized recommendations for multi-modal tasks, such as vision-language applications, by aligning prompts with user behavioral data. In contrast, \citet{shen2024heart} explores the use of LLMs to extract empathy and narrative styles from user-provided stories, but this work primarily focuses on style extraction and does not explicitly involve a retrieval component. 

\paragraph{\textbf{\textit{{(4). Others.}}}}
Reinforcement learning-based retrieval personalizes the process by optimizing retrieval policies based on user feedback, learning user preferences over time to adjust strategies. \citet{salemi2024optimization} combines models like BM25, RbR, and dense retrieval, refining them with reinforcement learning (RL) and knowledge distillation (KD) to adapt to user profiles for personalized outputs. Parameter-based retrieval leverages pre-trained model parameters to implicitly store and retrieve user-specific information, allowing direct retrieval from the model without traditional indices. PersonalTM \cite{lian2023personaltm} generates document identifiers (Document IDs) using a Transformer model, encoding query, history, and document relationships into its parameters for personalization. Similarly, \citet{zhang2024personalized} uses parameterized representations to integrate user queries and histories, tailoring responses to individual preferences.

\subsubsection{\textbf{Post-retrieval}}
Current Post-Retrieval methods primarily focus on refining retrieved documents or responses to improve relevance and coherence, current methodologies could be categorized into three parts (1) Re-ranking, (2) Summarization, and (3) Compression.

\paragraph{\textbf{\textit{{(1). Re-ranking.}}}}
Re-ranking enhances personalized content generation by prioritizing more relevant documents at the top. PersonaRAG~\cite{zerhoudi2024personarag} extends RAG by integrating user-centric agents, such as the Live Session Agent and the Document Ranking Agent, to refine document ranking and improve overall performance. \citet{pavliukevich2024improving} propose a cross-encoder BERT model for re-ranking external knowledge within a personalized context. UniMS-RAG~\cite{wang2024unims} introduces a scoring mechanism that evaluates retrieved documents and outputs by optimizing the retriever. Besides, it includes an evidence attention mask, enabling re-ranking during inference and applying it to personalized datasets. ~\citet{salemi2024learning} present an iterative approach to optimizing ranking results based on the expectation-maximization algorithm, with performance validated in personalized scenarios.

\paragraph{\textbf{\textit{{(2). Summarization.}}}}
Summarization refers to the process of summarizing retrieved information to enhance performance. For instance,~\citet{zhang2025rehearse} introduced a role-playing agent system to summarize retrieved history in order to improve the final Personalized Opinion Summarization process.

\paragraph{\textbf{\textit{{(3). Compression.}}}}
Compression involves condensing embeddings or retrieved content to enhance efficiency and effectiveness. Approaches like AutoCompressor~\cite{chevalier2023adapting} compress contextual embeddings into shorter semantic representations, and FIT-RAG~\cite{mao2024fit} introduces a self-knowledge recognizer along with a sub-document-level token reduction mechanism to minimize tokens within RAG pipeline. However, few studies have specifically explored personalized fields, highlighting a promising direction for future research.

\subsubsection{\textbf{Discussion}}
Indexing, retrieval, and post-retrieval methods each play a critical role in ensuring efficient and personalized information processing, with specific applications and trade-offs. Indexing focuses on organizing knowledge bases for efficient retrieval, using techniques such as sparse encoding methods like TF-IDF and BM25, which are efficient but limited in understanding semantics, and dense encoding methods like BERT and DeBERTa, which provide better semantic understanding but require significant computational resources. These methods are widely used in tasks like question answering and personalized recommendation systems. Retrieval involves matching user queries with relevant documents and can be categorized into dense retrieval, which provides high semantic understanding and personalization but is computationally expensive; sparse retrieval, which is efficient and interpretable but less capable of handling semantics; prompt-based retrieval, which is highly flexible and adaptable to user needs but requires careful engineering of prompts; and advanced methods like reinforcement learning-based approaches, which dynamically adapt to user feedback but are complex to implement. This step is essential in applications like personalized dialogue systems, search engines, and e-commerce. Post-retrieval methods refine retrieved results to enhance relevance and coherence through re-ranking, which improves personalization and prioritizes relevant content but increases computational overhead; summarization, which simplifies complex information for better user understanding but risks losing critical details; and compression, which reduces computational costs by condensing information but remains underexplored in personalized contexts. Together, these methods provide a comprehensive pipeline for delivering efficient, relevant, and personalized outputs, balancing their strengths in semantic understanding, relevance, and flexibility with challenges related to computational costs and implementation complexity.
\subsection{Generation}  \label{sec:Generation}
\subsubsection{\textbf{Definition}}
\begin{figure}[t]
    \centering
    \includegraphics[width=\linewidth]{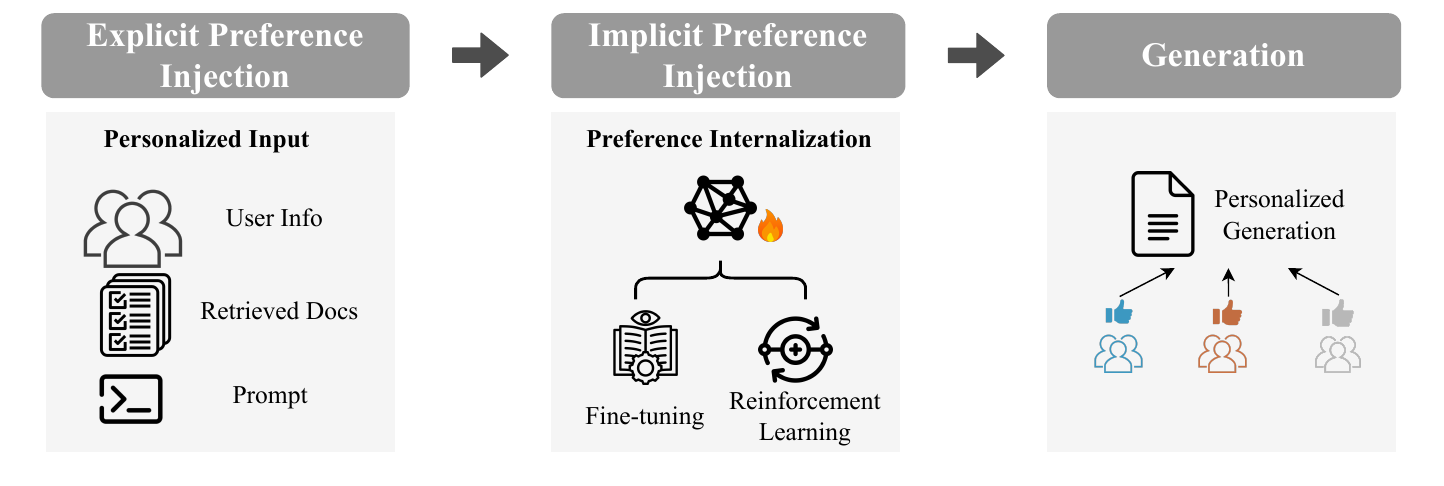}
    \caption{Overview of the personalized generation stage.}
    \label{fig:generation}
\end{figure}

Personalized generation incorporates user-specific retrieved documents $D^*$, task-specific prompt $prompt$, and user preference information $p$ via the generator $\mathcal{G}$ parameterized by $\theta$ to produce tailored content $g^*$ aligned with individual preference, where the flow is shown in Figure~\ref{fig:generation}. The generation process can be formulated as
\begin{equation}
g^* = \mathcal{G} \left(D^*,\text{prompt},p,\theta \right) .
\end{equation}
Personalized generation can be achieved by incorporating explicit and implicit preferences. Explicit preference-driven methodologies utilize direct input signals (\eg $D^*$, $\text{prompt}$, and $p$), to tailor outputs to specific user preferences. Conversely, implicit preference-encoded approaches embed personalized information within the parameters $\theta$ of the generator model, during training, thereby facilitating preference alignment without the necessity for explicit runtime inputs.

\subsubsection{\textbf{Generation from Explicit Preferences}}
Integrating explicit preferences into LLMs facilitates personalized content generation. Explicit preference information encompasses user demographic information (\eg age, occupation, gender, location), user behavior sequences (reflecting historical behavioral patterns), and user historical output texts (capturing writing style and tone preferences). The injection of explicit preferences for personalized generation can be categorized into three types: (1) Direct-integrated Prompting, (2) Summary-augmented Prompting, and (3) Adaptive Prompting.

\paragraph{\textbf{\textit{{(1). Direct-integrated Prompting.}}}}
Integrating user explicit preferences into language models through prompting enables the prediction of users' intent and behavioral patterns, facilitating personalized content generation. For instance, P$^2$~\cite{jiang2023evaluating}, Character Profiling~\cite{yuan2024evaluating}, and OpinionQA~\cite{santurkar2023whose} integrate personalized data into LLMs through prompting for role-playing task, thereby aligning the model's responses with specified user profiles. ~\citet{kang2023llms} and ~\citet{liu2023chatgpt} integrate interaction histories into LLMs via prompting to predict user rating for candidate items. Cue-CoT~\cite{wang2023cue} employs chain-of-thought reasoning to infer user needs from contextual cues, enabling personalized responses to in-depth dialogue questions. Additionally, TICL~\cite{cho2025tuning} proposes a trial-and-error framework that critiques initial LLM-generated responses, derives explanations and integrates these negative examples into prompts to improve personalization alignment.

\paragraph{\textbf{\textit{{(2). Summary-augmented Prompting.}}}}
Direct integration of personalized information via prompting struggles with ambiguous intent signals: Lengthy interaction histories introduce noise that obscures critical behavioral patterns~\cite{liu2024lost}, while sparse behavioral data lacks sufficient context for LLMs to derive meaningful user preferences. To address these issues, recent approaches focus on summarizing user personalized intents and integrating them into prompts. For instance, GPG~\cite{zhang2024guided} extracts key user habits and preferences from personal contexts, enabling fine-grained personalization. Similarly, LLMs are employed to generate task-specific summaries of user preferences, enhancing retrieval-augmented personalized generation capabilities~\cite{richardson2023integrating}. In recommendation systems, ONCE~\cite{liu2024once}, LLMTreeRec~\cite{zhang2025llmtreerec}, and KAR~\cite{xi2024towards} leverage historical user-item interactions to summarize user preferences. Furthermore, Matryoshka~\cite{li2024matryoshka} generates user preference summaries by dynamically retrieving and synthesizing historical data.

\paragraph{\textbf{\textit{{(3). Adaptive Prompting.}}}}
Manually designing personalized prompts demands both expert knowledge and significant labor, motivating the development of automated methods for personalized prompt generation. For example, \citet{li2024learning} trains a personalized prompt rewriter via supervised and reinforcement learning. RecGPT~\cite{zhang2024recgpt} and PEPLER-D~\cite{li2023personalized} leverage prompt tuning to generate personalized prompts, enhancing sequential and explainable recommendations, respectively. GRAPA~\cite{qu2024graph} integrates semantic and collaborative signals from user-item interaction graphs with graph neural networks to generate context-aware personalized prompts. SGPT~\cite{deng2024unlocking} employs prompt tuning
to jointly model common and group-specific patterns, bridging generalized and personalized federated learning paradigms. Furthermore, PFCL~\cite{yu2024personalized} achieves multi-granularity human preference modeling: coarse-grained prompts distill shared knowledge, while fine-grained prompts adapt to individual user characteristics.

\subsubsection{\textbf{Generation from Implicit Preferences}}
Unlike explicit preference modeling, which captures user preferences through textual input, implicit preference-based methods incorporate personalization through internal parameters. This personalization is achieved either through Parameter-Efficient Fine-tuning (PEFT) techniques, such as LoRA~\cite{hu2022lora}, or reinforcement learning-based approaches for preference alignment~\cite{li2024learning,chen2024pad}. Based on these strategies, we classify existing methods into two categories: (1) Fine-tuning-Based Methods and (2) Reinforcement Learning-Based Methods.

\paragraph{\textbf{\textit{{(1). Fine-tuning Based Methods.}}}}
For fine-tuning methods, LoRA is the most widely adopted since it is resource-efficient and enables rapid adaptation without compromising model performance. PLoRA~\cite{zhang2024personalized} introduces a personalized knowledge integration framework that combines task-specific LoRA with user-specific knowledge. Similarly, LM-P~\cite{wozniak2024personalized} personalizes information via LoRA by incorporating User ID as a personalization factor. MiLP~\cite{zhang2024personalized} employs Bayesian optimization to determine the optimal personalization injection configuration, including LoRA settings, to effectively capture and utilize user-specific information. OPPU~\cite{tan2025democratizing} and PER-PCS~\cite{tan2024personalized} follow a similar approach, leveraging user history data for fine-tuning LoRA-based personalization. However, PER-PCS differs by incorporating a gating module that selects the appropriate LoRA, enabling fine-grained personalization. Additionally, Review-LLM~\cite{peng2024reviewllm} integrates LoRA for supervised fine-tuning in the task of personalized review generation.

Beyond LoRA-based approaches, alternative pipelines have been proposed for personalized generation. UserIdentifier~\cite{mireshghallah2021useridentifier} introduces a user-specific identifier, significantly reducing training costs while enhancing personalized demonstration. UserAdapter~\cite{zhong2021useradapter} proposes user-independent prefix embeddings, leveraging prefix tuning for personalization. Meanwhile, HYDRA~\cite{zhuang2406hydra} achieves implicit personalization by training user-specific headers. Recently, researchers have also explored fine-tuning personalized model on edge devices~\cite{peng2024pocketllm} and collaborative learning between small and large language models to enable more personalized generation~\cite{zhang2024cogenesis}.

\paragraph{\textbf{\textit{{(2). Reinforcement Learning Based Methods.}}}}
Apart from fine-tuning based methods, recent research has explored reinforcement learning based techniques to personalize text generation by aligning outputs with user preferences. P-RLHF~\cite{li2024personalized} has been proposed to jointly learn a user-specific and reward model to enable text generation that aligns with a user's styles or criteria. P-SOUPS~\cite{jang2023personalized} models multiple user preferences as a Multi-Objective Reinforcement Learning (MORL) problem, decomposing preferences into multiple dimensions, each trained independently. PAD~\cite{chen2024pad} aligns text generation with human preferences during inference by utilizing token-level personalized rewards to guide the decoding process. REST-PG~\cite{salemi2025reasoning} introduces a framework that trains large language models to reason over personal data during response generation. This approach first generates reasoning paths to enhance the LLM's reasoning ability and then employs Expectation-Maximization Reinforced Self-Training to iteratively refine the model based on its high-reward outputs. Additionally,~\citet{salemi2024optimization} incorporate reinforcement learning into the RAG pipeline to improve retrieval accuracy, thereby enhancing the personalization of generated content. Other applications include RewriterSlRl~\cite{li2024learning}, which has been introduced to generate text via RL-based personalized prompt rewriting using API-based LLMs, and ~\citet{kulkarni2024reinforcement}, who explore the use of reinforcement learning to optimize RAG for improving the relevance and coherence of chatbot responses in specialized domains, ultimately enhancing user satisfaction and engagement.

\subsubsection{\textbf{Discussion}}
Personalized generation can be adopted via both explicit and implicit preference injection, yet they exhibit distinct characteristics that make them suitable for different scenarios.
In explicit preference-based generation, personalization is clearly defined through user profile descriptions, contextual information, and similar inputs, which are incorporated into generators via prompts. A key advantage of this approach is explainability, as the personalized information is explicitly provided and easily traceable. 
Despite leveraging provided preferences and internal knowledge, explicit preference injection's personalization is constrained by model capabilities and irrelevant information interference.
In contrast, implicit preference-based generation internalizes personalized information into the generator's parameters through scene-specific personalized data, thereby adapting the model for more fine-grained personalization. However, these methods typically incur substantial training and computational costs, as they require fine-tuning the generator’s internal parameters. Therefore, selecting between these approaches should be guided by the specific application scenario and resource constraints.
\subsection{From RAG to Agent}

\begin{figure*}[t]
    \centering
    \includegraphics[width=\linewidth]{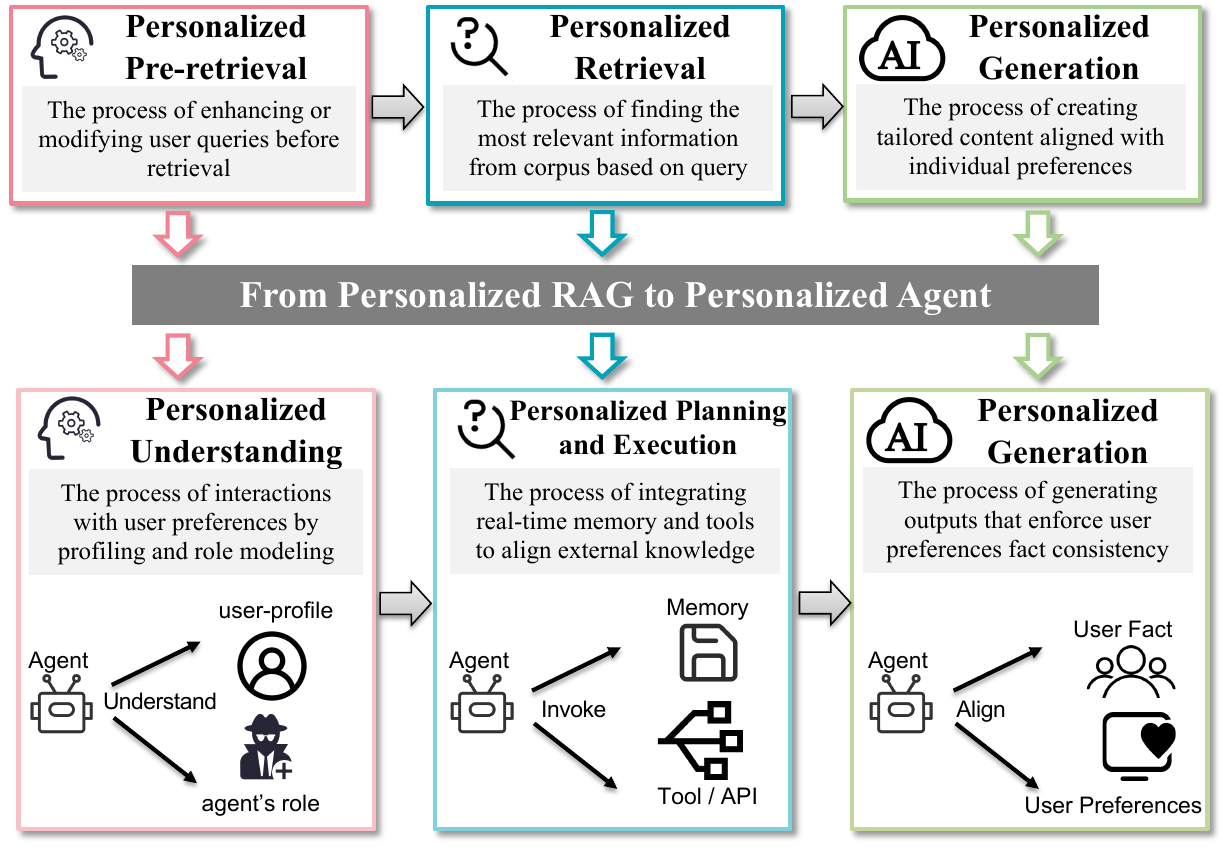}
    \caption{Overview of transition from personalized RAG to personalized agent.}
    \label{fig:Agent}
\end{figure*}

\subsubsection{\textbf{Definition}}
A personalized LLM-based agent is a system designed to dynamically incorporate user context, memory, and external tools or APIs to support highly personalized and goal-oriented interactions \cite{xi2025rise,huang2024understanding,chen2024persona}, and solve problems in a goal-oriented manner \cite{li2024personal,singh2025agentic}.
From the previously introduced stages of RAG, we observe that the evolution of personalized RAG reveals a structural convergence with agent architectures. We analyze them from three key perspectives: 
\begin{itemize}[leftmargin=*]
\item \textbf{Personalized Understanding}: This phase within the agent parallels the query understanding and rewriting process of RAG as outlined in Section \ref{sec:Pre-retrieval}. However, it extends beyond static semantic parsing by incorporating dynamic user profiling \cite{wang2023rolellm} and role modeling \cite{shao2023character}. This integration enables the agent to dynamically align interactions with implicit user preferences, facilitating personalized responses and task-specific adaptations \cite{ran2024capturing}.
\item \textbf{Personalized Planning and Execution}: This phase in agents mirrors RAG's retrieval process in Section \ref{sec:Retrieval} yet it advances beyond static document retrieval by incorporating real-time memory management \cite{park2023generative} and sophisticated tool and API calling \cite{wangvoyager}. This approach ensures the dynamic alignment of external knowledge with personalized constraints, such as integrating medical history in healthcare agents \cite{abbasian2023conversational}, to deliver context-aware and user-specific outcomes. 
\item \textbf{Personalized Generation}: This phase in agents mirrors RAG's generative process in Section \ref{sec:Generation} but transcends static template-based generation by integrating user preference and fact alignment. Agents dynamically enforce user preferences and ensure fact consistency through role-specific mechanisms (\eg social adaptability in conversational agents \cite{abbasian2023conversational}), enabling outputs to evolve in harmony with personalized and situational constraints rather than relying solely on predefined generative frameworks.

\end{itemize}
 In general we frame agent architectures as ``\textbf{personalized RAG++}'', where persistent memory \cite{wang2024crafting} replaces static indexes, and tool APIs \cite{cai2025large} serve as dynamic knowledge connectors, enabling complicated, human-aligned interactions beyond one-shot retrieval, as shown in Figure \ref{fig:Agent}. This progression highlights that as RAG systems incorporate deeper personalization—requiring user-state tracking, adaptive tool usage, and context-aware generation, they inherently adopt agent-like capabilities.

\subsubsection{\textbf{Personalized Understanding}}
Personalized understanding refers to an agent’s ability to accurately interpret user inputs by integrating user intent recognition and contextual analysis. This process ensures interactions that are both meaningful and contextually appropriate. The rationale behind this classification lies in its capacity to address three core aspects of understanding: recognizing user intent, analyzing context, and leveraging user profiles. Each of these aspects plays a distinct role in improving the agent's performance. 

\paragraph{\textbf{\textit{{(1). User-profile Understanding.}}}}
In user-profile understanding, an agent's personalized ability primarily depends on its capacity to accurately model and understand the user's preferences, context, and intentions. \citet{xu2024penetrative} proposes a framework in which LLMs are designed to understand the physical world, thereby facilitating a deeper connection between the agent and its environment, which is essential for accurate task execution. \citet{abbasian2023conversational} further expands this understanding by emphasizing the importance of personalization in health agents, where the user’s profile directly influences the behavior and decisions of the agent. This user understanding is foundational to ensuring that the AI agent performs tasks in a way that aligns with individual user needs.

\paragraph{\textbf{\textit{{(2). Role Understanding.}}}}
In agent's role understanding, the role of the agent within these environments is also crucial. Recent studies focus on enhancing role-playing capabilities within LLMs. \citet{wang2023rolellm} introduce RoleLLM, a benchmark that aims to elicit and refine the role-playing abilities of LLMs, demonstrating how role understanding influences agent performance in conversational tasks. Similarly, \citet{shao2023character} present Character-LLM, a trainable agent framework for role-playing, which tailors its responses based on predefined roles. \citet{wang2023incharacter} introduce a method for evaluating personality fidelity in role-playing agents through psychological interviews, aiming to enhance the realism and consistency of AI-driven characters. This role understanding allows for more contextually appropriate interactions, increasing the relevance and utility of AI agents across various applications. 

\paragraph{\textbf{\textit{{(3). User-role Joint Understanding.}}}}
In agent's user-role joint understanding, the intersection of user and role understanding is explored through frameworks that evaluate and enhance the social and personality aspects of LLMs. SocialBench \citet{chen2024socialbench} provides a sociality evaluation framework for role-playing agents. \citet{dai2024mmrole}, and \cite{ran2024capturing} extend this by incorporating multi-modal data and personality-indicative information, respectively, which allows agents to better adapt to both user and role understanding in dynamic environments. Furthermore, \citet{wang2023enabling} offers a perspective on how role and environment understanding can improve user experience. \citet{tu2024charactereval} contribute by providing a benchmark specifically for evaluating role-playing agents in the Chinese context, adding a cultural dimension to role understanding. Finally, Neeko \cite{yu2024neeko} further advances role-based interactions.

\subsubsection{\textbf{Personalized Planning and Execution}}
Personalized planning and execution refer to the process of designing and implementing strategies or actions that are specifically tailored to an individual’s unique context, and goals \cite{huang2022language,zhangbootstrap,park2023generative,singh2024personal}. It requires agents to dynamically integrate long-term memory, real-time reasoning, and external tool utilization \cite{hong2023metagpt,zheng2023agents,hongru2023large}, as demonstrated in healthcare decision support \cite{abbasian2023conversational} and travel planning scenarios \cite{cai2025large}. We analyze two fundamental components that enable this personalization in the following.

\paragraph{\textbf{\textit{{(1). Memory Management.}}}}
Effective memory systems allow agents to integrate users' historical preferences, behavioral patterns, and contextual habits, enhancing their ability to make planning and tailor interactions to user-specific needs \cite{wangvoyager,cai2025large,wangmemoryllm}. The EMG-RAG framework \cite{wang2024crafting} combines editable memory graphs with retrieval-augmented generation to maintain dynamic user profiles, while \citet{park2023generative} implements memory streams and periodic reflection mechanisms to simulate human-like behavior. In healthcare applications, \citet{abbasian2023conversational} integrates multimodal user data through specialized memory modules to optimize treatment recommendations. For recommendation systems, RecAgent \cite{wang2023user} employs hierarchical memory structures to model user interaction patterns across multiple domains. Recent advances like TravelPlanner+ \cite{singh2024personal} demonstrate how memory-augmented LLMs achieve higher relevance in personalized itinerary generation compared to generic planners. 

\paragraph{\textbf{\textit{{(2). Tool and API Calling.}}}}
The integration of external tools expands agents' capabilities beyond pure linguistic reasoning, enabling agents to interact with users and perform personalized tasks \cite{cai2025large,zhangbootstrap,xu2024penetrative,wangvoyager,wang2023enabling}. For instance, VOYAGER \cite{wangvoyager} establishes a paradigm for lifelong skill acquisition through automatic API curriculum learning and skill library construction. In robotics, \citet{zhangbootstrap} develops a bootstrapping framework where LLMs guide robots in tool-mediated skill discovery, enabling a high success rate in novel object manipulation tasks. The PUMA framework \cite{cai2025large} demonstrates how personalized web agents can achieve performance gains in e-commerce tasks through adaptive API orchestration. For mobile interaction, \citet{wang2023enabling} implements few-shot tool learning to handle diverse UI operations with minimal training data. These approaches highlight the importance of tool grounding mechanisms \cite{huang2022language} that translate linguistic plans into executable API sequences while maintaining personalization constraints.

This synthesis highlights that modern agent systems achieve enhanced personalization through two primary strategies: 1) Memory-augmented architectures, which leverage editable memory graphs \cite{wang2024crafting}, reflection mechanisms \cite{park2023generative}, and hierarchical memory structures \cite{wang2023user} to dynamically adapt to user preferences across various domains; and 2) Tool and API integration, which expand agent capabilities by balancing generalization  with specialization. Future work may explore improving the contextual relevance and adaptability of memory systems while optimizing real-time tool interaction for seamless task execution.

\subsubsection{\textbf{Personalized Generation}}
Based on the foundation of personalized planning and execution mechanisms, which enable agents to adapt strategies to user-specific contexts \cite{huang2022language,zhangbootstrap}, the next critical concern lies in personalized generation. This capability ensures that generated outputs not only align with factual correctness but also resonate with users' unique preferences, personality traits, and situational needs. Personalized generation bridges the gap between adaptive reasoning and human-aligned outcomes, allowing agents to produce contextually relevant and emotionally appropriate responses.

\paragraph{\textbf{\textit{{(1). Alignment with User Fact.}}}}
Alignment with User Fact emphasizes the accuracy, consistency, and factual grounding of personalized responses, ensuring they remain trustworthy across diverse user interactions. This is particularly challenging in personalized agents, where maintaining character authenticity while avoiding hallucinations requires balancing creativity with factual adherence.
Recent advances address these challenges through improved training frameworks and evaluation metrics. For instance, Character-LLM \cite{shao2023character} integrates memory-augmented architectures to reduce hallucinations while preserving character-specific traits. \citet{wang2024investigating} investigate quantization effects on personality consistency in edge-deployed agents and stabilize outputs under computational constraints. \citet{dai2024mmrole} ensures multimodal consistency (text-image) in role-playing. These works highlight the importance of architectural innovations and rigorous evaluation in achieving reliability.

\paragraph{\textbf{\textit{{(2). Alignment with User Preferences.}}}}
Alignment with user preferences ensures that generated outputs reflect individualized personalities, values, and interaction styles. This requires agents to dynamically interpret implicit user cues and adapt responses accordingly.
\citet{wang2023rolellm} benchmarks role-specific alignment. \citet{ran2024capturing} improves personality fidelity via psychological scale datasets. \citet{wang2023incharacter} quantifies alignment via psychological interviews. \citet{chen2024socialbench} evaluates social adaptability in conversations.

\subsubsection{\textbf{Discussion}}
The architectural evolution from RAG to personalized agents introduces significant advancements in human-AI interaction but also surfaces critical challenges that warrant further investigation.

\textbf{Personalized Understanding}, while enabling interpretation of user intent and context, faces limitations in real-time adaptability and generalization. Current approaches like RoleLLM \cite{wang2023rolellm} and Character-LLM \cite{shao2023character} demonstrate robust role-specific comprehension but struggle with dynamic user state tracking, particularly when handling evolving preferences or multi-session interactions. Furthermore, cultural specificity in benchmarks like CharacterEval \cite{tu2024charactereval} reveals gaps in global applicability, as agents trained on region-specific data often fail to generalize across diverse sociocultural contexts. Future work could explore hybrid architectures that combine continuous learning mechanisms with privacy-preserving federated learning to address these adaptability constraints while maintaining user trust.

\textbf{Personalized Planning and Execution}, achieves remarkable task specialization through memory management and tool integration, yet suffers from scalability issues in complex environments. While frameworks like EMG-RAG \cite{wang2024crafting} and VOYAGER \cite{wangvoyager} effectively manage user-specific constraints, their reliance on predefined API taxonomies limits emergent tool discovery in novel scenarios. The "cold-start" problem persists in domains requiring rapid skill acquisition, as seen in healthcare applications \cite{abbasian2023conversational}, where delayed API responses can compromise decision-making efficacy. A promising direction involves developing meta-reasoning architectures that dynamically prioritize memory recall versus tool invocation based on situational urgency and confidence thresholds.

\textbf{Personalized Generation} balances factual accuracy with preference alignment but risks over-fitting, where excessive finetuning to user profiles may reinforce cognitive biases. Techniques address surface-level alignment but lack mechanisms for ethical boundary detection. For instance, agents might inadvertently propagate harmful stereotypes when mirroring user preferences without critical oversight. Future systems could integrate value-aligned reinforcement learning with human-in-the-loop validation to preserve authenticity while preventing detrimental customization.

\section{Evaluation and Dataset}\label{sec:evaluation&dataset}
\begin{table*}[t]
\caption{Datasets and metrics for personalized RAG and Agent.}
\label{tab:datasets}
\centering
\resizebox{\textwidth}{!}{
\begin{tabular}{c|c|c|c} 
\toprule
\textbf{Field}                          & \begin{tabular}[c]{@{}c@{}}\textbf{Metrics}\\\textbf{Category}\end{tabular}      & \textbf{Metrics}                                                                                                                                                & \textbf{Datasets}                                                                                                                                                                                                                                                                                                                                                                                                                                                                                                                                                                                                                                                                                                                                                                                                                                                                                                                                                                                                                                                                                                                                     \\ 
\midrule
\multirow{8}{*}{\textbf{Pre-retrieval}} & \begin{tabular}[c]{@{}c@{}}Textual\\Quality\end{tabular}       & BLEU, ROUGE, EM                                                                                                                                        & Avocado Research Email Collection \cite{oard2015avocado,li2024learning}, Amazon review\cite{ni2019justifying,li2024learning}, Reddit comments\cite{reddit2015,li2024learning}, Amazon ESCI dataset\cite{reddy2022shopping,nguyen2025rl}, PIP                                                                                                                                                                                                                                                                                                                                                                                                                                                                                                                                                                                                                                                                                                                                                                                                                             \\ 
\cmidrule{2-4}
                               & \begin{tabular}[c]{@{}c@{}}Information\\Retrieval\end{tabular} & \begin{tabular}[c]{@{}c@{}}MAP, MRR, NDCG, \\Precision, Recall, RBP\end{tabular}                                                                       & \begin{tabular}[c]{@{}c@{}}AOL\cite{AOL,zhou2024cognitive}, WARRIORS\cite{ren2024bases}, Personalized Results Re-Ranking benchmark~\cite{bassani2022multi}, del.icio.us~\cite{bender2008exploiting, bouadjenek2011personalized,wu2017personalized, zhou2012improving},\\~Flickr~\cite{bender2008exploiting,schifanella2010folks}, CiteULike~\cite{bertier2009toward,bouadjenek2019personalized}, LRDP~\cite{biancalana2009social}, Delicious~\cite{wetzker2008analyzing}, Bibsonomy~\cite{mulhem2016axiomatic},\\~Wikipedia~\cite{baumann2024psqe, gao2010utilizing}\end{tabular}                                                                                                                                                                                                                                                  \\ 
\cmidrule{2-4}
                               & Classification                                                 & Accuracy, Macro-F1                                                                                                                                     & SCAN \cite{lake2018generalization,zhou2022least}, AITA WORKSM\cite{WORKSM,mysore2023pearl}, Robust04 \cite{li2023agent4ranking}                                                                                                                                                                                                                                                                                                                                                                                                                                                                                                                                                                                                                                                                                                                                                                                                                                                                                                                                                           \\ 
\cmidrule{2-4}
                               & Others                                                         & \begin{tabular}[c]{@{}c@{}}XEntropy, PMS, Image-Align,\\~PQEC, Prof$_{\text{overlap}}$\end{tabular}                                                      & Amazon ESCI dataset\cite{reddy2022shopping,nguyen2025rl}, PIP, Bibsonomy~\cite{mulhem2016axiomatic}                                                                                                                                                                                                                                                                                                                                                                                                                                                                                                                                                                                                                                                                                                                                                                                                                                                                                                                                                                                        \\ 
\midrule
\multirow{7}{*}{\textbf{Retrieval}}     & \begin{tabular}[c]{@{}c@{}}Textual\\Quality\end{tabular}       & BLEU, ROUGE, Dis, PPL                                                                                                                                  & TOPDIAL \cite{wang2023target}, Pchatbot \cite{qian2021pchatbot}, DuLemon \cite{xu2022long}                                                                                                                                                                                                                                                                                                                                                                                                                                                                                                                                                                                                                                                                                                                                                                                                                                                                                                                                                                                                \\ 
\cmidrule{2-4}
                               & \begin{tabular}[c]{@{}c@{}}Information\\Retrieval\end{tabular} & Recall, MRR, Precision, F1                                                                                                                             & LiveChat \cite{gao2023livechat}, Pchatbot \cite{qian2021pchatbot}, DuLemon \cite{xu2022long}                                                                                                                                                                                                                                                                                                                                                                                                                                                                                                                                                                                                                                                                                                                                                                                                                                                                                                                                                                                              \\ 
\cmidrule{2-4}
                               & Classification                                                 & Accuracy, Succ                                                                                                                                         & TOPDIAL \cite{wang2023target}, PersonalityEvd \cite{sun2024revealing}, DuLemon \cite{xu2022long}, PersonalityEdit \cite{mao2024editing}                                                                                                                                                                                                                                                                                                                                                                                                                                                                                                                                                                                                                                                                                                                                                                                                                                                                                                                                  \\ 
\cmidrule{2-4}
                               & Others                                                         & \begin{tabular}[c]{@{}c@{}}Fluency, Coherence,\\Plausibility, ES, DD, TPEI, PAE\end{tabular}                                                           & PersonalityEvd \cite{sun2024revealing}, PersonalityEdit \cite{mao2024editing}                                                                                                                                                                                                                                                                                                                                                                                                                                                                                                                                                                                                                                                                                                                                                                                                                                                                                                                                                                                                                              \\ 
\midrule
\multirow{6}{*}{\textbf{Generation}}    & \begin{tabular}[c]{@{}c@{}}Textual\\Quality\end{tabular}       & \begin{tabular}[c]{@{}c@{}}BLEU, ROUGE, Dis, \\PPL, METEOR\end{tabular}                                                                                & LaMP \cite{salemi2024lamp}, Long LaMP \cite{kumar2024longlamp}, Dulemon \cite{xu2022long}, PGraphRAG \cite{au2025personalized}, AmazonQA/Products~\cite{deng2022toward}, Reddit~\cite{zhong2022less}, MedicalDialogue~\cite{zhang2023memory}                                                                                                                                                                                                                                                                                                                                                                                                                                                                                                                                                                                                                                                                                                                          \\ 
\cmidrule{2-4}
                               & Classification                                                 & Accuracy, F1, Persona F1                                                                                                                               & LaMP \cite{salemi2024lamp}, Long LaMP \cite{kumar2024longlamp}, Dulemon \cite{xu2022long}, AmazonQA/Products~\cite{deng2022toward}, Reddit~\cite{zhong2022less}, MedicalDialogue~\cite{zhang2023memory}                                                                                                                                                                                                                                                                                                                                                                                                                                                                                                                                                                                                                                                                                                                                                                                \\ 
\cmidrule{2-4}
                               & Regression                                                     & MAE, RMSE                                                                                                                                              & LaMP \cite{salemi2024lamp}, Long LaMP \cite{kumar2024longlamp}, PGraphRAG \cite{au2025personalized}                                                                                                                                                                                                                                                                                                                                                                                                                                                                                                                                                                                                                                                                                                                                                                                                                                                                                                                                                                                       \\ 
\cmidrule{2-4}
                               & Others                                                         & \begin{tabular}[c]{@{}c@{}}Fluency, Mean Success Rate, \\Median Relative Improvements\end{tabular}                                                     & Personalized-Gen \cite{alhafni2024personalized}                                                                                                                                                                                                                                                                                                                                                                                                                                                                                                                                                                                                                                                                                                                                                                                                                                                                                                                                                                                                                                                                             \\ 
\midrule
\multirow{12}{*}{\textbf{Agent}}         & \begin{tabular}[c]{@{}c@{}}Textual\\Quality\end{tabular}       & \begin{tabular}[c]{@{}c@{}}BLEU, ROUGE, METEOR, \\CIDEr, EM, Fluency, \\Coherence, Instruction \\Adherence, Consistency \\related metrics\end{tabular} & \begin{tabular}[c]{@{}c@{}}RICO~\cite{wang2023enabling}, RoleBench~\cite{wang2023rolellm}, \citet{shao2023character}, Socialbench~\cite{chen2024socialbench}, MMRole-Data~\cite{dai2024mmrole}, ROLEPERSONALITY~\cite{ran2024capturing}, ChatHaruhi~\cite{wang2023incharacter}, \\Character-LLM-Data~\cite{yu2024neeko}, Knowledge Behind Persona~\cite{hongru2023large}, \citet{wang2024crafting}, \citet{wang2024investigating}, \citet{zheng2023agents}\end{tabular}                                                                                                                                                                                                                                                                                                                          \\ 
\cmidrule{2-4}
                               & \begin{tabular}[c]{@{}c@{}}Information\\Retrieval\end{tabular} & Recall, F1, Precision                                                                                                                                  & Knowledge Behind Persona~\cite{hongru2023large}                                                                                                                                                                                                                                                                                                                                                                                                                                                                                                                                                                                                                                                                                                                                                                                                                                                                                                                                                                                                                                                             \\ 
\cmidrule{2-4}
                               & Classification                                                 & \begin{tabular}[c]{@{}c@{}}Accuracy, Failure Rate, \\Classification Accuracy, \\Preference Rate, \\Correctness\end{tabular}                            & \begin{tabular}[c]{@{}c@{}}MIT-BIH Arrhythmia Database~\cite{xu2024penetrative}, VirtualHome~\cite{huang2022language}, Socialbench~\cite{chen2024socialbench}, ARC~\cite{renze2024self},AGIEval~\cite{renze2024self},HellaSwag~\cite{renze2024self},MedMCQA~\cite{renze2024self},\\AQUA-RAT~\cite{renze2024self},LogiQA~\cite{renze2024self},LSAT-AR~\cite{renze2024self},LSAT-LR~\cite{renze2024self},LSAT-RC~\cite{renze2024self},SAT-English~\cite{renze2024self},SAT-Math~\cite{renze2024self}, PersonalWAB~\cite{cai2025large},\\~TravelPlanner+~\cite{singh2024personal}\end{tabular}  \\ 
\cmidrule{2-4}
                               & Others                                                         & \begin{tabular}[c]{@{}c@{}}Pass@k, Executability, \\Productivity, Plausibility of the \\Story\end{tabular}                                             & \citet{hong2023metagpt}, \citet{zheng2023agents}                                                                                                                                                                                                                                                                                                                                                                                                                                                                                                                                                                                                                                                                                                                                                                                                                                                                                                                                                                                                                                                           \\
\bottomrule
\end{tabular}}
\end{table*}

In the evolving landscape of personalization, from RAG to advanced Agent-based systems, the evaluation of models relies heavily on diverse datasets and metrics tailored to specific tasks. This survey categorizes metrics into several key types: Textual Quality metrics (\eg BLEU, ROUGE, METEOR) assess the fluency and coherence of generated outputs; Information Retrieval metrics (\eg MAP, MRR, Recall) evaluate the accuracy and relevance of retrieved information; Classification metrics (\eg Accuracy, F1) measure task-specific correctness; Regression metrics (\eg MAE, RMSE) quantify prediction errors; and Other metrics (\eg Fluency, Pass@k) address domain-specific or task-unique aspects like plausibility or executability. These metrics span pre-retrieval, retrieval, generation, and agent-based personalization approaches, reflecting their varied objectives. To provide a comprehensive overview, we compile an extensive list of datasets across these fields, as detailed in Table~\ref{tab:datasets}. These datasets, paired with their respective metrics, enable researchers to benchmark and refine personalized systems, from enhancing query rewriting to enabling autonomous agents in physical and virtual environments.
\section{Challenges and Future Directions}
\label{sec:futuredirection}

Personalized RAG and agent-based systems still face several critical challenges that warrant further exploration. We list them as follows:
\begin{itemize}[leftmargin=*]
    \item \textbf{Balancing Personalization and Scalability}: Integrating personalization data (such as preferences, history, and contextual signals) into RAG processes often increases computational complexity,
    making it difficult to maintain efficiency and scalability across large-scale systems. Future work could explore lightweight, adaptive embeddings and hybrid frameworks that seamlessly fuse user profiles with real-time contexts.
    \item \textbf{Evaluating Personalization Effectively}: Current metrics like BLEU, ROUGE, and human evaluations fall short in capturing the nuanced alignment of outputs with dynamic user preferences, lacking tailored measures for personalization efficacy. Developing specialized benchmarks and metrics that assess long-term user satisfaction and adaptability is crucial for real-world applicability.
    \item \textbf{Preserving Privacy through Device–Cloud Collaboration:} Personalized retrieval often involves processing sensitive user data, raising privacy concerns, especially with the increased global emphasis on data protection regulations, such as the European Union’s General Data Protection Regulation (GDPR). Consequently, a promising approach is the collaborative integration of on-device small Language models which handle sensitive personal data locally, with cloud-based LLM, which provides broader contextual knowledge. 
    \item \textbf{Personalized Agent Planning:} Current research on agent planning remains mainly in its early stages, with much of the work focusing on building foundational frameworks such as GUI agents~\cite{nguyen2024gui} and the application of agents across diverse domains~\cite{wang2024survey}. Notably, the incorporation of personalized approaches has yet to be widely adopted. Exploring how to integrate personalized support into existing frameworks to enhance user experience represents a promising and valuable direction for future investigation.
    \item \textbf{Ensuring Ethical and Coherent Systems}: Bias in data processing, privacy concerns in user profiling, and coherence across retrieval and generation stages remain unresolved. Future directions should prioritize ethical safeguards, privacy-preserving techniques, and cross-stage optimization to build trustworthy, unified personalized systems.
\end{itemize}

\section{Conclusion}
In this paper, we explore the landscape of personalization from Retrieval-Augmented Generation (RAG) to advanced LLM-based Agents, detailing adaptations across pre-retrieval, retrieval, and generation stages while extending into agentic capabilities. By reviewing recent literature, datasets, and metrics, we highlight the progress and diversity in enhancing user satisfaction through tailored AI systems. 
However, challenges such as scalability, effective evaluation, and ethical concerns underscore the need for innovative solutions. Future research should focus on lightweight frameworks, specialized benchmarks, and privacy-preserving techniques to advance personalized AI. 
Relevant papers and resources are also compiled online for ease of future research.

\bibliographystyle{ACM-Reference-Format}
\bibliography{sample-base}


\begin{thebibliography}{175}


\ifx \showCODEN    \undefined \def \showCODEN     #1{\unskip}     \fi
\ifx \showDOI      \undefined \def \showDOI       #1{#1}\fi
\ifx \showISBNx    \undefined \def \showISBNx     #1{\unskip}     \fi
\ifx \showISBNxiii \undefined \def \showISBNxiii  #1{\unskip}     \fi
\ifx \showISSN     \undefined \def \showISSN      #1{\unskip}     \fi
\ifx \showLCCN     \undefined \def \showLCCN      #1{\unskip}     \fi
\ifx \shownote     \undefined \def \shownote      #1{#1}          \fi
\ifx \showarticletitle \undefined \def \showarticletitle #1{#1}   \fi
\ifx \showURL      \undefined \def \showURL       {\relax}        \fi
\providecommand\bibfield[2]{#2}
\providecommand\bibinfo[2]{#2}
\providecommand\natexlab[1]{#1}
\providecommand\showeprint[2][]{arXiv:#2}

\bibitem[kor(2021)]%
        {koroteev2021bert}
 \bibinfo{year}{2021}\natexlab{}.
\newblock \showarticletitle{BERT: a review of applications in natural language processing and understanding}.
\newblock \bibinfo{journal}{\emph{arXiv preprint arXiv:2103.11943}} (\bibinfo{year}{2021}).
\newblock


\bibitem[Abbasian et~al\mbox{.}(2023)]%
        {abbasian2023conversational}
\bibfield{author}{\bibinfo{person}{Mahyar Abbasian}, \bibinfo{person}{Iman Azimi}, \bibinfo{person}{Amir~M Rahmani}, {and} \bibinfo{person}{Ramesh Jain}.} \bibinfo{year}{2023}\natexlab{}.
\newblock \showarticletitle{Conversational health agents: A personalized llm-powered agent framework}.
\newblock \bibinfo{journal}{\emph{arXiv preprint arXiv:2310.02374}} (\bibinfo{year}{2023}).
\newblock


\bibitem[Alhafni et~al\mbox{.}(2024)]%
        {alhafni2024personalized}
\bibfield{author}{\bibinfo{person}{Bashar Alhafni}, \bibinfo{person}{Vivek Kulkarni}, \bibinfo{person}{Dhruv Kumar}, {and} \bibinfo{person}{Vipul Raheja}.} \bibinfo{year}{2024}\natexlab{}.
\newblock \showarticletitle{Personalized Text Generation with Fine-Grained Linguistic Control}. In \bibinfo{booktitle}{\emph{Proceedings of the 1st Workshop on Personalization of Generative AI Systems (PERSONALIZE 2024)}}. \bibinfo{pages}{88--101}.
\newblock


\bibitem[Amazon({[n.\,d.]})]%
        {amazon_dataset}
\bibfield{author}{\bibinfo{person}{Amazon}.} \bibinfo{year}{[n.\,d.]}\natexlab{}.
\newblock \bibinfo{title}{Amazon Customer Review Dataset}.
\newblock \bibinfo{howpublished}{Online dataset}.
\newblock
\urldef\tempurl%
\url{https://nijianmo.github.io/amazon/}
\showURL{%
\tempurl}


\bibitem[Au et~al\mbox{.}(2025)]%
        {au2025personalized}
\bibfield{author}{\bibinfo{person}{Steven Au}, \bibinfo{person}{Cameron~J Dimacali}, \bibinfo{person}{Ojasmitha Pedirappagari}, \bibinfo{person}{Namyong Park}, \bibinfo{person}{Franck Dernoncourt}, \bibinfo{person}{Yu Wang}, \bibinfo{person}{Nikos Kanakaris}, \bibinfo{person}{Hanieh Deilamsalehy}, \bibinfo{person}{Ryan~A Rossi}, {and} \bibinfo{person}{Nesreen~K Ahmed}.} \bibinfo{year}{2025}\natexlab{}.
\newblock \showarticletitle{Personalized Graph-Based Retrieval for Large Language Models}.
\newblock \bibinfo{journal}{\emph{arXiv preprint arXiv:2501.02157}} (\bibinfo{year}{2025}).
\newblock


\bibitem[Bassani et~al\mbox{.}(2022)]%
        {bassani2022multi}
\bibfield{author}{\bibinfo{person}{Elias Bassani}, \bibinfo{person}{Pranav Kasela}, \bibinfo{person}{Alessandro Raganato}, {and} \bibinfo{person}{Gabriella Pasi}.} \bibinfo{year}{2022}\natexlab{}.
\newblock \showarticletitle{A multi-domain benchmark for personalized search evaluation}. In \bibinfo{booktitle}{\emph{Proceedings of the 31st ACM International Conference on Information \& Knowledge Management}}. \bibinfo{pages}{3822--3827}.
\newblock


\bibitem[Bassani et~al\mbox{.}(2023)]%
        {bassani2023personalized}
\bibfield{author}{\bibinfo{person}{Elias Bassani}, \bibinfo{person}{Nicola Tonellotto}, {and} \bibinfo{person}{Gabriella Pasi}.} \bibinfo{year}{2023}\natexlab{}.
\newblock \showarticletitle{Personalized query expansion with contextual word embeddings}.
\newblock \bibinfo{journal}{\emph{ACM Transactions on Information Systems}} \bibinfo{volume}{42}, \bibinfo{number}{2} (\bibinfo{year}{2023}), \bibinfo{pages}{1--35}.
\newblock


\bibitem[Baumann and Schoenfeld(2024)]%
        {baumann2024psqe}
\bibfield{author}{\bibinfo{person}{Oliver Baumann} {and} \bibinfo{person}{Mirco Schoenfeld}.} \bibinfo{year}{2024}\natexlab{}.
\newblock \showarticletitle{PSQE: Personalized Semantic Query Expansion for user-centric query disambiguation}.
\newblock  (\bibinfo{year}{2024}).
\newblock


\bibitem[Bender et~al\mbox{.}(2008)]%
        {bender2008exploiting}
\bibfield{author}{\bibinfo{person}{Matthias Bender}, \bibinfo{person}{Tom Crecelius}, \bibinfo{person}{Mouna Kacimi}, \bibinfo{person}{Sebastian Michel}, \bibinfo{person}{Thomas Neumann}, \bibinfo{person}{Josiane~Xavier Parreira}, \bibinfo{person}{Ralf Schenkel}, {and} \bibinfo{person}{Gerhard Weikum}.} \bibinfo{year}{2008}\natexlab{}.
\newblock \showarticletitle{Exploiting social relations for query expansion and result ranking}. In \bibinfo{booktitle}{\emph{2008 IEEE 24th International Conference on Data Engineering Workshop}}. IEEE, \bibinfo{pages}{501--506}.
\newblock


\bibitem[Bertier et~al\mbox{.}(2009)]%
        {bertier2009toward}
\bibfield{author}{\bibinfo{person}{Marin Bertier}, \bibinfo{person}{Rachid Guerraoui}, \bibinfo{person}{Vincent Leroy}, {and} \bibinfo{person}{Anne-Marie Kermarrec}.} \bibinfo{year}{2009}\natexlab{}.
\newblock \showarticletitle{Toward personalized query expansion}. In \bibinfo{booktitle}{\emph{Proceedings of the Second ACM EuroSys Workshop on Social Network Systems}}. \bibinfo{pages}{7--12}.
\newblock


\bibitem[Bi et~al\mbox{.}(2021)]%
        {bi2021learning}
\bibfield{author}{\bibinfo{person}{Keping Bi}, \bibinfo{person}{Qingyao Ai}, {and} \bibinfo{person}{W~Bruce Croft}.} \bibinfo{year}{2021}\natexlab{}.
\newblock \showarticletitle{Learning a fine-grained review-based transformer model for personalized product search}. In \bibinfo{booktitle}{\emph{Proceedings of the 44th international ACM SIGIR conference on research and development in information retrieval}}. \bibinfo{pages}{123--132}.
\newblock


\bibitem[Biancalana and Micarelli(2009)]%
        {biancalana2009social}
\bibfield{author}{\bibinfo{person}{Claudio Biancalana} {and} \bibinfo{person}{Alessandro Micarelli}.} \bibinfo{year}{2009}\natexlab{}.
\newblock \showarticletitle{Social tagging in query expansion: A new way for personalized web search}. In \bibinfo{booktitle}{\emph{2009 International Conference on Computational Science and Engineering}}, Vol.~\bibinfo{volume}{4}. IEEE, \bibinfo{pages}{1060--1065}.
\newblock


\bibitem[Bing({[n.\,d.]})]%
        {bing}
\bibfield{author}{\bibinfo{person}{Microsoft Bing}.} \bibinfo{year}{[n.\,d.]}\natexlab{}.
\newblock \bibinfo{booktitle}{\emph{Bing Search Engine}}.
\newblock
\urldef\tempurl%
\url{https://www.bing.com}
\showURL{%
\tempurl}


\bibitem[Bouadjenek et~al\mbox{.}(2019)]%
        {bouadjenek2019personalized}
\bibfield{author}{\bibinfo{person}{Mohamed~Reda Bouadjenek}, \bibinfo{person}{Hakim Hacid}, {and} \bibinfo{person}{Mokrane Bouzeghoub}.} \bibinfo{year}{2019}\natexlab{}.
\newblock \showarticletitle{Personalized social query expansion using social annotations}.
\newblock \bibinfo{journal}{\emph{Transactions on Large-Scale Data-and Knowledge-Centered Systems XL}} (\bibinfo{year}{2019}), \bibinfo{pages}{1--25}.
\newblock


\bibitem[Bouadjenek et~al\mbox{.}(2011)]%
        {bouadjenek2011personalized}
\bibfield{author}{\bibinfo{person}{Mohamed~Reda Bouadjenek}, \bibinfo{person}{Hakim Hacid}, \bibinfo{person}{Mokrane Bouzeghoub}, {and} \bibinfo{person}{Johann Daigremont}.} \bibinfo{year}{2011}\natexlab{}.
\newblock \showarticletitle{Personalized social query expansion using social bookmarking systems}. In \bibinfo{booktitle}{\emph{Proceedings of the 34th international ACM SIGIR conference on Research and development in Information Retrieval}}. \bibinfo{pages}{1113--1114}.
\newblock


\bibitem[Bulfamante(2023)]%
        {bulfamante2023generative}
\bibfield{author}{\bibinfo{person}{Domenico Bulfamante}.} \bibinfo{year}{2023}\natexlab{}.
\newblock \emph{\bibinfo{title}{Generative enterprise search with extensible knowledge base using ai}}.
\newblock \bibinfo{thesistype}{Ph.\,D. Dissertation}. \bibinfo{school}{Politecnico di Torino}.
\newblock


\bibitem[Cai et~al\mbox{.}({[n.\,d.]})]%
        {cai2025large}
\bibfield{author}{\bibinfo{person}{Hongru Cai}, \bibinfo{person}{Yongqi Li}, \bibinfo{person}{Wenjie Wang}, \bibinfo{person}{ZHU Fengbin}, \bibinfo{person}{Xiaoyu Shen}, \bibinfo{person}{Wenjie Li}, {and} \bibinfo{person}{Tat-Seng Chua}.} \bibinfo{year}{[n.\,d.]}\natexlab{}.
\newblock \showarticletitle{Large Language Models Empowered Personalized Web Agents}. In \bibinfo{booktitle}{\emph{THE WEB CONFERENCE 2025}}.
\newblock


\bibitem[Chen et~al\mbox{.}(2024b)]%
        {chen2024socialbench}
\bibfield{author}{\bibinfo{person}{Hongzhan Chen}, \bibinfo{person}{Hehong Chen}, \bibinfo{person}{Ming Yan}, \bibinfo{person}{Wenshen Xu}, \bibinfo{person}{Xing Gao}, \bibinfo{person}{Weizhou Shen}, \bibinfo{person}{Xiaojun Quan}, \bibinfo{person}{Chenliang Li}, \bibinfo{person}{Ji Zhang}, \bibinfo{person}{Fei Huang}, {et~al\mbox{.}}} \bibinfo{year}{2024}\natexlab{b}.
\newblock \showarticletitle{Socialbench: Sociality evaluation of role-playing conversational agents}.
\newblock \bibinfo{journal}{\emph{arXiv preprint arXiv:2403.13679}} (\bibinfo{year}{2024}).
\newblock


\bibitem[Chen et~al\mbox{.}(2024c)]%
        {chen2024persona}
\bibfield{author}{\bibinfo{person}{Jiangjie Chen}, \bibinfo{person}{Xintao Wang}, \bibinfo{person}{Rui Xu}, \bibinfo{person}{Siyu Yuan}, \bibinfo{person}{Yikai Zhang}, \bibinfo{person}{Wei Shi}, \bibinfo{person}{Jian Xie}, \bibinfo{person}{Shuang Li}, \bibinfo{person}{Ruihan Yang}, \bibinfo{person}{Tinghui Zhu}, {et~al\mbox{.}}} \bibinfo{year}{2024}\natexlab{c}.
\newblock \showarticletitle{From persona to personalization: A survey on role-playing language agents}.
\newblock \bibinfo{journal}{\emph{arXiv preprint arXiv:2404.18231}} (\bibinfo{year}{2024}).
\newblock


\bibitem[Chen et~al\mbox{.}(2024d)]%
        {chen2024pad}
\bibfield{author}{\bibinfo{person}{Ruizhe Chen}, \bibinfo{person}{Xiaotian Zhang}, \bibinfo{person}{Meng Luo}, \bibinfo{person}{Wenhao Chai}, {and} \bibinfo{person}{Zuozhu Liu}.} \bibinfo{year}{2024}\natexlab{d}.
\newblock \showarticletitle{Pad: Personalized alignment of llms at decoding-time}.
\newblock \bibinfo{journal}{\emph{arXiv preprint arXiv:2410.04070}} (\bibinfo{year}{2024}).
\newblock


\bibitem[Chen et~al\mbox{.}(2024a)]%
        {chen2024kg}
\bibfield{author}{\bibinfo{person}{Weijie Chen}, \bibinfo{person}{Ting Bai}, \bibinfo{person}{Jinbo Su}, \bibinfo{person}{Jian Luan}, \bibinfo{person}{Wei Liu}, {and} \bibinfo{person}{Chuan Shi}.} \bibinfo{year}{2024}\natexlab{a}.
\newblock \showarticletitle{Kg-retriever: Efficient knowledge indexing for retrieval-augmented large language models}.
\newblock \bibinfo{journal}{\emph{arXiv preprint arXiv:2412.05547}} (\bibinfo{year}{2024}).
\newblock


\bibitem[Chen et~al\mbox{.}(2023)]%
        {chen2023graph}
\bibfield{author}{\bibinfo{person}{Zheng Chen}, \bibinfo{person}{Ziyan Jiang}, \bibinfo{person}{Fan Yang}, \bibinfo{person}{Eunah Cho}, \bibinfo{person}{Xing Fan}, \bibinfo{person}{Xiaojiang Huang}, \bibinfo{person}{Yanbin Lu}, {and} \bibinfo{person}{Aram Galstyan}.} \bibinfo{year}{2023}\natexlab{}.
\newblock \showarticletitle{Graph meets LLM: A novel approach to collaborative filtering for robust conversational understanding}.
\newblock \bibinfo{journal}{\emph{arXiv preprint arXiv:2305.14449}} (\bibinfo{year}{2023}).
\newblock


\bibitem[Cheng et~al\mbox{.}(2023)]%
        {cheng2023explainable}
\bibfield{author}{\bibinfo{person}{Hao Cheng}, \bibinfo{person}{Shuo Wang}, \bibinfo{person}{Wensheng Lu}, \bibinfo{person}{Wei Zhang}, \bibinfo{person}{Mingyang Zhou}, \bibinfo{person}{Kezhong Lu}, {and} \bibinfo{person}{Hao Liao}.} \bibinfo{year}{2023}\natexlab{}.
\newblock \showarticletitle{Explainable recommendation with personalized review retrieval and aspect learning}.
\newblock \bibinfo{journal}{\emph{arXiv preprint arXiv:2306.12657}} (\bibinfo{year}{2023}).
\newblock


\bibitem[Chevalier et~al\mbox{.}(2023)]%
        {chevalier2023adapting}
\bibfield{author}{\bibinfo{person}{Alexis Chevalier}, \bibinfo{person}{Alexander Wettig}, \bibinfo{person}{Anirudh Ajith}, {and} \bibinfo{person}{Danqi Chen}.} \bibinfo{year}{2023}\natexlab{}.
\newblock \showarticletitle{Adapting language models to compress contexts}.
\newblock \bibinfo{journal}{\emph{arXiv preprint arXiv:2305.14788}} (\bibinfo{year}{2023}).
\newblock


\bibitem[Cho et~al\mbox{.}(2021)]%
        {cho2021personalized}
\bibfield{author}{\bibinfo{person}{Eunah Cho}, \bibinfo{person}{Ziyan Jiang}, \bibinfo{person}{Jie Hao}, \bibinfo{person}{Zheng Chen}, \bibinfo{person}{Saurabh Gupta}, \bibinfo{person}{Xing Fan}, {and} \bibinfo{person}{Chenlei Guo}.} \bibinfo{year}{2021}\natexlab{}.
\newblock \showarticletitle{Personalized search-based query rewrite system for conversational ai}. In \bibinfo{booktitle}{\emph{Proceedings of the 3rd Workshop on Natural Language Processing for Conversational AI}}. \bibinfo{pages}{179--188}.
\newblock


\bibitem[Cho et~al\mbox{.}(2025)]%
        {cho2025tuning}
\bibfield{author}{\bibinfo{person}{Hyundong Cho}, \bibinfo{person}{Karishma Sharma}, \bibinfo{person}{Nicolaas Jedema}, \bibinfo{person}{Leonardo~FR Ribeiro}, \bibinfo{person}{Alessandro Moschitti}, \bibinfo{person}{Ravi Krishnan}, {and} \bibinfo{person}{Jonathan May}.} \bibinfo{year}{2025}\natexlab{}.
\newblock \showarticletitle{Tuning-Free Personalized Alignment via Trial-Error-Explain In-Context Learning}.
\newblock \bibinfo{journal}{\emph{arXiv preprint arXiv:2502.08972}} (\bibinfo{year}{2025}).
\newblock


\bibitem[Dai et~al\mbox{.}(2024)]%
        {dai2024mmrole}
\bibfield{author}{\bibinfo{person}{Yanqi Dai}, \bibinfo{person}{Huanran Hu}, \bibinfo{person}{Lei Wang}, \bibinfo{person}{Shengjie Jin}, \bibinfo{person}{Xu Chen}, {and} \bibinfo{person}{Zhiwu Lu}.} \bibinfo{year}{2024}\natexlab{}.
\newblock \showarticletitle{Mmrole: A comprehensive framework for developing and evaluating multimodal role-playing agents}.
\newblock \bibinfo{journal}{\emph{arXiv preprint arXiv:2408.04203}} (\bibinfo{year}{2024}).
\newblock


\bibitem[Deng et~al\mbox{.}(2024)]%
        {deng2024unlocking}
\bibfield{author}{\bibinfo{person}{Wenlong Deng}, \bibinfo{person}{Christos Thrampoulidis}, {and} \bibinfo{person}{Xiaoxiao Li}.} \bibinfo{year}{2024}\natexlab{}.
\newblock \showarticletitle{Unlocking the potential of prompt-tuning in bridging generalized and personalized federated learning}. In \bibinfo{booktitle}{\emph{Proceedings of the IEEE/CVF Conference on Computer Vision and Pattern Recognition}}. \bibinfo{pages}{6087--6097}.
\newblock


\bibitem[Deng et~al\mbox{.}(2022)]%
        {deng2022toward}
\bibfield{author}{\bibinfo{person}{Yang Deng}, \bibinfo{person}{Yaliang Li}, \bibinfo{person}{Wenxuan Zhang}, \bibinfo{person}{Bolin Ding}, {and} \bibinfo{person}{Wai Lam}.} \bibinfo{year}{2022}\natexlab{}.
\newblock \showarticletitle{Toward personalized answer generation in e-commerce via multi-perspective preference modeling}.
\newblock \bibinfo{journal}{\emph{ACM Transactions on Information Systems (TOIS)}} \bibinfo{volume}{40}, \bibinfo{number}{4} (\bibinfo{year}{2022}), \bibinfo{pages}{1--28}.
\newblock


\bibitem[Douze et~al\mbox{.}(2024)]%
        {douze2024faiss}
\bibfield{author}{\bibinfo{person}{Matthijs Douze}, \bibinfo{person}{Alexandr Guzhva}, \bibinfo{person}{Chengqi Deng}, \bibinfo{person}{Jeff Johnson}, \bibinfo{person}{Gergely Szilvasy}, \bibinfo{person}{Pierre-Emmanuel Mazaré}, \bibinfo{person}{Maria Lomeli}, \bibinfo{person}{Lucas Hosseini}, {and} \bibinfo{person}{Hervé Jégou}.} \bibinfo{year}{2024}\natexlab{}.
\newblock \showarticletitle{The Faiss library}.
\newblock  (\bibinfo{year}{2024}).
\newblock
\showeprint[arxiv]{2401.08281}~[cs.LG]


\bibitem[ESPN({[n.\,d.]})]%
        {espn_dataset}
\bibfield{author}{\bibinfo{person}{ESPN}.} \bibinfo{year}{[n.\,d.]}\natexlab{}.
\newblock \bibinfo{title}{ESPN Sports Statistics Dataset}.
\newblock \bibinfo{howpublished}{Online dataset}.
\newblock


\bibitem[Fan et~al\mbox{.}(2024)]%
        {fan2024survey}
\bibfield{author}{\bibinfo{person}{Wenqi Fan}, \bibinfo{person}{Yujuan Ding}, \bibinfo{person}{Liangbo Ning}, \bibinfo{person}{Shijie Wang}, \bibinfo{person}{Hengyun Li}, \bibinfo{person}{Dawei Yin}, \bibinfo{person}{Tat-Seng Chua}, {and} \bibinfo{person}{Qing Li}.} \bibinfo{year}{2024}\natexlab{}.
\newblock \showarticletitle{A survey on rag meeting llms: Towards retrieval-augmented large language models}. In \bibinfo{booktitle}{\emph{Proceedings of the 30th ACM SIGKDD Conference on Knowledge Discovery and Data Mining}}. \bibinfo{pages}{6491--6501}.
\newblock


\bibitem[Gao et~al\mbox{.}(2010)]%
        {gao2010utilizing}
\bibfield{author}{\bibinfo{person}{Byron~J Gao}, \bibinfo{person}{David~C Anastasiu}, {and} \bibinfo{person}{Xing Jiang}.} \bibinfo{year}{2010}\natexlab{}.
\newblock \showarticletitle{Utilizing user-input contextual terms for query disambiguation}. In \bibinfo{booktitle}{\emph{Coling 2010: Posters}}. \bibinfo{pages}{329--337}.
\newblock


\bibitem[Gao et~al\mbox{.}(2023a)]%
        {gao2023livechat}
\bibfield{author}{\bibinfo{person}{Jingsheng Gao}, \bibinfo{person}{Yixin Lian}, \bibinfo{person}{Ziyi Zhou}, \bibinfo{person}{Yuzhuo Fu}, {and} \bibinfo{person}{Baoyuan Wang}.} \bibinfo{year}{2023}\natexlab{a}.
\newblock \showarticletitle{LiveChat: A large-scale personalized dialogue dataset automatically constructed from live streaming}.
\newblock \bibinfo{journal}{\emph{arXiv preprint arXiv:2306.08401}} (\bibinfo{year}{2023}).
\newblock


\bibitem[Gao et~al\mbox{.}(2023b)]%
        {gao2023retrieval}
\bibfield{author}{\bibinfo{person}{Yunfan Gao}, \bibinfo{person}{Yun Xiong}, \bibinfo{person}{Xinyu Gao}, \bibinfo{person}{Kangxiang Jia}, \bibinfo{person}{Jinliu Pan}, \bibinfo{person}{Yuxi Bi}, \bibinfo{person}{Yi Dai}, \bibinfo{person}{Jiawei Sun}, \bibinfo{person}{Haofen Wang}, {and} \bibinfo{person}{Haofen Wang}.} \bibinfo{year}{2023}\natexlab{b}.
\newblock \showarticletitle{Retrieval-augmented generation for large language models: A survey}.
\newblock \bibinfo{journal}{\emph{arXiv preprint arXiv:2312.10997}}  \bibinfo{volume}{2} (\bibinfo{year}{2023}).
\newblock


\bibitem[Google({[n.\,d.]})]%
        {google}
\bibfield{author}{\bibinfo{person}{Google}.} \bibinfo{year}{[n.\,d.]}\natexlab{}.
\newblock \bibinfo{booktitle}{\emph{Google Search}}.
\newblock
\urldef\tempurl%
\url{https://www.google.com}
\showURL{%
\tempurl}


\bibitem[Gu et~al\mbox{.}(2021)]%
        {gu2021partner}
\bibfield{author}{\bibinfo{person}{Jia-Chen Gu}, \bibinfo{person}{Hui Liu}, \bibinfo{person}{Zhen-Hua Ling}, \bibinfo{person}{Quan Liu}, \bibinfo{person}{Zhigang Chen}, {and} \bibinfo{person}{Xiaodan Zhu}.} \bibinfo{year}{2021}\natexlab{}.
\newblock \showarticletitle{Partner matters! an empirical study on fusing personas for personalized response selection in retrieval-based chatbots}. In \bibinfo{booktitle}{\emph{Proceedings of the 44th International ACM SIGIR Conference on Research and Development in Information Retrieval}}. \bibinfo{pages}{565--574}.
\newblock


\bibitem[Hao et~al\mbox{.}(2022)]%
        {hao2022cgf}
\bibfield{author}{\bibinfo{person}{Jie Hao}, \bibinfo{person}{Yang Liu}, \bibinfo{person}{Xing Fan}, \bibinfo{person}{Saurabh Gupta}, \bibinfo{person}{Saleh Soltan}, \bibinfo{person}{Rakesh Chada}, \bibinfo{person}{Pradeep Natarajan}, \bibinfo{person}{Chenlei Guo}, {and} \bibinfo{person}{G{\"o}khan T{\"u}r}.} \bibinfo{year}{2022}\natexlab{}.
\newblock \showarticletitle{CGF: Constrained generation framework for query rewriting in conversational AI}. In \bibinfo{booktitle}{\emph{Proceedings of the 2022 Conference on Empirical Methods in Natural Language Processing: Industry Track}}. \bibinfo{pages}{475--483}.
\newblock


\bibitem[Henze et~al\mbox{.}(2004)]%
        {henze2004reasoning}
\bibfield{author}{\bibinfo{person}{Nicola Henze}, \bibinfo{person}{Peter Dolog}, {and} \bibinfo{person}{Wolfgang Nejdl}.} \bibinfo{year}{2004}\natexlab{}.
\newblock \showarticletitle{Reasoning and ontologies for personalized e-learning in the semantic web}.
\newblock \bibinfo{journal}{\emph{Journal of Educational Technology \& Society}} \bibinfo{volume}{7}, \bibinfo{number}{4} (\bibinfo{year}{2004}), \bibinfo{pages}{82--97}.
\newblock


\bibitem[Hong et~al\mbox{.}(2023)]%
        {hong2023metagpt}
\bibfield{author}{\bibinfo{person}{Sirui Hong}, \bibinfo{person}{Xiawu Zheng}, \bibinfo{person}{Jonathan Chen}, \bibinfo{person}{Yuheng Cheng}, \bibinfo{person}{Jinlin Wang}, \bibinfo{person}{Ceyao Zhang}, \bibinfo{person}{Zili Wang}, \bibinfo{person}{Steven Ka~Shing Yau}, \bibinfo{person}{Zijuan Lin}, \bibinfo{person}{Liyang Zhou}, {et~al\mbox{.}}} \bibinfo{year}{2023}\natexlab{}.
\newblock \showarticletitle{Metagpt: Meta programming for multi-agent collaborative framework}.
\newblock \bibinfo{journal}{\emph{arXiv preprint arXiv:2308.00352}} \bibinfo{volume}{3}, \bibinfo{number}{4} (\bibinfo{year}{2023}), \bibinfo{pages}{6}.
\newblock


\bibitem[Hongru et~al\mbox{.}({[n.\,d.]})]%
        {hongru2023large}
\bibfield{author}{\bibinfo{person}{WANG Hongru}, \bibinfo{person}{Minda Hu}, \bibinfo{person}{Yang Deng}, \bibinfo{person}{Rui Wang}, \bibinfo{person}{Fei Mi}, \bibinfo{person}{Weichao Wang}, \bibinfo{person}{Yasheng Wang}, \bibinfo{person}{Wai-Chung Kwan}, \bibinfo{person}{Irwin King}, {and} \bibinfo{person}{Kam-Fai Wong}.} \bibinfo{year}{[n.\,d.]}\natexlab{}.
\newblock \showarticletitle{Large Language Models as Source Planner for Personalized Knowledge-grounded Dialogues}. In \bibinfo{booktitle}{\emph{The 2023 Conference on Empirical Methods in Natural Language Processing}}.
\newblock


\bibitem[Hu et~al\mbox{.}(2022)]%
        {hu2022lora}
\bibfield{author}{\bibinfo{person}{Edward~J Hu}, \bibinfo{person}{Yelong Shen}, \bibinfo{person}{Phillip Wallis}, \bibinfo{person}{Zeyuan Allen-Zhu}, \bibinfo{person}{Yuanzhi Li}, \bibinfo{person}{Shean Wang}, \bibinfo{person}{Lu Wang}, \bibinfo{person}{Weizhu Chen}, {et~al\mbox{.}}} \bibinfo{year}{2022}\natexlab{}.
\newblock \showarticletitle{Lora: Low-rank adaptation of large language models.}
\newblock \bibinfo{journal}{\emph{ICLR}} \bibinfo{volume}{1}, \bibinfo{number}{2} (\bibinfo{year}{2022}), \bibinfo{pages}{3}.
\newblock


\bibitem[Huang et~al\mbox{.}(2024a)]%
        {huang2024learning}
\bibfield{author}{\bibinfo{person}{Qiushi Huang}, \bibinfo{person}{Shuai Fu}, \bibinfo{person}{Xubo Liu}, \bibinfo{person}{Wenwu Wang}, \bibinfo{person}{Tom Ko}, \bibinfo{person}{Yu Zhang}, {and} \bibinfo{person}{Lilian Tang}.} \bibinfo{year}{2024}\natexlab{a}.
\newblock \showarticletitle{Learning retrieval augmentation for personalized dialogue generation}.
\newblock \bibinfo{journal}{\emph{arXiv preprint arXiv:2406.18847}} (\bibinfo{year}{2024}).
\newblock


\bibitem[Huang et~al\mbox{.}(2022)]%
        {huang2022language}
\bibfield{author}{\bibinfo{person}{Wenlong Huang}, \bibinfo{person}{Pieter Abbeel}, \bibinfo{person}{Deepak Pathak}, {and} \bibinfo{person}{Igor Mordatch}.} \bibinfo{year}{2022}\natexlab{}.
\newblock \showarticletitle{Language models as zero-shot planners: Extracting actionable knowledge for embodied agents}. In \bibinfo{booktitle}{\emph{International conference on machine learning}}. PMLR, \bibinfo{pages}{9118--9147}.
\newblock


\bibitem[Huang et~al\mbox{.}(2024b)]%
        {huang2024understanding}
\bibfield{author}{\bibinfo{person}{Xu Huang}, \bibinfo{person}{Weiwen Liu}, \bibinfo{person}{Xiaolong Chen}, \bibinfo{person}{Xingmei Wang}, \bibinfo{person}{Hao Wang}, \bibinfo{person}{Defu Lian}, \bibinfo{person}{Yasheng Wang}, \bibinfo{person}{Ruiming Tang}, {and} \bibinfo{person}{Enhong Chen}.} \bibinfo{year}{2024}\natexlab{b}.
\newblock \showarticletitle{Understanding the planning of LLM agents: A survey}.
\newblock \bibinfo{journal}{\emph{arXiv preprint arXiv:2402.02716}} (\bibinfo{year}{2024}).
\newblock


\bibitem[Jagerman et~al\mbox{.}(2023)]%
        {jagerman2023query}
\bibfield{author}{\bibinfo{person}{Rolf Jagerman}, \bibinfo{person}{Honglei Zhuang}, \bibinfo{person}{Zhen Qin}, \bibinfo{person}{Xuanhui Wang}, {and} \bibinfo{person}{Michael Bendersky}.} \bibinfo{year}{2023}\natexlab{}.
\newblock \showarticletitle{Query expansion by prompting large language models}.
\newblock \bibinfo{journal}{\emph{arXiv preprint arXiv:2305.03653}} (\bibinfo{year}{2023}).
\newblock


\bibitem[Jang et~al\mbox{.}(2023)]%
        {jang2023personalized}
\bibfield{author}{\bibinfo{person}{Joel Jang}, \bibinfo{person}{Seungone Kim}, \bibinfo{person}{Bill~Yuchen Lin}, \bibinfo{person}{Yizhong Wang}, \bibinfo{person}{Jack Hessel}, \bibinfo{person}{Luke Zettlemoyer}, \bibinfo{person}{Hannaneh Hajishirzi}, \bibinfo{person}{Yejin Choi}, {and} \bibinfo{person}{Prithviraj Ammanabrolu}.} \bibinfo{year}{2023}\natexlab{}.
\newblock \showarticletitle{Personalized soups: Personalized large language model alignment via post-hoc parameter merging}.
\newblock \bibinfo{journal}{\emph{arXiv preprint arXiv:2310.11564}} (\bibinfo{year}{2023}).
\newblock


\bibitem[Jia et~al\mbox{.}(2024)]%
        {jia2024mill}
\bibfield{author}{\bibinfo{person}{Pengyue Jia}, \bibinfo{person}{Yiding Liu}, \bibinfo{person}{Xiangyu Zhao}, \bibinfo{person}{Xiaopeng Li}, \bibinfo{person}{Changying Hao}, \bibinfo{person}{Shuaiqiang Wang}, {and} \bibinfo{person}{Dawei Yin}.} \bibinfo{year}{2024}\natexlab{}.
\newblock \showarticletitle{MILL: Mutual Verification with Large Language Models for Zero-Shot Query Expansion}. In \bibinfo{booktitle}{\emph{Proceedings of the 2024 Conference of the North American Chapter of the Association for Computational Linguistics: Human Language Technologies (Volume 1: Long Papers)}}. \bibinfo{pages}{2498--2518}.
\newblock


\bibitem[Jiang et~al\mbox{.}(2023)]%
        {jiang2023evaluating}
\bibfield{author}{\bibinfo{person}{Guangyuan Jiang}, \bibinfo{person}{Manjie Xu}, \bibinfo{person}{Song-Chun Zhu}, \bibinfo{person}{Wenjuan Han}, \bibinfo{person}{Chi Zhang}, {and} \bibinfo{person}{Yixin Zhu}.} \bibinfo{year}{2023}\natexlab{}.
\newblock \showarticletitle{Evaluating and inducing personality in pre-trained language models}.
\newblock \bibinfo{journal}{\emph{Advances in Neural Information Processing Systems}}  \bibinfo{volume}{36} (\bibinfo{year}{2023}), \bibinfo{pages}{10622--10643}.
\newblock


\bibitem[Joko et~al\mbox{.}(2024)]%
        {joko2024doing}
\bibfield{author}{\bibinfo{person}{Hideaki Joko}, \bibinfo{person}{Shubham Chatterjee}, \bibinfo{person}{Andrew Ramsay}, \bibinfo{person}{Arjen~P De~Vries}, \bibinfo{person}{Jeff Dalton}, {and} \bibinfo{person}{Faegheh Hasibi}.} \bibinfo{year}{2024}\natexlab{}.
\newblock \showarticletitle{Doing personal laps: Llm-augmented dialogue construction for personalized multi-session conversational search}. In \bibinfo{booktitle}{\emph{Proceedings of the 47th International ACM SIGIR Conference on Research and Development in Information Retrieval}}. \bibinfo{pages}{796--806}.
\newblock


\bibitem[Kang et~al\mbox{.}(2023)]%
        {kang2023llms}
\bibfield{author}{\bibinfo{person}{Wang-Cheng Kang}, \bibinfo{person}{Jianmo Ni}, \bibinfo{person}{Nikhil Mehta}, \bibinfo{person}{Maheswaran Sathiamoorthy}, \bibinfo{person}{Lichan Hong}, \bibinfo{person}{Ed Chi}, {and} \bibinfo{person}{Derek~Zhiyuan Cheng}.} \bibinfo{year}{2023}\natexlab{}.
\newblock \showarticletitle{Do llms understand user preferences? evaluating llms on user rating prediction}.
\newblock \bibinfo{journal}{\emph{arXiv preprint arXiv:2305.06474}} (\bibinfo{year}{2023}).
\newblock


\bibitem[Kannadasan and Aslanyan(2019)]%
        {kannadasan2019personalized}
\bibfield{author}{\bibinfo{person}{Manojkumar~Rangasamy Kannadasan} {and} \bibinfo{person}{Grigor Aslanyan}.} \bibinfo{year}{2019}\natexlab{}.
\newblock \showarticletitle{Personalized query auto-completion through a lightweight representation of the user context}.
\newblock \bibinfo{journal}{\emph{arXiv preprint arXiv:1905.01386}} (\bibinfo{year}{2019}).
\newblock


\bibitem[Kannan et~al\mbox{.}(2016)]%
        {WORKSM}
\bibfield{author}{\bibinfo{person}{Anjuli Kannan}, \bibinfo{person}{Karol Kurach}, \bibinfo{person}{Sujith Ravi}, \bibinfo{person}{Tobias Kaufmann}, \bibinfo{person}{Andrew Tomkins}, \bibinfo{person}{Balint Miklos}, \bibinfo{person}{Greg Corrado}, \bibinfo{person}{Laszlo Lukacs}, \bibinfo{person}{Marina Ganea}, \bibinfo{person}{Peter Young}, {and} \bibinfo{person}{Vivek Ramavajjala}.} \bibinfo{year}{2016}\natexlab{}.
\newblock \showarticletitle{Smart Reply: Automated Response Suggestion for Email}. In \bibinfo{booktitle}{\emph{Proceedings of the 22nd ACM SIGKDD International Conference on Knowledge Discovery and Data Mining}} (San Francisco, California, USA) \emph{(\bibinfo{series}{KDD '16})}. \bibinfo{publisher}{Association for Computing Machinery}, \bibinfo{address}{New York, NY, USA}, \bibinfo{pages}{955–964}.
\newblock
\showISBNx{9781450342322}
\urldef\tempurl%
\url{https://doi.org/10.1145/2939672.2939801}
\showDOI{\tempurl}


\bibitem[Kulkarni et~al\mbox{.}(2024)]%
        {kulkarni2024reinforcement}
\bibfield{author}{\bibinfo{person}{Mandar Kulkarni}, \bibinfo{person}{Praveen Tangarajan}, \bibinfo{person}{Kyung Kim}, {and} \bibinfo{person}{Anusua Trivedi}.} \bibinfo{year}{2024}\natexlab{}.
\newblock \showarticletitle{Reinforcement learning for optimizing rag for domain chatbots}.
\newblock \bibinfo{journal}{\emph{arXiv preprint arXiv:2401.06800}} (\bibinfo{year}{2024}).
\newblock


\bibitem[Kumar et~al\mbox{.}(2024)]%
        {kumar2024longlamp}
\bibfield{author}{\bibinfo{person}{Ishita Kumar}, \bibinfo{person}{Snigdha Viswanathan}, \bibinfo{person}{Sushrita Yerra}, \bibinfo{person}{Alireza Salemi}, \bibinfo{person}{Ryan~A Rossi}, \bibinfo{person}{Franck Dernoncourt}, \bibinfo{person}{Hanieh Deilamsalehy}, \bibinfo{person}{Xiang Chen}, \bibinfo{person}{Ruiyi Zhang}, \bibinfo{person}{Shubham Agarwal}, {et~al\mbox{.}}} \bibinfo{year}{2024}\natexlab{}.
\newblock \showarticletitle{Longlamp: A benchmark for personalized long-form text generation}.
\newblock \bibinfo{journal}{\emph{arXiv preprint arXiv:2407.11016}} (\bibinfo{year}{2024}).
\newblock


\bibitem[Lake and Baroni(2018)]%
        {lake2018generalization}
\bibfield{author}{\bibinfo{person}{Brenden Lake} {and} \bibinfo{person}{Marco Baroni}.} \bibinfo{year}{2018}\natexlab{}.
\newblock \showarticletitle{Generalization without systematicity: On the compositional skills of sequence-to-sequence recurrent networks}. In \bibinfo{booktitle}{\emph{International conference on machine learning}}. PMLR, \bibinfo{pages}{2873--2882}.
\newblock


\bibitem[Li et~al\mbox{.}(2024b)]%
        {li2024learning}
\bibfield{author}{\bibinfo{person}{Cheng Li}, \bibinfo{person}{Mingyang Zhang}, \bibinfo{person}{Qiaozhu Mei}, \bibinfo{person}{Weize Kong}, {and} \bibinfo{person}{Michael Bendersky}.} \bibinfo{year}{2024}\natexlab{b}.
\newblock \showarticletitle{Learning to rewrite prompts for personalized text generation}. In \bibinfo{booktitle}{\emph{Proceedings of the ACM Web Conference 2024}}. \bibinfo{pages}{3367--3378}.
\newblock


\bibitem[Li et~al\mbox{.}(2024d)]%
        {li2024matryoshka}
\bibfield{author}{\bibinfo{person}{Changhao Li}, \bibinfo{person}{Yuchen Zhuang}, \bibinfo{person}{Rushi Qiang}, \bibinfo{person}{Haotian Sun}, \bibinfo{person}{Hanjun Dai}, \bibinfo{person}{Chao Zhang}, {and} \bibinfo{person}{Bo Dai}.} \bibinfo{year}{2024}\natexlab{d}.
\newblock \showarticletitle{Matryoshka: Learning to Drive Black-Box LLMs with LLMs}.
\newblock \bibinfo{journal}{\emph{arXiv preprint arXiv:2410.20749}} (\bibinfo{year}{2024}).
\newblock


\bibitem[Li et~al\mbox{.}(2023b)]%
        {li2023personalized}
\bibfield{author}{\bibinfo{person}{Lei Li}, \bibinfo{person}{Yongfeng Zhang}, {and} \bibinfo{person}{Li Chen}.} \bibinfo{year}{2023}\natexlab{b}.
\newblock \showarticletitle{Personalized prompt learning for explainable recommendation}.
\newblock \bibinfo{journal}{\emph{ACM Transactions on Information Systems}} \bibinfo{volume}{41}, \bibinfo{number}{4} (\bibinfo{year}{2023}), \bibinfo{pages}{1--26}.
\newblock


\bibitem[Li et~al\mbox{.}(2022)]%
        {li2022query}
\bibfield{author}{\bibinfo{person}{Sen Li}, \bibinfo{person}{Fuyu Lv}, \bibinfo{person}{Taiwei Jin}, \bibinfo{person}{Guiyang Li}, \bibinfo{person}{Yukun Zheng}, \bibinfo{person}{Tao Zhuang}, \bibinfo{person}{Qingwen Liu}, \bibinfo{person}{Xiaoyi Zeng}, \bibinfo{person}{James Kwok}, {and} \bibinfo{person}{Qianli Ma}.} \bibinfo{year}{2022}\natexlab{}.
\newblock \showarticletitle{Query rewriting in taobao search}. In \bibinfo{booktitle}{\emph{Proceedings of the 31st ACM International Conference on Information \& Knowledge Management}}. \bibinfo{pages}{3262--3271}.
\newblock


\bibitem[Li et~al\mbox{.}(2023a)]%
        {li2023agent4ranking}
\bibfield{author}{\bibinfo{person}{Xiaopeng Li}, \bibinfo{person}{Lixin Su}, \bibinfo{person}{Pengyue Jia}, \bibinfo{person}{Xiangyu Zhao}, \bibinfo{person}{Suqi Cheng}, \bibinfo{person}{Junfeng Wang}, {and} \bibinfo{person}{Dawei Yin}.} \bibinfo{year}{2023}\natexlab{a}.
\newblock \showarticletitle{Agent4ranking: Semantic robust ranking via personalized query rewriting using multi-agent llm}.
\newblock \bibinfo{journal}{\emph{arXiv preprint arXiv:2312.15450}} (\bibinfo{year}{2023}).
\newblock


\bibitem[Li et~al\mbox{.}(2024c)]%
        {li2024personalized}
\bibfield{author}{\bibinfo{person}{Xinyu Li}, \bibinfo{person}{Ruiyang Zhou}, \bibinfo{person}{Zachary~C Lipton}, {and} \bibinfo{person}{Liu Leqi}.} \bibinfo{year}{2024}\natexlab{c}.
\newblock \showarticletitle{Personalized language modeling from personalized human feedback}.
\newblock \bibinfo{journal}{\emph{arXiv preprint arXiv:2402.05133}} (\bibinfo{year}{2024}).
\newblock


\bibitem[Li et~al\mbox{.}(2024a)]%
        {li2024personal}
\bibfield{author}{\bibinfo{person}{Yuanchun Li}, \bibinfo{person}{Hao Wen}, \bibinfo{person}{Weijun Wang}, \bibinfo{person}{Xiangyu Li}, \bibinfo{person}{Yizhen Yuan}, \bibinfo{person}{Guohong Liu}, \bibinfo{person}{Jiacheng Liu}, \bibinfo{person}{Wenxing Xu}, \bibinfo{person}{Xiang Wang}, \bibinfo{person}{Yi Sun}, {et~al\mbox{.}}} \bibinfo{year}{2024}\natexlab{a}.
\newblock \showarticletitle{Personal llm agents: Insights and survey about the capability, efficiency and security}.
\newblock \bibinfo{journal}{\emph{arXiv preprint arXiv:2401.05459}} (\bibinfo{year}{2024}).
\newblock


\bibitem[Li et~al\mbox{.}(2023c)]%
        {li2023towards}
\bibfield{author}{\bibinfo{person}{Zehan Li}, \bibinfo{person}{Xin Zhang}, \bibinfo{person}{Yanzhao Zhang}, \bibinfo{person}{Dingkun Long}, \bibinfo{person}{Pengjun Xie}, {and} \bibinfo{person}{Meishan Zhang}.} \bibinfo{year}{2023}\natexlab{c}.
\newblock \showarticletitle{Towards general text embeddings with multi-stage contrastive learning}.
\newblock \bibinfo{journal}{\emph{arXiv preprint arXiv:2308.03281}} (\bibinfo{year}{2023}).
\newblock


\bibitem[Lian et~al\mbox{.}(2023)]%
        {lian2023personaltm}
\bibfield{author}{\bibinfo{person}{Ruixue Lian}, \bibinfo{person}{Sixing Lu}, \bibinfo{person}{Clint Solomon}, \bibinfo{person}{Gustavo Aguilar}, \bibinfo{person}{Pragaash Ponnusamy}, \bibinfo{person}{Jialong Han}, \bibinfo{person}{Chengyuan Ma}, {and} \bibinfo{person}{Chenlei Guo}.} \bibinfo{year}{2023}\natexlab{}.
\newblock \showarticletitle{PersonalTM: Transformer memory for personalized retrieval}. In \bibinfo{booktitle}{\emph{Proceedings of the 46th International ACM SIGIR Conference on Research and Development in Information Retrieval}}. \bibinfo{pages}{2256--2260}.
\newblock


\bibitem[Lin and Huang(2006)]%
        {lin2006personalized}
\bibfield{author}{\bibinfo{person}{Shan-Mu Lin} {and} \bibinfo{person}{Chuen-Min Huang}.} \bibinfo{year}{2006}\natexlab{}.
\newblock \showarticletitle{Personalized optimal search in local query expansion}. In \bibinfo{booktitle}{\emph{Proceedings of the 18th Conference on Computational Linguistics and Speech Processing}}. \bibinfo{pages}{221--236}.
\newblock


\bibitem[Liu et~al\mbox{.}(2023b)]%
        {liu2023chatgpt}
\bibfield{author}{\bibinfo{person}{Junling Liu}, \bibinfo{person}{Chao Liu}, \bibinfo{person}{Peilin Zhou}, \bibinfo{person}{Renjie Lv}, \bibinfo{person}{Kang Zhou}, {and} \bibinfo{person}{Yan Zhang}.} \bibinfo{year}{2023}\natexlab{b}.
\newblock \showarticletitle{Is chatgpt a good recommender? a preliminary study}.
\newblock \bibinfo{journal}{\emph{arXiv preprint arXiv:2304.10149}} (\bibinfo{year}{2023}).
\newblock


\bibitem[Liu et~al\mbox{.}(2025)]%
        {liu2025survey}
\bibfield{author}{\bibinfo{person}{Jiahong Liu}, \bibinfo{person}{Zexuan Qiu}, \bibinfo{person}{Zhongyang Li}, \bibinfo{person}{Quanyu Dai}, \bibinfo{person}{Jieming Zhu}, \bibinfo{person}{Minda Hu}, \bibinfo{person}{Menglin Yang}, {and} \bibinfo{person}{Irwin King}.} \bibinfo{year}{2025}\natexlab{}.
\newblock \showarticletitle{A Survey of Personalized Large Language Models: Progress and Future Directions}.
\newblock \bibinfo{journal}{\emph{arXiv preprint arXiv:2502.11528}} (\bibinfo{year}{2025}).
\newblock


\bibitem[Liu et~al\mbox{.}(2024b)]%
        {liu2024lost}
\bibfield{author}{\bibinfo{person}{Nelson~F Liu}, \bibinfo{person}{Kevin Lin}, \bibinfo{person}{John Hewitt}, \bibinfo{person}{Ashwin Paranjape}, \bibinfo{person}{Michele Bevilacqua}, \bibinfo{person}{Fabio Petroni}, {and} \bibinfo{person}{Percy Liang}.} \bibinfo{year}{2024}\natexlab{b}.
\newblock \showarticletitle{Lost in the middle: How language models use long contexts}.
\newblock \bibinfo{journal}{\emph{Transactions of the Association for Computational Linguistics}}  \bibinfo{volume}{12} (\bibinfo{year}{2024}), \bibinfo{pages}{157--173}.
\newblock


\bibitem[Liu et~al\mbox{.}(2024a)]%
        {liu2024once}
\bibfield{author}{\bibinfo{person}{Qijiong Liu}, \bibinfo{person}{Nuo Chen}, \bibinfo{person}{Tetsuya Sakai}, {and} \bibinfo{person}{Xiao-Ming Wu}.} \bibinfo{year}{2024}\natexlab{a}.
\newblock \showarticletitle{Once: Boosting content-based recommendation with both open-and closed-source large language models}. In \bibinfo{booktitle}{\emph{Proceedings of the 17th ACM International Conference on Web Search and Data Mining}}. \bibinfo{pages}{452--461}.
\newblock


\bibitem[Liu et~al\mbox{.}(2023a)]%
        {liu2023recap}
\bibfield{author}{\bibinfo{person}{Shuai Liu}, \bibinfo{person}{Hyundong~J Cho}, \bibinfo{person}{Marjorie Freedman}, \bibinfo{person}{Xuezhe Ma}, {and} \bibinfo{person}{Jonathan May}.} \bibinfo{year}{2023}\natexlab{a}.
\newblock \showarticletitle{RECAP: retrieval-enhanced context-aware prefix encoder for personalized dialogue response generation}.
\newblock \bibinfo{journal}{\emph{arXiv preprint arXiv:2306.07206}} (\bibinfo{year}{2023}).
\newblock


\bibitem[Lu and Boutilier(2011)]%
        {lu2011budgeted}
\bibfield{author}{\bibinfo{person}{Tyler Lu} {and} \bibinfo{person}{Craig Boutilier}.} \bibinfo{year}{2011}\natexlab{}.
\newblock \showarticletitle{Budgeted social choice: From consensus to personalized decision making}. In \bibinfo{booktitle}{\emph{IJCAI}}, Vol.~\bibinfo{volume}{11}. \bibinfo{pages}{280--286}.
\newblock


\bibitem[Ma et~al\mbox{.}(2021)]%
        {ma2021one}
\bibfield{author}{\bibinfo{person}{Zhengyi Ma}, \bibinfo{person}{Zhicheng Dou}, \bibinfo{person}{Yutao Zhu}, \bibinfo{person}{Hanxun Zhong}, {and} \bibinfo{person}{Ji-Rong Wen}.} \bibinfo{year}{2021}\natexlab{}.
\newblock \showarticletitle{One chatbot per person: Creating personalized chatbots based on implicit user profiles}. In \bibinfo{booktitle}{\emph{Proceedings of the 44th international ACM SIGIR conference on research and development in information retrieval}}. \bibinfo{pages}{555--564}.
\newblock


\bibitem[Madaan et~al\mbox{.}(2022)]%
        {madaan2022memory}
\bibfield{author}{\bibinfo{person}{Aman Madaan}, \bibinfo{person}{Niket Tandon}, \bibinfo{person}{Peter Clark}, {and} \bibinfo{person}{Yiming Yang}.} \bibinfo{year}{2022}\natexlab{}.
\newblock \showarticletitle{Memory-assisted prompt editing to improve GPT-3 after deployment}.
\newblock \bibinfo{journal}{\emph{arXiv preprint arXiv:2201.06009}} (\bibinfo{year}{2022}).
\newblock


\bibitem[Mao et~al\mbox{.}(2024b)]%
        {mao2024editing}
\bibfield{author}{\bibinfo{person}{Shengyu Mao}, \bibinfo{person}{Xiaohan Wang}, \bibinfo{person}{Mengru Wang}, \bibinfo{person}{Yong Jiang}, \bibinfo{person}{Pengjun Xie}, \bibinfo{person}{Fei Huang}, {and} \bibinfo{person}{Ningyu Zhang}.} \bibinfo{year}{2024}\natexlab{b}.
\newblock \showarticletitle{Editing Personality for Large Language Models}. In \bibinfo{booktitle}{\emph{CCF International Conference on Natural Language Processing and Chinese Computing}}. Springer, \bibinfo{pages}{241--254}.
\newblock


\bibitem[Mao et~al\mbox{.}(2024a)]%
        {mao2024fit}
\bibfield{author}{\bibinfo{person}{Yuren Mao}, \bibinfo{person}{Xuemei Dong}, \bibinfo{person}{Wenyi Xu}, \bibinfo{person}{Yunjun Gao}, \bibinfo{person}{Bin Wei}, {and} \bibinfo{person}{Ying Zhang}.} \bibinfo{year}{2024}\natexlab{a}.
\newblock \showarticletitle{Fit-rag: black-box rag with factual information and token reduction}.
\newblock \bibinfo{journal}{\emph{arXiv preprint arXiv:2403.14374}} (\bibinfo{year}{2024}).
\newblock


\bibitem[Mathur et~al\mbox{.}(2023)]%
        {mathur2023personalm}
\bibfield{author}{\bibinfo{person}{Puneet Mathur}, \bibinfo{person}{Zhe Liu}, \bibinfo{person}{Ke Li}, \bibinfo{person}{Yingyi Ma}, \bibinfo{person}{Gil Keren}, \bibinfo{person}{Zeeshan Ahmed}, \bibinfo{person}{Dinesh Manocha}, {and} \bibinfo{person}{Xuedong Zhang}.} \bibinfo{year}{2023}\natexlab{}.
\newblock \showarticletitle{Personalm: Language model personalization via domain-distributed span aggregated k-nearest n-gram retrieval augmentation}. In \bibinfo{booktitle}{\emph{Findings of the Association for Computational Linguistics: EMNLP 2023}}. \bibinfo{pages}{11314--11328}.
\newblock


\bibitem[Mireshghallah et~al\mbox{.}(2021)]%
        {mireshghallah2021useridentifier}
\bibfield{author}{\bibinfo{person}{Fatemehsadat Mireshghallah}, \bibinfo{person}{Vaishnavi Shrivastava}, \bibinfo{person}{Milad Shokouhi}, \bibinfo{person}{Taylor Berg-Kirkpatrick}, \bibinfo{person}{Robert Sim}, {and} \bibinfo{person}{Dimitrios Dimitriadis}.} \bibinfo{year}{2021}\natexlab{}.
\newblock \showarticletitle{Useridentifier: Implicit user representations for simple and effective personalized sentiment analysis}.
\newblock \bibinfo{journal}{\emph{arXiv preprint arXiv:2110.00135}} (\bibinfo{year}{2021}).
\newblock


\bibitem[Mulhem et~al\mbox{.}(2016)]%
        {mulhem2016axiomatic}
\bibfield{author}{\bibinfo{person}{Philippe Mulhem}, \bibinfo{person}{Nawal~Ould Amer}, {and} \bibinfo{person}{Mathias G{\'e}ry}.} \bibinfo{year}{2016}\natexlab{}.
\newblock \showarticletitle{Axiomatic term-based personalized query expansion using bookmarking system}. In \bibinfo{booktitle}{\emph{International Conference on Database and Expert Systems Applications}}. Springer, \bibinfo{pages}{235--243}.
\newblock


\bibitem[Mysore et~al\mbox{.}(2023)]%
        {mysore2023pearl}
\bibfield{author}{\bibinfo{person}{Sheshera Mysore}, \bibinfo{person}{Zhuoran Lu}, \bibinfo{person}{Mengting Wan}, \bibinfo{person}{Longqi Yang}, \bibinfo{person}{Steve Menezes}, \bibinfo{person}{Tina Baghaee}, \bibinfo{person}{Emmanuel~Barajas Gonzalez}, \bibinfo{person}{Jennifer Neville}, {and} \bibinfo{person}{Tara Safavi}.} \bibinfo{year}{2023}\natexlab{}.
\newblock \showarticletitle{Pearl: Personalizing large language model writing assistants with generation-calibrated retrievers}.
\newblock \bibinfo{journal}{\emph{arXiv preprint arXiv:2311.09180}} (\bibinfo{year}{2023}).
\newblock


\bibitem[Nguyen et~al\mbox{.}(2024)]%
        {nguyen2024gui}
\bibfield{author}{\bibinfo{person}{Dang Nguyen}, \bibinfo{person}{Jian Chen}, \bibinfo{person}{Yu Wang}, \bibinfo{person}{Gang Wu}, \bibinfo{person}{Namyong Park}, \bibinfo{person}{Zhengmian Hu}, \bibinfo{person}{Hanjia Lyu}, \bibinfo{person}{Junda Wu}, \bibinfo{person}{Ryan Aponte}, \bibinfo{person}{Yu Xia}, {et~al\mbox{.}}} \bibinfo{year}{2024}\natexlab{}.
\newblock \showarticletitle{Gui agents: A survey}.
\newblock \bibinfo{journal}{\emph{arXiv preprint arXiv:2412.13501}} (\bibinfo{year}{2024}).
\newblock


\bibitem[Nguyen et~al\mbox{.}(2025)]%
        {nguyen2025rl}
\bibfield{author}{\bibinfo{person}{Duy~A Nguyen}, \bibinfo{person}{Rishi~Kesav Mohan}, \bibinfo{person}{Van Yang}, \bibinfo{person}{Pritom~Saha Akash}, {and} \bibinfo{person}{Kevin Chen-Chuan Chang}.} \bibinfo{year}{2025}\natexlab{}.
\newblock \showarticletitle{RL-based Query Rewriting with Distilled LLM for online E-Commerce Systems}.
\newblock \bibinfo{journal}{\emph{arXiv preprint arXiv:2501.18056}} (\bibinfo{year}{2025}).
\newblock


\bibitem[Ni et~al\mbox{.}(2019)]%
        {ni2019justifying}
\bibfield{author}{\bibinfo{person}{Jianmo Ni}, \bibinfo{person}{Jiacheng Li}, {and} \bibinfo{person}{Julian McAuley}.} \bibinfo{year}{2019}\natexlab{}.
\newblock \showarticletitle{Justifying recommendations using distantly-labeled reviews and fine-grained aspects}. In \bibinfo{booktitle}{\emph{Proceedings of the 2019 conference on empirical methods in natural language processing and the 9th international joint conference on natural language processing (EMNLP-IJCNLP)}}. \bibinfo{pages}{188--197}.
\newblock


\bibitem[Ning et~al\mbox{.}(2024)]%
        {ning2024user}
\bibfield{author}{\bibinfo{person}{Lin Ning}, \bibinfo{person}{Luyang Liu}, \bibinfo{person}{Jiaxing Wu}, \bibinfo{person}{Neo Wu}, \bibinfo{person}{Devora Berlowitz}, \bibinfo{person}{Sushant Prakash}, \bibinfo{person}{Bradley Green}, \bibinfo{person}{Shawn O'Banion}, {and} \bibinfo{person}{Jun Xie}.} \bibinfo{year}{2024}\natexlab{}.
\newblock \showarticletitle{User-llm: Efficient llm contextualization with user embeddings}.
\newblock \bibinfo{journal}{\emph{arXiv preprint arXiv:2402.13598}} (\bibinfo{year}{2024}).
\newblock


\bibitem[Oard et~al\mbox{.}(2015)]%
        {oard2015avocado}
\bibfield{author}{\bibinfo{person}{Douglas Oard}, \bibinfo{person}{William Webber}, \bibinfo{person}{David Kirsch}, {and} \bibinfo{person}{Sergey Golitsynskiy}.} \bibinfo{year}{2015}\natexlab{}.
\newblock \showarticletitle{Avocado research email collection}.
\newblock \bibinfo{journal}{\emph{Philadelphia: Linguistic Data Consortium}} (\bibinfo{year}{2015}).
\newblock


\bibitem[of~Medicine({[n.\,d.]})]%
        {pubmed}
\bibfield{author}{\bibinfo{person}{U.S. National~Library of Medicine}.} \bibinfo{year}{[n.\,d.]}\natexlab{}.
\newblock \bibinfo{booktitle}{\emph{PubMed: A Free Resource for Biomedical Literature}}.
\newblock
\urldef\tempurl%
\url{https://pubmed.ncbi.nlm.nih.gov/}
\showURL{%
\tempurl}


\bibitem[Park et~al\mbox{.}(2023)]%
        {park2023generative}
\bibfield{author}{\bibinfo{person}{Joon~Sung Park}, \bibinfo{person}{Joseph O'Brien}, \bibinfo{person}{Carrie~Jun Cai}, \bibinfo{person}{Meredith~Ringel Morris}, \bibinfo{person}{Percy Liang}, {and} \bibinfo{person}{Michael~S Bernstein}.} \bibinfo{year}{2023}\natexlab{}.
\newblock \showarticletitle{Generative agents: Interactive simulacra of human behavior}. In \bibinfo{booktitle}{\emph{Proceedings of the 36th annual acm symposium on user interface software and technology}}. \bibinfo{pages}{1--22}.
\newblock


\bibitem[Pass et~al\mbox{.}(2006)]%
        {AOL}
\bibfield{author}{\bibinfo{person}{Greg Pass}, \bibinfo{person}{Abdur Chowdhury}, {and} \bibinfo{person}{Cayley Torgeson}.} \bibinfo{year}{2006}\natexlab{}.
\newblock \showarticletitle{A picture of search}. In \bibinfo{booktitle}{\emph{Proceedings of the 1st International Conference on Scalable Information Systems}} (Hong Kong) \emph{(\bibinfo{series}{InfoScale '06})}. \bibinfo{publisher}{Association for Computing Machinery}, \bibinfo{address}{New York, NY, USA}, \bibinfo{pages}{1–es}.
\newblock
\showISBNx{1595934286}
\urldef\tempurl%
\url{https://doi.org/10.1145/1146847.1146848}
\showDOI{\tempurl}


\bibitem[Pavliukevich et~al\mbox{.}({[n.\,d.]})]%
        {pavliukevich2024improving}
\bibfield{author}{\bibinfo{person}{Vadim~Igorevich Pavliukevich}, \bibinfo{person}{Alina~Khasanovna Zherdeva}, \bibinfo{person}{Olesya~Vladimirovna Makhnytkina}, {and} \bibinfo{person}{Dmitriy~Viktorovich Dyrmovskiy}.} \bibinfo{year}{[n.\,d.]}\natexlab{}.
\newblock \showarticletitle{Improving RAG with LoRA finetuning for persona text generation}.
\newblock  (\bibinfo{year}{[n.\,d.]}).
\newblock


\bibitem[Peng et~al\mbox{.}(2024a)]%
        {peng2024pocketllm}
\bibfield{author}{\bibinfo{person}{Dan Peng}, \bibinfo{person}{Zhihui Fu}, {and} \bibinfo{person}{Jun Wang}.} \bibinfo{year}{2024}\natexlab{a}.
\newblock \showarticletitle{Pocketllm: Enabling on-device fine-tuning for personalized llms}.
\newblock \bibinfo{journal}{\emph{arXiv preprint arXiv:2407.01031}} (\bibinfo{year}{2024}).
\newblock


\bibitem[Peng et~al\mbox{.}(2024b)]%
        {peng2024reviewllm}
\bibfield{author}{\bibinfo{person}{Qiyao Peng}, \bibinfo{person}{Hongtao Liu}, \bibinfo{person}{Hongyan Xu}, \bibinfo{person}{Qing Yang}, \bibinfo{person}{Minglai Shao}, {and} \bibinfo{person}{Wenjun Wang}.} \bibinfo{year}{2024}\natexlab{b}.
\newblock \bibinfo{title}{Review-LLM: Harnessing Large Language Models for Personalized Review Generation}.
\newblock
\newblock
\showeprint[arxiv]{2407.07487}~[cs.CL]
\urldef\tempurl%
\url{https://arxiv.org/abs/2407.07487}
\showURL{%
\tempurl}


\bibitem[Qian et~al\mbox{.}(2021a)]%
        {qian2021learning}
\bibfield{author}{\bibinfo{person}{Hongjin Qian}, \bibinfo{person}{Zhicheng Dou}, \bibinfo{person}{Yutao Zhu}, \bibinfo{person}{Yueyuan Ma}, {and} \bibinfo{person}{Ji-Rong Wen}.} \bibinfo{year}{2021}\natexlab{a}.
\newblock \showarticletitle{Learning implicit user profile for personalized retrieval-based chatbot}. In \bibinfo{booktitle}{\emph{proceedings of the 30th ACM international conference on Information \& Knowledge Management}}. \bibinfo{pages}{1467--1477}.
\newblock


\bibitem[Qian et~al\mbox{.}(2021b)]%
        {qian2021pchatbot}
\bibfield{author}{\bibinfo{person}{Hongjin Qian}, \bibinfo{person}{Xiaohe Li}, \bibinfo{person}{Hanxun Zhong}, \bibinfo{person}{Yu Guo}, \bibinfo{person}{Yueyuan Ma}, \bibinfo{person}{Yutao Zhu}, \bibinfo{person}{Zhanliang Liu}, \bibinfo{person}{Zhicheng Dou}, {and} \bibinfo{person}{Ji-Rong Wen}.} \bibinfo{year}{2021}\natexlab{b}.
\newblock \showarticletitle{Pchatbot: a large-scale dataset for personalized chatbot}. In \bibinfo{booktitle}{\emph{Proceedings of the 44th international ACM SIGIR conference on research and development in information retrieval}}. \bibinfo{pages}{2470--2477}.
\newblock


\bibitem[Qu et~al\mbox{.}(2024)]%
        {qu2024graph}
\bibfield{author}{\bibinfo{person}{Xiaoru Qu}, \bibinfo{person}{Yifan Wang}, \bibinfo{person}{Zhao Li}, {and} \bibinfo{person}{Jun Gao}.} \bibinfo{year}{2024}\natexlab{}.
\newblock \showarticletitle{Graph-enhanced prompt learning for personalized review generation}.
\newblock \bibinfo{journal}{\emph{Data Science and Engineering}} \bibinfo{volume}{9}, \bibinfo{number}{3} (\bibinfo{year}{2024}), \bibinfo{pages}{309--324}.
\newblock


\bibitem[Rajaraman and Ullman(2011)]%
        {rajaraman2011mining}
\bibfield{author}{\bibinfo{person}{A. Rajaraman} {and} \bibinfo{person}{J.D. Ullman}.} \bibinfo{year}{2011}\natexlab{}.
\newblock \bibinfo{booktitle}{\emph{Mining of Massive Datasets}}.
\newblock \bibinfo{publisher}{Cambridge University Press}.
\newblock
\showISBNx{9781139505345}
\urldef\tempurl%
\url{https://books.google.co.uk/books?id=OefRhZyYOb0C}
\showURL{%
\tempurl}


\bibitem[Ran et~al\mbox{.}(2024)]%
        {ran2024capturing}
\bibfield{author}{\bibinfo{person}{Yiting Ran}, \bibinfo{person}{Xintao Wang}, \bibinfo{person}{Rui Xu}, \bibinfo{person}{Xinfeng Yuan}, \bibinfo{person}{Jiaqing Liang}, \bibinfo{person}{Deqing Yang}, {and} \bibinfo{person}{Yanghua Xiao}.} \bibinfo{year}{2024}\natexlab{}.
\newblock \showarticletitle{Capturing minds, not just words: Enhancing role-playing language models with personality-indicative data}.
\newblock \bibinfo{journal}{\emph{arXiv preprint arXiv:2406.18921}} (\bibinfo{year}{2024}).
\newblock


\bibitem[Reddy et~al\mbox{.}(2022)]%
        {reddy2022shopping}
\bibfield{author}{\bibinfo{person}{Chandan~K. Reddy}, \bibinfo{person}{Lluís Màrquez}, \bibinfo{person}{Fran Valero}, \bibinfo{person}{Nikhil Rao}, \bibinfo{person}{Hugo Zaragoza}, \bibinfo{person}{Sambaran Bandyopadhyay}, \bibinfo{person}{Arnab Biswas}, \bibinfo{person}{Anlu Xing}, {and} \bibinfo{person}{Karthik Subbian}.} \bibinfo{year}{2022}\natexlab{}.
\newblock \showarticletitle{Shopping Queries Dataset: A Large-Scale {ESCI} Benchmark for Improving Product Search}.
\newblock  (\bibinfo{year}{2022}).
\newblock
\showeprint[arxiv]{2206.06588}


\bibitem[Reimers and Gurevych(2019)]%
        {reimers2019sentence}
\bibfield{author}{\bibinfo{person}{Nils Reimers} {and} \bibinfo{person}{Iryna Gurevych}.} \bibinfo{year}{2019}\natexlab{}.
\newblock \showarticletitle{Sentence-bert: Sentence embeddings using siamese bert-networks}.
\newblock \bibinfo{journal}{\emph{arXiv preprint arXiv:1908.10084}} (\bibinfo{year}{2019}).
\newblock


\bibitem[Ren et~al\mbox{.}(2024)]%
        {ren2024bases}
\bibfield{author}{\bibinfo{person}{Ruiyang Ren}, \bibinfo{person}{Peng Qiu}, \bibinfo{person}{Yingqi Qu}, \bibinfo{person}{Jing Liu}, \bibinfo{person}{Wayne~Xin Zhao}, \bibinfo{person}{Hua Wu}, \bibinfo{person}{Ji-Rong Wen}, {and} \bibinfo{person}{Haifeng Wang}.} \bibinfo{year}{2024}\natexlab{}.
\newblock \showarticletitle{Bases: Large-scale web search user simulation with large language model based agents}.
\newblock \bibinfo{journal}{\emph{arXiv preprint arXiv:2402.17505}} (\bibinfo{year}{2024}).
\newblock


\bibitem[Renze and Guven(2024)]%
        {renze2024self}
\bibfield{author}{\bibinfo{person}{Matthew Renze} {and} \bibinfo{person}{Erhan Guven}.} \bibinfo{year}{2024}\natexlab{}.
\newblock \showarticletitle{Self-reflection in llm agents: Effects on problem-solving performance}.
\newblock \bibinfo{journal}{\emph{arXiv preprint arXiv:2405.06682}} (\bibinfo{year}{2024}).
\newblock


\bibitem[Richardson et~al\mbox{.}(2023)]%
        {richardson2023integrating}
\bibfield{author}{\bibinfo{person}{Chris Richardson}, \bibinfo{person}{Yao Zhang}, \bibinfo{person}{Kellen Gillespie}, \bibinfo{person}{Sudipta Kar}, \bibinfo{person}{Arshdeep Singh}, \bibinfo{person}{Zeynab Raeesy}, \bibinfo{person}{Omar~Zia Khan}, {and} \bibinfo{person}{Abhinav Sethy}.} \bibinfo{year}{2023}\natexlab{}.
\newblock \showarticletitle{Integrating summarization and retrieval for enhanced personalization via large language models}.
\newblock \bibinfo{journal}{\emph{arXiv preprint arXiv:2310.20081}} (\bibinfo{year}{2023}).
\newblock


\bibitem[Robertson et~al\mbox{.}(2009)]%
        {robertson2009probabilistic}
\bibfield{author}{\bibinfo{person}{Stephen Robertson}, \bibinfo{person}{Hugo Zaragoza}, {et~al\mbox{.}}} \bibinfo{year}{2009}\natexlab{}.
\newblock \showarticletitle{The probabilistic relevance framework: BM25 and beyond}.
\newblock \bibinfo{journal}{\emph{Foundations and Trends{\textregistered} in Information Retrieval}} \bibinfo{volume}{3}, \bibinfo{number}{4} (\bibinfo{year}{2009}), \bibinfo{pages}{333--389}.
\newblock


\bibitem[Salemi et~al\mbox{.}(2024a)]%
        {salemi2024optimization}
\bibfield{author}{\bibinfo{person}{Alireza Salemi}, \bibinfo{person}{Surya Kallumadi}, {and} \bibinfo{person}{Hamed Zamani}.} \bibinfo{year}{2024}\natexlab{a}.
\newblock \showarticletitle{Optimization methods for personalizing large language models through retrieval augmentation}. In \bibinfo{booktitle}{\emph{Proceedings of the 47th International ACM SIGIR Conference on Research and Development in Information Retrieval}}. \bibinfo{pages}{752--762}.
\newblock


\bibitem[Salemi et~al\mbox{.}(2025)]%
        {salemi2025reasoning}
\bibfield{author}{\bibinfo{person}{Alireza Salemi}, \bibinfo{person}{Cheng Li}, \bibinfo{person}{Mingyang Zhang}, \bibinfo{person}{Qiaozhu Mei}, \bibinfo{person}{Weize Kong}, \bibinfo{person}{Tao Chen}, \bibinfo{person}{Zhuowan Li}, \bibinfo{person}{Michael Bendersky}, {and} \bibinfo{person}{Hamed Zamani}.} \bibinfo{year}{2025}\natexlab{}.
\newblock \showarticletitle{Reasoning-Enhanced Self-Training for Long-Form Personalized Text Generation}.
\newblock \bibinfo{journal}{\emph{arXiv preprint arXiv:2501.04167}} (\bibinfo{year}{2025}).
\newblock


\bibitem[Salemi et~al\mbox{.}(2024b)]%
        {salemi2024lamp}
\bibfield{author}{\bibinfo{person}{Alireza Salemi}, \bibinfo{person}{Sheshera Mysore}, \bibinfo{person}{Michael Bendersky}, {and} \bibinfo{person}{Hamed Zamani}.} \bibinfo{year}{2024}\natexlab{b}.
\newblock \showarticletitle{LaMP: When Large Language Models Meet Personalization}. In \bibinfo{booktitle}{\emph{Proceedings of the 62nd Annual Meeting of the Association for Computational Linguistics (Volume 1: Long Papers)}}. \bibinfo{pages}{7370--7392}.
\newblock


\bibitem[Salemi and Zamani(2024)]%
        {salemi2024learning}
\bibfield{author}{\bibinfo{person}{Alireza Salemi} {and} \bibinfo{person}{Hamed Zamani}.} \bibinfo{year}{2024}\natexlab{}.
\newblock \showarticletitle{Learning to Rank for Multiple Retrieval-Augmented Models through Iterative Utility Maximization}.
\newblock \bibinfo{journal}{\emph{arXiv preprint arXiv:2410.09942}} (\bibinfo{year}{2024}).
\newblock


\bibitem[Santurkar et~al\mbox{.}(2023)]%
        {santurkar2023whose}
\bibfield{author}{\bibinfo{person}{Shibani Santurkar}, \bibinfo{person}{Esin Durmus}, \bibinfo{person}{Faisal Ladhak}, \bibinfo{person}{Cinoo Lee}, \bibinfo{person}{Percy Liang}, {and} \bibinfo{person}{Tatsunori Hashimoto}.} \bibinfo{year}{2023}\natexlab{}.
\newblock \showarticletitle{Whose opinions do language models reflect?}. In \bibinfo{booktitle}{\emph{International Conference on Machine Learning}}. PMLR, \bibinfo{pages}{29971--30004}.
\newblock


\bibitem[Schifanella et~al\mbox{.}(2010)]%
        {schifanella2010folks}
\bibfield{author}{\bibinfo{person}{Rossano Schifanella}, \bibinfo{person}{Alain Barrat}, \bibinfo{person}{Ciro Cattuto}, \bibinfo{person}{Benjamin Markines}, {and} \bibinfo{person}{Filippo Menczer}.} \bibinfo{year}{2010}\natexlab{}.
\newblock \showarticletitle{Folks in folksonomies: social link prediction from shared metadata}. In \bibinfo{booktitle}{\emph{Proceedings of the third ACM international conference on Web search and data mining}}. \bibinfo{pages}{271--280}.
\newblock


\bibitem[Shaker et~al\mbox{.}(2010)]%
        {shaker2010towards}
\bibfield{author}{\bibinfo{person}{Noor Shaker}, \bibinfo{person}{Georgios Yannakakis}, {and} \bibinfo{person}{Julian Togelius}.} \bibinfo{year}{2010}\natexlab{}.
\newblock \showarticletitle{Towards automatic personalized content generation for platform games}. In \bibinfo{booktitle}{\emph{Proceedings of the AAAI Conference on Artificial Intelligence and Interactive Digital Entertainment}}, Vol.~\bibinfo{volume}{6}. \bibinfo{pages}{63--68}.
\newblock


\bibitem[Shao et~al\mbox{.}(2023)]%
        {shao2023character}
\bibfield{author}{\bibinfo{person}{Yunfan Shao}, \bibinfo{person}{Linyang Li}, \bibinfo{person}{Junqi Dai}, {and} \bibinfo{person}{Xipeng Qiu}.} \bibinfo{year}{2023}\natexlab{}.
\newblock \showarticletitle{Character-llm: A trainable agent for role-playing}.
\newblock \bibinfo{journal}{\emph{arXiv preprint arXiv:2310.10158}} (\bibinfo{year}{2023}).
\newblock


\bibitem[Shen et~al\mbox{.}(2024)]%
        {shen2024heart}
\bibfield{author}{\bibinfo{person}{Jocelyn Shen}, \bibinfo{person}{Joel Mire}, \bibinfo{person}{Hae~Won Park}, \bibinfo{person}{Cynthia Breazeal}, {and} \bibinfo{person}{Maarten Sap}.} \bibinfo{year}{2024}\natexlab{}.
\newblock \showarticletitle{HEART-felt Narratives: Tracing Empathy and Narrative Style in Personal Stories with LLMs}.
\newblock \bibinfo{journal}{\emph{arXiv preprint arXiv:2405.17633}} (\bibinfo{year}{2024}).
\newblock


\bibitem[Shi et~al\mbox{.}(2024)]%
        {shi2024eragent}
\bibfield{author}{\bibinfo{person}{Yunxiao Shi}, \bibinfo{person}{Xing Zi}, \bibinfo{person}{Zijing Shi}, \bibinfo{person}{Haimin Zhang}, \bibinfo{person}{Qiang Wu}, {and} \bibinfo{person}{Min Xu}.} \bibinfo{year}{2024}\natexlab{}.
\newblock \showarticletitle{Eragent: Enhancing retrieval-augmented language models with improved accuracy, efficiency, and personalization}.
\newblock \bibinfo{journal}{\emph{arXiv preprint arXiv:2405.06683}} (\bibinfo{year}{2024}).
\newblock


\bibitem[Singh et~al\mbox{.}(2025)]%
        {singh2025agentic}
\bibfield{author}{\bibinfo{person}{Aditi Singh}, \bibinfo{person}{Abul Ehtesham}, \bibinfo{person}{Saket Kumar}, {and} \bibinfo{person}{Tala~Talaei Khoei}.} \bibinfo{year}{2025}\natexlab{}.
\newblock \showarticletitle{Agentic Retrieval-Augmented Generation: A Survey on Agentic RAG}.
\newblock \bibinfo{journal}{\emph{arXiv preprint arXiv:2501.09136}} (\bibinfo{year}{2025}).
\newblock


\bibitem[Singh et~al\mbox{.}(2024)]%
        {singh2024personal}
\bibfield{author}{\bibinfo{person}{Harmanpreet Singh}, \bibinfo{person}{Nikhil Verma}, \bibinfo{person}{Yixiao Wang}, \bibinfo{person}{Manasa Bharadwaj}, \bibinfo{person}{Homa Fashandi}, \bibinfo{person}{Kevin Ferreira}, {and} \bibinfo{person}{Chul Lee}.} \bibinfo{year}{2024}\natexlab{}.
\newblock \showarticletitle{Personal Large Language Model Agents: A Case Study on Tailored Travel Planning}. In \bibinfo{booktitle}{\emph{Proceedings of the 2024 Conference on Empirical Methods in Natural Language Processing: Industry Track}}. \bibinfo{pages}{486--514}.
\newblock


\bibitem[Siriwardhana et~al\mbox{.}(2023)]%
        {siriwardhana2023improving}
\bibfield{author}{\bibinfo{person}{Shamane Siriwardhana}, \bibinfo{person}{Rivindu Weerasekera}, \bibinfo{person}{Elliott Wen}, \bibinfo{person}{Tharindu Kaluarachchi}, \bibinfo{person}{Rajib Rana}, {and} \bibinfo{person}{Suranga Nanayakkara}.} \bibinfo{year}{2023}\natexlab{}.
\newblock \showarticletitle{Improving the domain adaptation of retrieval augmented generation (RAG) models for open domain question answering}.
\newblock \bibinfo{journal}{\emph{Transactions of the Association for Computational Linguistics}}  \bibinfo{volume}{11} (\bibinfo{year}{2023}), \bibinfo{pages}{1--17}.
\newblock


\bibitem[Song and Zheng(2024)]%
        {song2024survey}
\bibfield{author}{\bibinfo{person}{Mingyang Song} {and} \bibinfo{person}{Mao Zheng}.} \bibinfo{year}{2024}\natexlab{}.
\newblock \showarticletitle{A Survey of Query Optimization in Large Language Models}.
\newblock \bibinfo{journal}{\emph{arXiv preprint arXiv:2412.17558}} (\bibinfo{year}{2024}).
\newblock


\bibitem[Spotify(2023)]%
        {annoy}
\bibfield{author}{\bibinfo{person}{Spotify}.} \bibinfo{year}{2023}\natexlab{}.
\newblock \bibinfo{title}{Annoy: Approximate Nearest Neighbors in C++/Python}.
\newblock
\newblock
\urldef\tempurl%
\url{https://github.com/spotify/annoy}
\showURL{%
\tempurl}


\bibitem[Stuck\_In\_the\_Matrix(2015)]%
        {reddit2015}
\bibfield{author}{\bibinfo{person}{Stuck\_In\_the\_Matrix}.} \bibinfo{year}{2015}\natexlab{}.
\newblock \showarticletitle{Reddit Public Comments (2007-10 through 2015-05)}.
\newblock  (\bibinfo{year}{2015}).
\newblock
\urldef\tempurl%
\url{https://www.reddit.com/r/datasets/comments/3bxlg7/i\_have\_every\_publicly\_available\_reddit\_comment/}
\showURL{%
\tempurl}


\bibitem[Sun et~al\mbox{.}(2024)]%
        {sun2024revealing}
\bibfield{author}{\bibinfo{person}{Lei Sun}, \bibinfo{person}{Jinming Zhao}, {and} \bibinfo{person}{Qin Jin}.} \bibinfo{year}{2024}\natexlab{}.
\newblock \showarticletitle{Revealing Personality Traits: A New Benchmark Dataset for Explainable Personality Recognition on Dialogues}.
\newblock \bibinfo{journal}{\emph{arXiv preprint arXiv:2409.19723}} (\bibinfo{year}{2024}).
\newblock


\bibitem[Tan et~al\mbox{.}(2024a)]%
        {tan2024personalized}
\bibfield{author}{\bibinfo{person}{Zhaoxuan Tan}, \bibinfo{person}{Zheyuan Liu}, {and} \bibinfo{person}{Meng Jiang}.} \bibinfo{year}{2024}\natexlab{a}.
\newblock \showarticletitle{Personalized pieces: Efficient personalized large language models through collaborative efforts}.
\newblock \bibinfo{journal}{\emph{arXiv preprint arXiv:2406.10471}} (\bibinfo{year}{2024}).
\newblock


\bibitem[Tan et~al\mbox{.}(2024b)]%
        {tan2024democratizing}
\bibfield{author}{\bibinfo{person}{Zhaoxuan Tan}, \bibinfo{person}{Qingkai Zeng}, \bibinfo{person}{Yijun Tian}, \bibinfo{person}{Zheyuan Liu}, \bibinfo{person}{Bing Yin}, {and} \bibinfo{person}{Meng Jiang}.} \bibinfo{year}{2024}\natexlab{b}.
\newblock \showarticletitle{Democratizing large language models via personalized parameter-efficient fine-tuning}.
\newblock \bibinfo{journal}{\emph{arXiv preprint arXiv:2402.04401}} (\bibinfo{year}{2024}).
\newblock


\bibitem[Tan et~al\mbox{.}(2025)]%
        {tan2025democratizing}
\bibfield{author}{\bibinfo{person}{Zhaoxuan Tan}, \bibinfo{person}{Qingkai Zeng}, \bibinfo{person}{Yijun Tian}, \bibinfo{person}{Zheyuan Liu}, \bibinfo{person}{Bing Yin}, {and} \bibinfo{person}{Meng Jiang}.} \bibinfo{year}{2025}\natexlab{}.
\newblock \bibinfo{title}{Democratizing Large Language Models via Personalized Parameter-Efficient Fine-tuning}.
\newblock
\newblock
\showeprint[arxiv]{2402.04401}~[cs.CL]
\urldef\tempurl%
\url{https://arxiv.org/abs/2402.04401}
\showURL{%
\tempurl}


\bibitem[Tu et~al\mbox{.}(2024)]%
        {tu2024charactereval}
\bibfield{author}{\bibinfo{person}{Quan Tu}, \bibinfo{person}{Shilong Fan}, \bibinfo{person}{Zihang Tian}, {and} \bibinfo{person}{Rui Yan}.} \bibinfo{year}{2024}\natexlab{}.
\newblock \showarticletitle{Charactereval: A chinese benchmark for role-playing conversational agent evaluation}.
\newblock \bibinfo{journal}{\emph{arXiv preprint arXiv:2401.01275}} (\bibinfo{year}{2024}).
\newblock


\bibitem[University({[n.\,d.]})]%
        {arxiv}
\bibfield{author}{\bibinfo{person}{Cornell University}.} \bibinfo{year}{[n.\,d.]}\natexlab{}.
\newblock \bibinfo{booktitle}{\emph{arXiv: An Open Access Repository for Research}}.
\newblock
\urldef\tempurl%
\url{https://arxiv.org/}
\showURL{%
\tempurl}


\bibitem[Vemuri et~al\mbox{.}(2023)]%
        {vemuri2023personalized}
\bibfield{author}{\bibinfo{person}{Hemanth Vemuri}, \bibinfo{person}{Sheshansh Agrawal}, \bibinfo{person}{Shivam Mittal}, \bibinfo{person}{Deepak Saini}, \bibinfo{person}{Akshay Soni}, \bibinfo{person}{Abhinav~V Sambasivan}, \bibinfo{person}{Wenhao Lu}, \bibinfo{person}{Yajun Wang}, \bibinfo{person}{Mehul Parsana}, \bibinfo{person}{Purushottam Kar}, {et~al\mbox{.}}} \bibinfo{year}{2023}\natexlab{}.
\newblock \showarticletitle{Personalized retrieval over millions of items}. In \bibinfo{booktitle}{\emph{Proceedings of the 46th International ACM SIGIR Conference on Research and Development in Information Retrieval}}. \bibinfo{pages}{1014--1022}.
\newblock


\bibitem[Wang et~al\mbox{.}(2023b)]%
        {wang2023enabling}
\bibfield{author}{\bibinfo{person}{Bryan Wang}, \bibinfo{person}{Gang Li}, {and} \bibinfo{person}{Yang Li}.} \bibinfo{year}{2023}\natexlab{b}.
\newblock \showarticletitle{Enabling conversational interaction with mobile ui using large language models}. In \bibinfo{booktitle}{\emph{Proceedings of the 2023 CHI Conference on Human Factors in Computing Systems}}. \bibinfo{pages}{1--17}.
\newblock


\bibitem[Wang et~al\mbox{.}(2023f)]%
        {wangvoyager}
\bibfield{author}{\bibinfo{person}{Guanzhi Wang}, \bibinfo{person}{Yuqi Xie}, \bibinfo{person}{Yunfan Jiang}, \bibinfo{person}{Ajay Mandlekar}, \bibinfo{person}{Chaowei Xiao}, \bibinfo{person}{Yuke Zhu}, \bibinfo{person}{Linxi Fan}, {and} \bibinfo{person}{Anima Anandkumar}.} \bibinfo{year}{2023}\natexlab{f}.
\newblock \showarticletitle{Voyager: An open-ended embodied agent with large language models}.
\newblock \bibinfo{journal}{\emph{arXiv preprint arXiv:2305.16291}} (\bibinfo{year}{2023}).
\newblock


\bibitem[Wang et~al\mbox{.}(2024b)]%
        {wang2024unims}
\bibfield{author}{\bibinfo{person}{Hongru Wang}, \bibinfo{person}{Wenyu Huang}, \bibinfo{person}{Yang Deng}, \bibinfo{person}{Rui Wang}, \bibinfo{person}{Zezhong Wang}, \bibinfo{person}{Yufei Wang}, \bibinfo{person}{Fei Mi}, \bibinfo{person}{Jeff~Z Pan}, {and} \bibinfo{person}{Kam-Fai Wong}.} \bibinfo{year}{2024}\natexlab{b}.
\newblock \showarticletitle{Unims-rag: A unified multi-source retrieval-augmented generation for personalized dialogue systems}.
\newblock \bibinfo{journal}{\emph{arXiv preprint arXiv:2401.13256}} (\bibinfo{year}{2024}).
\newblock


\bibitem[Wang et~al\mbox{.}(2023d)]%
        {wang2023cue}
\bibfield{author}{\bibinfo{person}{Hongru Wang}, \bibinfo{person}{Rui Wang}, \bibinfo{person}{Fei Mi}, \bibinfo{person}{Yang Deng}, \bibinfo{person}{Zezhong Wang}, \bibinfo{person}{Bin Liang}, \bibinfo{person}{Ruifeng Xu}, {and} \bibinfo{person}{Kam-Fai Wong}.} \bibinfo{year}{2023}\natexlab{d}.
\newblock \showarticletitle{Cue-CoT: Chain-of-thought prompting for responding to in-depth dialogue questions with LLMs}.
\newblock \bibinfo{journal}{\emph{arXiv preprint arXiv:2305.11792}} (\bibinfo{year}{2023}).
\newblock


\bibitem[Wang et~al\mbox{.}(2023a)]%
        {wang2023target}
\bibfield{author}{\bibinfo{person}{Jian Wang}, \bibinfo{person}{Yi Cheng}, \bibinfo{person}{Dongding Lin}, \bibinfo{person}{Chak~Tou Leong}, {and} \bibinfo{person}{Wenjie Li}.} \bibinfo{year}{2023}\natexlab{a}.
\newblock \showarticletitle{Target-oriented proactive dialogue systems with personalization: Problem formulation and dataset curation}.
\newblock \bibinfo{journal}{\emph{arXiv preprint arXiv:2310.07397}} (\bibinfo{year}{2023}).
\newblock


\bibitem[Wang et~al\mbox{.}(2024d)]%
        {wang2024survey}
\bibfield{author}{\bibinfo{person}{Lei Wang}, \bibinfo{person}{Chen Ma}, \bibinfo{person}{Xueyang Feng}, \bibinfo{person}{Zeyu Zhang}, \bibinfo{person}{Hao Yang}, \bibinfo{person}{Jingsen Zhang}, \bibinfo{person}{Zhiyuan Chen}, \bibinfo{person}{Jiakai Tang}, \bibinfo{person}{Xu Chen}, \bibinfo{person}{Yankai Lin}, {et~al\mbox{.}}} \bibinfo{year}{2024}\natexlab{d}.
\newblock \showarticletitle{A survey on large language model based autonomous agents}.
\newblock \bibinfo{journal}{\emph{Frontiers of Computer Science}} \bibinfo{volume}{18}, \bibinfo{number}{6} (\bibinfo{year}{2024}), \bibinfo{pages}{186345}.
\newblock


\bibitem[Wang et~al\mbox{.}(2023g)]%
        {wang2023query2doc}
\bibfield{author}{\bibinfo{person}{Liang Wang}, \bibinfo{person}{Nan Yang}, {and} \bibinfo{person}{Furu Wei}.} \bibinfo{year}{2023}\natexlab{g}.
\newblock \showarticletitle{Query2doc: Query expansion with large language models}.
\newblock \bibinfo{journal}{\emph{arXiv preprint arXiv:2303.07678}} (\bibinfo{year}{2023}).
\newblock


\bibitem[Wang et~al\mbox{.}(2023h)]%
        {wang2023user}
\bibfield{author}{\bibinfo{person}{Lei Wang}, \bibinfo{person}{Jingsen Zhang}, \bibinfo{person}{Hao Yang}, \bibinfo{person}{Zhiyuan Chen}, \bibinfo{person}{Jiakai Tang}, \bibinfo{person}{Zeyu Zhang}, \bibinfo{person}{Xu Chen}, \bibinfo{person}{Yankai Lin}, \bibinfo{person}{Ruihua Song}, \bibinfo{person}{Wayne~Xin Zhao}, {et~al\mbox{.}}} \bibinfo{year}{2023}\natexlab{h}.
\newblock \showarticletitle{User behavior simulation with large language model based agents}.
\newblock \bibinfo{journal}{\emph{arXiv preprint arXiv:2306.02552}} (\bibinfo{year}{2023}).
\newblock


\bibitem[Wang et~al\mbox{.}(2023e)]%
        {wang2023incharacter}
\bibfield{author}{\bibinfo{person}{Xintao Wang}, \bibinfo{person}{Yunze Xiao}, \bibinfo{person}{Jen-tse Huang}, \bibinfo{person}{Siyu Yuan}, \bibinfo{person}{Rui Xu}, \bibinfo{person}{Haoran Guo}, \bibinfo{person}{Quan Tu}, \bibinfo{person}{Yaying Fei}, \bibinfo{person}{Ziang Leng}, \bibinfo{person}{Wei Wang}, {et~al\mbox{.}}} \bibinfo{year}{2023}\natexlab{e}.
\newblock \showarticletitle{Incharacter: Evaluating personality fidelity in role-playing agents through psychological interviews}.
\newblock \bibinfo{journal}{\emph{arXiv preprint arXiv:2310.17976}} (\bibinfo{year}{2023}).
\newblock


\bibitem[Wang et~al\mbox{.}(2024a)]%
        {wang2024investigating}
\bibfield{author}{\bibinfo{person}{Yixiao Wang}, \bibinfo{person}{Homa Fashandi}, {and} \bibinfo{person}{Kevin Ferreira}.} \bibinfo{year}{2024}\natexlab{a}.
\newblock \showarticletitle{Investigating the Personality Consistency in Quantized Role-Playing Dialogue Agents}. In \bibinfo{booktitle}{\emph{Proceedings of the 2024 Conference on Empirical Methods in Natural Language Processing: Industry Track}}. \bibinfo{pages}{239--255}.
\newblock


\bibitem[Wang et~al\mbox{.}({[n.\,d.]})]%
        {wangmemoryllm}
\bibfield{author}{\bibinfo{person}{Yu Wang}, \bibinfo{person}{Yifan Gao}, \bibinfo{person}{Xiusi Chen}, \bibinfo{person}{Haoming Jiang}, \bibinfo{person}{Shiyang Li}, \bibinfo{person}{Jingfeng Yang}, \bibinfo{person}{Qingyu Yin}, \bibinfo{person}{Zheng Li}, \bibinfo{person}{Xian Li}, \bibinfo{person}{Bing Yin}, {et~al\mbox{.}}} \bibinfo{year}{[n.\,d.]}\natexlab{}.
\newblock \showarticletitle{MEMORYLLM: Towards Self-Updatable Large Language Models}. In \bibinfo{booktitle}{\emph{Forty-first International Conference on Machine Learning}}.
\newblock


\bibitem[Wang et~al\mbox{.}(2024c)]%
        {wang2024crafting}
\bibfield{author}{\bibinfo{person}{Zheng Wang}, \bibinfo{person}{Zhongyang Li}, \bibinfo{person}{Zeren Jiang}, \bibinfo{person}{Dandan Tu}, {and} \bibinfo{person}{Wei Shi}.} \bibinfo{year}{2024}\natexlab{c}.
\newblock \showarticletitle{Crafting Personalized Agents through Retrieval-Augmented Generation on Editable Memory Graphs}. In \bibinfo{booktitle}{\emph{Proceedings of the 2024 Conference on Empirical Methods in Natural Language Processing}}. \bibinfo{pages}{4891--4906}.
\newblock


\bibitem[Wang and Chau(2024)]%
        {wang2024mememo}
\bibfield{author}{\bibinfo{person}{Zijie~J Wang} {and} \bibinfo{person}{Duen~Horng Chau}.} \bibinfo{year}{2024}\natexlab{}.
\newblock \showarticletitle{MeMemo: On-device Retrieval Augmentation for Private and Personalized Text Generation}. In \bibinfo{booktitle}{\emph{Proceedings of the 47th International ACM SIGIR Conference on Research and Development in Information Retrieval}}. \bibinfo{pages}{2765--2770}.
\newblock


\bibitem[Wang et~al\mbox{.}(2023c)]%
        {wang2023rolellm}
\bibfield{author}{\bibinfo{person}{Zekun~Moore Wang}, \bibinfo{person}{Zhongyuan Peng}, \bibinfo{person}{Haoran Que}, \bibinfo{person}{Jiaheng Liu}, \bibinfo{person}{Wangchunshu Zhou}, \bibinfo{person}{Yuhan Wu}, \bibinfo{person}{Hongcheng Guo}, \bibinfo{person}{Ruitong Gan}, \bibinfo{person}{Zehao Ni}, \bibinfo{person}{Jian Yang}, {et~al\mbox{.}}} \bibinfo{year}{2023}\natexlab{c}.
\newblock \showarticletitle{Rolellm: Benchmarking, eliciting, and enhancing role-playing abilities of large language models}.
\newblock \bibinfo{journal}{\emph{arXiv preprint arXiv:2310.00746}} (\bibinfo{year}{2023}).
\newblock


\bibitem[Wei et~al\mbox{.}(2024)]%
        {wei2024towards}
\bibfield{author}{\bibinfo{person}{Tianxin Wei}, \bibinfo{person}{Bowen Jin}, \bibinfo{person}{Ruirui Li}, \bibinfo{person}{Hansi Zeng}, \bibinfo{person}{Zhengyang Wang}, \bibinfo{person}{Jianhui Sun}, \bibinfo{person}{Qingyu Yin}, \bibinfo{person}{Hanqing Lu}, \bibinfo{person}{Suhang Wang}, \bibinfo{person}{Jingrui He}, {et~al\mbox{.}}} \bibinfo{year}{2024}\natexlab{}.
\newblock \showarticletitle{Towards unified multi-modal personalization: Large vision-language models for generative recommendation and beyond}.
\newblock \bibinfo{journal}{\emph{arXiv preprint arXiv:2403.10667}} (\bibinfo{year}{2024}).
\newblock


\bibitem[Wetzker et~al\mbox{.}(2008)]%
        {wetzker2008analyzing}
\bibfield{author}{\bibinfo{person}{Robert Wetzker}, \bibinfo{person}{Carsten Zimmermann}, {and} \bibinfo{person}{Christian Bauckhage}.} \bibinfo{year}{2008}\natexlab{}.
\newblock \showarticletitle{Analyzing social bookmarking systems: A del. icio. us cookbook}. In \bibinfo{booktitle}{\emph{Proceedings of the ECAI 2008 Mining Social Data Workshop}}. \bibinfo{pages}{26--30}.
\newblock


\bibitem[Wo{\'z}niak et~al\mbox{.}(2024)]%
        {wozniak2024personalized}
\bibfield{author}{\bibinfo{person}{Stanis{\l}aw Wo{\'z}niak}, \bibinfo{person}{Bart{\l}omiej Koptyra}, \bibinfo{person}{Arkadiusz Janz}, \bibinfo{person}{Przemys{\l}aw Kazienko}, {and} \bibinfo{person}{Jan Koco{\'n}}.} \bibinfo{year}{2024}\natexlab{}.
\newblock \showarticletitle{Personalized large language models}.
\newblock \bibinfo{journal}{\emph{arXiv preprint arXiv:2402.09269}} (\bibinfo{year}{2024}).
\newblock


\bibitem[Wu et~al\mbox{.}(2024)]%
        {wu2024medical}
\bibfield{author}{\bibinfo{person}{Junde Wu}, \bibinfo{person}{Jiayuan Zhu}, \bibinfo{person}{Yunli Qi}, \bibinfo{person}{Jingkun Chen}, \bibinfo{person}{Min Xu}, \bibinfo{person}{Filippo Menolascina}, {and} \bibinfo{person}{Vicente Grau}.} \bibinfo{year}{2024}\natexlab{}.
\newblock \showarticletitle{Medical graph rag: Towards safe medical large language model via graph retrieval-augmented generation}.
\newblock \bibinfo{journal}{\emph{arXiv preprint arXiv:2408.04187}} (\bibinfo{year}{2024}).
\newblock


\bibitem[Wu et~al\mbox{.}(2017)]%
        {wu2017personalized}
\bibfield{author}{\bibinfo{person}{Xuan Wu}, \bibinfo{person}{Dong Zhou}, \bibinfo{person}{Yu Xu}, {and} \bibinfo{person}{S{\'e}amus Lawless}.} \bibinfo{year}{2017}\natexlab{}.
\newblock \showarticletitle{Personalized query expansion utilizing multi-relational social data}. In \bibinfo{booktitle}{\emph{2017 12th International Workshop on Semantic and Social Media Adaptation and Personalization (SMAP)}}. IEEE, \bibinfo{pages}{65--70}.
\newblock


\bibitem[Xi et~al\mbox{.}(2024)]%
        {xi2024towards}
\bibfield{author}{\bibinfo{person}{Yunjia Xi}, \bibinfo{person}{Weiwen Liu}, \bibinfo{person}{Jianghao Lin}, \bibinfo{person}{Xiaoling Cai}, \bibinfo{person}{Hong Zhu}, \bibinfo{person}{Jieming Zhu}, \bibinfo{person}{Bo Chen}, \bibinfo{person}{Ruiming Tang}, \bibinfo{person}{Weinan Zhang}, {and} \bibinfo{person}{Yong Yu}.} \bibinfo{year}{2024}\natexlab{}.
\newblock \showarticletitle{Towards open-world recommendation with knowledge augmentation from large language models}. In \bibinfo{booktitle}{\emph{Proceedings of the 18th ACM Conference on Recommender Systems}}. \bibinfo{pages}{12--22}.
\newblock


\bibitem[Xi et~al\mbox{.}(2025)]%
        {xi2025rise}
\bibfield{author}{\bibinfo{person}{Zhiheng Xi}, \bibinfo{person}{Wenxiang Chen}, \bibinfo{person}{Xin Guo}, \bibinfo{person}{Wei He}, \bibinfo{person}{Yiwen Ding}, \bibinfo{person}{Boyang Hong}, \bibinfo{person}{Ming Zhang}, \bibinfo{person}{Junzhe Wang}, \bibinfo{person}{Senjie Jin}, \bibinfo{person}{Enyu Zhou}, {et~al\mbox{.}}} \bibinfo{year}{2025}\natexlab{}.
\newblock \showarticletitle{The rise and potential of large language model based agents: A survey}.
\newblock \bibinfo{journal}{\emph{Science China Information Sciences}} \bibinfo{volume}{68}, \bibinfo{number}{2} (\bibinfo{year}{2025}), \bibinfo{pages}{121101}.
\newblock


\bibitem[Xiao et~al\mbox{.}(2024)]%
        {xiao2024c}
\bibfield{author}{\bibinfo{person}{Shitao Xiao}, \bibinfo{person}{Zheng Liu}, \bibinfo{person}{Peitian Zhang}, \bibinfo{person}{Niklas Muennighoff}, \bibinfo{person}{Defu Lian}, {and} \bibinfo{person}{Jian-Yun Nie}.} \bibinfo{year}{2024}\natexlab{}.
\newblock \showarticletitle{C-pack: Packed resources for general chinese embeddings}. In \bibinfo{booktitle}{\emph{Proceedings of the 47th international ACM SIGIR conference on research and development in information retrieval}}. \bibinfo{pages}{641--649}.
\newblock


\bibitem[Xu et~al\mbox{.}(2024)]%
        {xu2024penetrative}
\bibfield{author}{\bibinfo{person}{Huatao Xu}, \bibinfo{person}{Liying Han}, \bibinfo{person}{Qirui Yang}, \bibinfo{person}{Mo Li}, {and} \bibinfo{person}{Mani Srivastava}.} \bibinfo{year}{2024}\natexlab{}.
\newblock \showarticletitle{Penetrative ai: Making llms comprehend the physical world}. In \bibinfo{booktitle}{\emph{Proceedings of the 25th International Workshop on Mobile Computing Systems and Applications}}. \bibinfo{pages}{1--7}.
\newblock


\bibitem[Xu et~al\mbox{.}(2021)]%
        {xu2021transformer}
\bibfield{author}{\bibinfo{person}{Hongyan Xu}, \bibinfo{person}{Hongtao Liu}, \bibinfo{person}{Pengfei Jiao}, {and} \bibinfo{person}{Wenjun Wang}.} \bibinfo{year}{2021}\natexlab{}.
\newblock \showarticletitle{Transformer reasoning network for personalized review summarization}. In \bibinfo{booktitle}{\emph{Proceedings of the 44th International ACM SIGIR Conference on Research and Development in Information Retrieval}}. \bibinfo{pages}{1452--1461}.
\newblock


\bibitem[Xu et~al\mbox{.}(2022)]%
        {xu2022long}
\bibfield{author}{\bibinfo{person}{Xinchao Xu}, \bibinfo{person}{Zhibin Gou}, \bibinfo{person}{Wenquan Wu}, \bibinfo{person}{Zheng-Yu Niu}, \bibinfo{person}{Hua Wu}, \bibinfo{person}{Haifeng Wang}, {and} \bibinfo{person}{Shihang Wang}.} \bibinfo{year}{2022}\natexlab{}.
\newblock \showarticletitle{Long time no see! open-domain conversation with long-term persona memory}.
\newblock \bibinfo{journal}{\emph{arXiv preprint arXiv:2203.05797}} (\bibinfo{year}{2022}).
\newblock


\bibitem[Xu et~al\mbox{.}(2025)]%
        {xu2025personalized}
\bibfield{author}{\bibinfo{person}{Yiyan Xu}, \bibinfo{person}{Jinghao Zhang}, \bibinfo{person}{Alireza Salemi}, \bibinfo{person}{Xinting Hu}, \bibinfo{person}{Wenjie Wang}, \bibinfo{person}{Fuli Feng}, \bibinfo{person}{Hamed Zamani}, \bibinfo{person}{Xiangnan He}, {and} \bibinfo{person}{Tat-Seng Chua}.} \bibinfo{year}{2025}\natexlab{}.
\newblock \showarticletitle{Personalized Generation In Large Model Era: A Survey}.
\newblock \bibinfo{journal}{\emph{arXiv preprint arXiv:2503.02614}} (\bibinfo{year}{2025}).
\newblock


\bibitem[Yu et~al\mbox{.}(2024b)]%
        {yu2024personalized}
\bibfield{author}{\bibinfo{person}{Hao Yu}, \bibinfo{person}{Xin Yang}, \bibinfo{person}{Xin Gao}, \bibinfo{person}{Yan Kang}, \bibinfo{person}{Hao Wang}, \bibinfo{person}{Junbo Zhang}, {and} \bibinfo{person}{Tianrui Li}.} \bibinfo{year}{2024}\natexlab{b}.
\newblock \showarticletitle{Personalized federated continual learning via multi-granularity prompt}. In \bibinfo{booktitle}{\emph{Proceedings of the 30th ACM SIGKDD Conference on Knowledge Discovery and Data Mining}}. \bibinfo{pages}{4023--4034}.
\newblock


\bibitem[Yu et~al\mbox{.}(2024a)]%
        {yu2024neeko}
\bibfield{author}{\bibinfo{person}{Xiaoyan Yu}, \bibinfo{person}{Tongxu Luo}, \bibinfo{person}{Yifan Wei}, \bibinfo{person}{Fangyu Lei}, \bibinfo{person}{Yiming Huang}, \bibinfo{person}{Hao Peng}, {and} \bibinfo{person}{Liehuang Zhu}.} \bibinfo{year}{2024}\natexlab{a}.
\newblock \showarticletitle{Neeko: Leveraging dynamic lora for efficient multi-character role-playing agent}.
\newblock \bibinfo{journal}{\emph{arXiv preprint arXiv:2402.13717}} (\bibinfo{year}{2024}).
\newblock


\bibitem[Yuan et~al\mbox{.}(2024)]%
        {yuan2024evaluating}
\bibfield{author}{\bibinfo{person}{Xinfeng Yuan}, \bibinfo{person}{Siyu Yuan}, \bibinfo{person}{Yuhan Cui}, \bibinfo{person}{Tianhe Lin}, \bibinfo{person}{Xintao Wang}, \bibinfo{person}{Rui Xu}, \bibinfo{person}{Jiangjie Chen}, {and} \bibinfo{person}{Deqing Yang}.} \bibinfo{year}{2024}\natexlab{}.
\newblock \showarticletitle{Evaluating character understanding of large language models via character profiling from fictional works}.
\newblock \bibinfo{journal}{\emph{arXiv preprint arXiv:2404.12726}} (\bibinfo{year}{2024}).
\newblock


\bibitem[Zeng et~al\mbox{.}(2023)]%
        {zeng2023personalized}
\bibfield{author}{\bibinfo{person}{Hansi Zeng}, \bibinfo{person}{Surya Kallumadi}, \bibinfo{person}{Zaid Alibadi}, \bibinfo{person}{Rodrigo Nogueira}, {and} \bibinfo{person}{Hamed Zamani}.} \bibinfo{year}{2023}\natexlab{}.
\newblock \showarticletitle{A personalized dense retrieval framework for unified information access}. In \bibinfo{booktitle}{\emph{Proceedings of the 46th International ACM SIGIR Conference on Research and Development in Information Retrieval}}. \bibinfo{pages}{121--130}.
\newblock


\bibitem[Zerhoudi and Granitzer(2024)]%
        {zerhoudi2024personarag}
\bibfield{author}{\bibinfo{person}{Saber Zerhoudi} {and} \bibinfo{person}{Michael Granitzer}.} \bibinfo{year}{2024}\natexlab{}.
\newblock \showarticletitle{PersonaRAG: Enhancing Retrieval-Augmented Generation Systems with User-Centric Agents}.
\newblock \bibinfo{journal}{\emph{arXiv preprint arXiv:2407.09394}} (\bibinfo{year}{2024}).
\newblock


\bibitem[Zhang et~al\mbox{.}(2020)]%
        {zhang2020towards}
\bibfield{author}{\bibinfo{person}{Han Zhang}, \bibinfo{person}{Songlin Wang}, \bibinfo{person}{Kang Zhang}, \bibinfo{person}{Zhiling Tang}, \bibinfo{person}{Yunjiang Jiang}, \bibinfo{person}{Yun Xiao}, \bibinfo{person}{Weipeng Yan}, {and} \bibinfo{person}{Wen-Yun Yang}.} \bibinfo{year}{2020}\natexlab{}.
\newblock \showarticletitle{Towards personalized and semantic retrieval: An end-to-end solution for e-commerce search via embedding learning}. In \bibinfo{booktitle}{\emph{Proceedings of the 43rd International ACM SIGIR Conference on Research and Development in Information Retrieval}}. \bibinfo{pages}{2407--2416}.
\newblock


\bibitem[Zhang(2024)]%
        {zhang2024guided}
\bibfield{author}{\bibinfo{person}{Jiarui Zhang}.} \bibinfo{year}{2024}\natexlab{}.
\newblock \showarticletitle{Guided profile generation improves personalization with llms}.
\newblock \bibinfo{journal}{\emph{arXiv preprint arXiv:2409.13093}} (\bibinfo{year}{2024}).
\newblock


\bibitem[Zhang et~al\mbox{.}({[n.\,d.]})]%
        {zhangbootstrap}
\bibfield{author}{\bibinfo{person}{Jesse Zhang}, \bibinfo{person}{Jiahui Zhang}, \bibinfo{person}{Karl Pertsch}, \bibinfo{person}{Ziyi Liu}, \bibinfo{person}{Xiang Ren}, \bibinfo{person}{Minsuk Chang}, \bibinfo{person}{Shao-Hua Sun}, {and} \bibinfo{person}{Joseph~J Lim}.} \bibinfo{year}{[n.\,d.]}\natexlab{}.
\newblock \showarticletitle{Bootstrap Your Own Skills: Learning to Solve New Tasks with Large Language Model Guidance}. In \bibinfo{booktitle}{\emph{7th Annual Conference on Robot Learning}}.
\newblock


\bibitem[Zhang et~al\mbox{.}(2023a)]%
        {zhang2023llm}
\bibfield{author}{\bibinfo{person}{Kai Zhang}, \bibinfo{person}{Yangyang Kang}, \bibinfo{person}{Fubang Zhao}, {and} \bibinfo{person}{Xiaozhong Liu}.} \bibinfo{year}{2023}\natexlab{a}.
\newblock \showarticletitle{LLM-based medical assistant personalization with short-and long-term memory coordination}.
\newblock \bibinfo{journal}{\emph{arXiv preprint arXiv:2309.11696}} (\bibinfo{year}{2023}).
\newblock


\bibitem[Zhang et~al\mbox{.}(2024c)]%
        {zhang2024cogenesis}
\bibfield{author}{\bibinfo{person}{Kaiyan Zhang}, \bibinfo{person}{Jianyu Wang}, \bibinfo{person}{Ermo Hua}, \bibinfo{person}{Biqing Qi}, \bibinfo{person}{Ning Ding}, {and} \bibinfo{person}{Bowen Zhou}.} \bibinfo{year}{2024}\natexlab{c}.
\newblock \showarticletitle{Cogenesis: A framework collaborating large and small language models for secure context-aware instruction following}.
\newblock \bibinfo{journal}{\emph{arXiv preprint arXiv:2403.03129}} (\bibinfo{year}{2024}).
\newblock


\bibitem[Zhang et~al\mbox{.}(2023b)]%
        {zhang2023memory}
\bibfield{author}{\bibinfo{person}{Kai Zhang}, \bibinfo{person}{Fubang Zhao}, \bibinfo{person}{Yangyang Kang}, {and} \bibinfo{person}{Xiaozhong Liu}.} \bibinfo{year}{2023}\natexlab{b}.
\newblock \showarticletitle{Memory-augmented llm personalization with short-and long-term memory coordination}.
\newblock \bibinfo{journal}{\emph{arXiv preprint arXiv:2309.11696}} (\bibinfo{year}{2023}).
\newblock


\bibitem[Zhang et~al\mbox{.}(2025b)]%
        {zhang2025llmtreerec}
\bibfield{author}{\bibinfo{person}{Wenlin Zhang}, \bibinfo{person}{Chuhan Wu}, \bibinfo{person}{Xiangyang Li}, \bibinfo{person}{Yuhao Wang}, \bibinfo{person}{Kuicai Dong}, \bibinfo{person}{Yichao Wang}, \bibinfo{person}{Xinyi Dai}, \bibinfo{person}{Xiangyu Zhao}, \bibinfo{person}{Huifeng Guo}, {and} \bibinfo{person}{Ruiming Tang}.} \bibinfo{year}{2025}\natexlab{b}.
\newblock \showarticletitle{LLMTreeRec: Unleashing the Power of Large Language Models for Cold-Start Recommendations}. In \bibinfo{booktitle}{\emph{Proceedings of the 31st International Conference on Computational Linguistics}}. \bibinfo{pages}{886--896}.
\newblock


\bibitem[Zhang et~al\mbox{.}(2025a)]%
        {zhang2025rehearse}
\bibfield{author}{\bibinfo{person}{Yanyue Zhang}, \bibinfo{person}{Yulan He}, {and} \bibinfo{person}{Deyu Zhou}.} \bibinfo{year}{2025}\natexlab{a}.
\newblock \showarticletitle{Rehearse With User: Personalized Opinion Summarization via Role-Playing based on Large Language Models}.
\newblock \bibinfo{journal}{\emph{arXiv preprint arXiv:2503.00449}} (\bibinfo{year}{2025}).
\newblock


\bibitem[Zhang et~al\mbox{.}(2024d)]%
        {zhang2024personalized}
\bibfield{author}{\bibinfo{person}{You Zhang}, \bibinfo{person}{Jin Wang}, \bibinfo{person}{Liang-Chih Yu}, \bibinfo{person}{Dan Xu}, {and} \bibinfo{person}{Xuejie Zhang}.} \bibinfo{year}{2024}\natexlab{d}.
\newblock \showarticletitle{Personalized LoRA for human-centered text understanding}. In \bibinfo{booktitle}{\emph{Proceedings of the AAAI Conference on Artificial Intelligence}}, Vol.~\bibinfo{volume}{38}. \bibinfo{pages}{19588--19596}.
\newblock


\bibitem[Zhang et~al\mbox{.}(2024e)]%
        {zhang2024recgpt}
\bibfield{author}{\bibinfo{person}{Yabin Zhang}, \bibinfo{person}{Wenhui Yu}, \bibinfo{person}{Erhan Zhang}, \bibinfo{person}{Xu Chen}, \bibinfo{person}{Lantao Hu}, \bibinfo{person}{Peng Jiang}, {and} \bibinfo{person}{Kun Gai}.} \bibinfo{year}{2024}\natexlab{e}.
\newblock \showarticletitle{Recgpt: Generative personalized prompts for sequential recommendation via chatgpt training paradigm}.
\newblock \bibinfo{journal}{\emph{arXiv preprint arXiv:2404.08675}} (\bibinfo{year}{2024}).
\newblock


\bibitem[Zhang et~al\mbox{.}(2024a)]%
        {zhang2024survey}
\bibfield{author}{\bibinfo{person}{Zeyu Zhang}, \bibinfo{person}{Xiaohe Bo}, \bibinfo{person}{Chen Ma}, \bibinfo{person}{Rui Li}, \bibinfo{person}{Xu Chen}, \bibinfo{person}{Quanyu Dai}, \bibinfo{person}{Jieming Zhu}, \bibinfo{person}{Zhenhua Dong}, {and} \bibinfo{person}{Ji-Rong Wen}.} \bibinfo{year}{2024}\natexlab{a}.
\newblock \showarticletitle{A survey on the memory mechanism of large language model based agents}.
\newblock \bibinfo{journal}{\emph{arXiv preprint arXiv:2404.13501}} (\bibinfo{year}{2024}).
\newblock


\bibitem[Zhang et~al\mbox{.}(2024b)]%
        {zhang2024personalization}
\bibfield{author}{\bibinfo{person}{Zhehao Zhang}, \bibinfo{person}{Ryan~A Rossi}, \bibinfo{person}{Branislav Kveton}, \bibinfo{person}{Yijia Shao}, \bibinfo{person}{Diyi Yang}, \bibinfo{person}{Hamed Zamani}, \bibinfo{person}{Franck Dernoncourt}, \bibinfo{person}{Joe Barrow}, \bibinfo{person}{Tong Yu}, \bibinfo{person}{Sungchul Kim}, {et~al\mbox{.}}} \bibinfo{year}{2024}\natexlab{b}.
\newblock \showarticletitle{Personalization of large language models: A survey}.
\newblock \bibinfo{journal}{\emph{arXiv preprint arXiv:2411.00027}} (\bibinfo{year}{2024}).
\newblock


\bibitem[Zheng et~al\mbox{.}(2023)]%
        {zheng2023agents}
\bibfield{author}{\bibinfo{person}{Yi Zheng}, \bibinfo{person}{Chongyang Ma}, \bibinfo{person}{Kanle Shi}, {and} \bibinfo{person}{Haibin Huang}.} \bibinfo{year}{2023}\natexlab{}.
\newblock \showarticletitle{Agents meet okr: An object and key results driven agent system with hierarchical self-collaboration and self-evaluation}.
\newblock \bibinfo{journal}{\emph{arXiv preprint arXiv:2311.16542}} (\bibinfo{year}{2023}).
\newblock


\bibitem[Zhong et~al\mbox{.}(2022)]%
        {zhong2022less}
\bibfield{author}{\bibinfo{person}{Hanxun Zhong}, \bibinfo{person}{Zhicheng Dou}, \bibinfo{person}{Yutao Zhu}, \bibinfo{person}{Hongjin Qian}, {and} \bibinfo{person}{Ji-Rong Wen}.} \bibinfo{year}{2022}\natexlab{}.
\newblock \showarticletitle{Less is more: Learning to refine dialogue history for personalized dialogue generation}.
\newblock \bibinfo{journal}{\emph{arXiv preprint arXiv:2204.08128}} (\bibinfo{year}{2022}).
\newblock


\bibitem[Zhong et~al\mbox{.}(2021)]%
        {zhong2021useradapter}
\bibfield{author}{\bibinfo{person}{Wanjun Zhong}, \bibinfo{person}{Duyu Tang}, \bibinfo{person}{Jiahai Wang}, \bibinfo{person}{Jian Yin}, {and} \bibinfo{person}{Nan Duan}.} \bibinfo{year}{2021}\natexlab{}.
\newblock \showarticletitle{UserAdapter: Few-shot user learning in sentiment analysis}. In \bibinfo{booktitle}{\emph{Findings of the Association for Computational Linguistics: ACL-IJCNLP 2021}}. \bibinfo{pages}{1484--1488}.
\newblock


\bibitem[Zhou et~al\mbox{.}(2012)]%
        {zhou2012improving}
\bibfield{author}{\bibinfo{person}{Dong Zhou}, \bibinfo{person}{S{\'e}amus Lawless}, {and} \bibinfo{person}{Vincent Wade}.} \bibinfo{year}{2012}\natexlab{}.
\newblock \showarticletitle{Improving search via personalized query expansion using social media}.
\newblock \bibinfo{journal}{\emph{Information retrieval}}  \bibinfo{volume}{15} (\bibinfo{year}{2012}), \bibinfo{pages}{218--242}.
\newblock


\bibitem[Zhou et~al\mbox{.}(2022)]%
        {zhou2022least}
\bibfield{author}{\bibinfo{person}{Denny Zhou}, \bibinfo{person}{Nathanael Sch{\"a}rli}, \bibinfo{person}{Le Hou}, \bibinfo{person}{Jason Wei}, \bibinfo{person}{Nathan Scales}, \bibinfo{person}{Xuezhi Wang}, \bibinfo{person}{Dale Schuurmans}, \bibinfo{person}{Claire Cui}, \bibinfo{person}{Olivier Bousquet}, \bibinfo{person}{Quoc Le}, {et~al\mbox{.}}} \bibinfo{year}{2022}\natexlab{}.
\newblock \showarticletitle{Least-to-most prompting enables complex reasoning in large language models}.
\newblock \bibinfo{journal}{\emph{arXiv preprint arXiv:2205.10625}} (\bibinfo{year}{2022}).
\newblock


\bibitem[Zhou et~al\mbox{.}(2024)]%
        {zhou2024cognitive}
\bibfield{author}{\bibinfo{person}{Yujia Zhou}, \bibinfo{person}{Qiannan Zhu}, \bibinfo{person}{Jiajie Jin}, {and} \bibinfo{person}{Zhicheng Dou}.} \bibinfo{year}{2024}\natexlab{}.
\newblock \showarticletitle{Cognitive personalized search integrating large language models with an efficient memory mechanism}. In \bibinfo{booktitle}{\emph{Proceedings of the ACM Web Conference 2024}}. \bibinfo{pages}{1464--1473}.
\newblock


\bibitem[Zhuang et~al\mbox{.}({[n.\,d.]})]%
        {zhuang2406hydra}
\bibfield{author}{\bibinfo{person}{Yuchen Zhuang}, \bibinfo{person}{Haotian Sun}, \bibinfo{person}{Yue Yu}, \bibinfo{person}{Rushi Qiang}, \bibinfo{person}{Qifan Wang}, \bibinfo{person}{Chao Zhang}, {and} \bibinfo{person}{Bo Dai}.} \bibinfo{year}{[n.\,d.]}\natexlab{}.
\newblock \showarticletitle{Hydra: Model factorization framework for black-box llm personalization, 2024}.
\newblock \bibinfo{journal}{\emph{URL https://arxiv. org/abs/2406.02888}} (\bibinfo{year}{[n.\,d.]}).
\newblock


\end{thebibliography}

\end{document}